\documentclass[prd,aps,floats,epsfig,eqsecnum,nofootinbib]{revtex4}
\usepackage{array,amsmath,amssymb,verbatim,epsfig,rotating}
\usepackage{bm}
\begin{document}
\def\E{{\cal E}}
\def\H{{\hat{\cal H}}}
\def\hphi{{\hat\phi}}
\def\dm{\partial_\mu}
\def\udm{\partial^\mu}
\def\dk{\frac{dk}{2\pi}}
\def\bk{\overline{k}}
\def\Teff{T_{\rm eff}(t)}
\def\Tpeff{T_{\rm p,eff}(t)}
\newcommand{\be}{\begin{equation}}
\newcommand{\ee}{\end{equation}}
\newcommand{\bea}{\begin{eqnarray}}
\newcommand{\eea}{\end{eqnarray}}
\newcommand{\avg}[1]{\langle #1 \rangle}
\newcommand{\Avg}[1]{\Big\langle #1 \Big\rangle}
\newcommand{\rotatefigure}[2]
\date{}
\title{The approach to thermalization in the classical $\phi^4$ theory in $
1+1 $ dimensions:\\ energy cascades and universal scaling.}
\author{D. Boyanovsky$^{(a)}$}
\email{boyan@pitt.edu}
\author{C. Destri$^{(b)}$}
\email{Claudio.Destri@mib.infn.it}
\author{H. J. de Vega$^{(c)}$}
\email{devega@lpthe.jussieu.fr} \affiliation{ $^{(a)}$ Department
of Physics and Astronomy,~~University of Pittsburgh,
Pittsburgh, Pennsylvania 15260, USA\\
$^{(b)}$ Dipartimento di Fisica G. Occhialini, Universit\`a
Milano-Bicocca Piazza della Scienza 3, 20126 Milano and
INFN, sezione di Milano, via Celoria 16, Milano Italia\\
$^{(c)}$ LPTHE, Universit\'e Pierre~et~Marie Curie,~~Paris VI et
Denis Diderot, Paris VII, Tour 16, 1er.~~\'etage, 4, Place
Jussieu, 75252 Paris, Cedex 05, France}
\begin{abstract}
  We study the dynamics of thermalization and the approach to equilibrium
  in the classical $\phi^4$ theory in $1+1$ spacetime dimensions. At
  thermal equilibrium we exploit the equivalence between the classical
  canonical averages and transfer matrix quantum traces of the anharmonic
  oscillator to obtain {\bf exact} results for the temperature dependence
  of several observables, which provide a set of criteria for
  thermalization.  In this context, comparing to the \emph{exact
  results} we find that the Hartree approximation is remarkably
  accurate in equilibrium. The non-equilibrium dynamics is studied by
  numerically solving the equations of motion in light--cone coordinates
  for a broad range of initial conditions and energy densities. The long
  time evolution is described by several distinct stages, all characterized
  by a cascade of energy towards the ultraviolet. After an initial
  transient stage, the spatio--temporal gradient terms become larger than
  the nonlinear term, and there emerges a stage of universal cascade. This
  cascade starts at a time scale $t_0$ independent of the initial
  conditions (except for very low energy density). During this stage the
  power spectra feature universal scaling behavior and the front of the
  cascade $\bk(t)$ moves to the ultraviolet as a power law $\bk(t) \sim
  t^{\alpha}$ with $\alpha \lesssim 0.25$ an exponent weakly dependent on
  the energy density alone. The wake behind the cascade is described as a
  state of {\bf Local Thermodynamic Equilibrium} (LTE) with all correlations
  being determined by the equilibrium functional form with an effective
  time dependent temperature $\Teff$, which slowly decreases with time
  as $\sim t^{-\alpha}$. Two well separated time scales emerge: while
  $\Teff$ varies slowly, the wavectors in the wake with $k < \bk(t)$ attain
  LTE on much shorter time scales.  This universal scaling stage ends when
  the front of the ultraviolet cascade reaches the cutoff at a time scale
  $t_1 \sim a^{-\frac{1}{\alpha}}$. Virialization starts to set much
  earlier than LTE.  We find that strict thermalization is achieved
  only for an infinite time scale.
\end{abstract}
\date{\today}
\maketitle \tableofcontents
\section{Introduction}
The dynamics of thermalization and relaxation in a field theory is
a topic of much current interest both in early cosmology as well
as in ultrarelativistic heavy ion collisions, where the current
experimental program at RHIC and the forthcoming LHC will search
for a new state of matter, the quark gluon plasma.

Such current interest on the non-equilibrium dynamics of
relaxation and thermalization within the settings of cosmology and
heavy ion collisions in fact rekindles the issue of thermalization
in non-linear field theories. Pioneering work in this area was
initiated by  Fermi, Pasta and Ulam\cite{FPU} (FPU) who were the
first to address the question of ergodicity focusing on the
non-linear dynamics in discrete chains of anharmonic oscillators
using one of the first computers. Since then this problem has been
studied within a variety of models\cite{varios,gior} with the goal
of answering fundamental questions on ergodicity, equipartition
and in general the approach to equilibrium in non-linear theories
with a large but finite number of degrees of freedom.

Within both contexts a detailed understanding of the dynamics and
the time scales for thermalization are extremely important. In
cosmology the inflationary paradigm currently being tested by
precise observations of the cosmic microwave background
\emph{assumes} that after the inflationary stage a period of
particle production and relaxation leads to a state of local
thermal equilibrium thus merging inflation with the standard hot
big bang cosmology\cite{kolbook,lidlyt}.

In ultrarelativistic heavy ion collisions, the current theoretical
understanding suggests that after the nucleus-nucleus collision,
partons (mainly gluons) are liberated and parton-parton scattering
leads to a state of thermal equilibrium\cite{mueller}. It has been
recently argued\cite{baier} that there are several different
stages that ultimately lead to thermalization after a
nucleus-nucleus collision. During most of the non-equilibrium
evolution the dynamics is determined by the \emph{classical}
Yang-Mills equations because the gluon distribution function is
non-perturbatively large $\sim 1/\alpha_s$  up to a saturation
scale\cite{mueller,baier}. This observation motivates a numerical
program to study the early stages of ultrarelativistic heavy ion
collisions in terms of the non-equilibrium dynamics of
\emph{classical} Yang-Mills fields\cite{venugopalan}.

Inflationary scenarios lead to particle production either via
parametric amplification of fluctuations (in the case of an
oscillating inflaton)  of the inflaton or by spinodal
instabilities during phase transitions\cite{linde,ours,patkos}. In
both cases the non-equilibrium dynamics is non-perturbative and
results in a large population of soft quanta whose dynamics is
nearly classical. The non-equilibrium dynamics of particle
production and eventual thermalization are non-perturbative and
the resulting fluctuations contribute to the evolution of the
scale factor, namely the backreaction from the fluctuations
becomes important in the evolution of the cosmological
space-time\cite{ours}.  Both parametric amplification or spinodal
decomposition lead to non-perturbative particle production of a
band of wavevectors, typically for soft
momenta\cite{linde,ours,patkos}. This non-perturbatively large
population allows a classical treatment of the non-equilibrium
evolution.

More recently and again motivated by cosmology and heavy ion
collisions a fairly intense effort has been devoted to studying
thermalization and in general relaxation in classical as well as
quantum field theories\cite{smitaarts}-\cite{marcelo}. Several
approximate methods have been implemented, from homogeneous and
inhomogeneous mean field (Hartree)\cite{smitaarts} to a variety of
schemes that include quantum correlations after suitable
truncations of the hierarchy of
equations\cite{wett,aarts,berges,cooper,baacke} and comparisons
between them. A numerical study of the non-equilibrium dynamics in
the \emph{classical} $\phi^4$ field theory in $1+1$ dimensions and
a comparison of the classical evolution to that obtained from
several approximate schemes has been reported in
refs.\cite{wett,cooper,baacke}.

In this article we study the non-equilibrium dynamics leading to
thermalization in the classical $\phi^4$ theory in $1+1$
dimensions. As mentioned above the initial stages of
non-equilibrium dynamics either in cosmology or in
ultrarelativistic heavy ion collisions is mainly \emph{classical}.
Classical field theory must be understood with an ultraviolet
cutoff, or equivalently an underlying lattice spacing, to avoid
the Rayleigh-Jeans catastrophe. Both in cosmology as well as in
heavy ion collisions there are natural ultraviolet cutoffs: in
cosmology it is the Planck scale (although the distribution of
particles produced during inflation is typically concentrated at
much smaller comoving scales in wavenumbers\cite{ours}), in
ultrarelativistic heavy ion collision it is the saturation
scale\cite{mueller}.

Our study of the non-equilibrium dynamics in $1+1$ dimensional
classical field theory is \emph{inspired} by the description of
the early stages of the non-equilibrium dynamics in
ultrarelativistic heavy ion collisions. The study of the
non-equilibrium dynamics of a classical field theory is
interesting and important on its own right and the $1+1$
dimensional case is the simplest scenario.

 The advantage of studying
the classical field theory is that the equation of motion can be
solved exactly.

In fact this is the main reason why it has been studied previously
as a testing ground for different approximation schemes.

\bigskip

\textbf{The goals of this article:}

\bigskip

Our motivation, focus and goals in this article are very different
from those of previous studies of the classical $\phi^4$ theory in
$1+1$ dimensions reported in
refs.\cite{varios,gior,wett,cooper,baacke}. While
refs.\cite{varios,gior,wett,cooper,baacke} studied whether
ergodicity and thermalization are achieved and tested a variety of
approximate schemes useful in the quantum
case\cite{wett,cooper,baacke}, we seek to provide a more detailed
understanding of the \emph{main dynamical mechanisms} that lead to
thermalization and to study the approach to thermalization by
several different observables. We address several questions on the
equilibrium and non-equilibrium aspects: i) what are the criteria
for thermalization in an interacting theory?, ii) what is the
mechanism that leads to thermalization?, iii) what is the dynamics
for different observables and how they reach thermal equilibrium?.

We focus our study on these issues within the context of classical
field theory, which is an interesting and timely problem all by
itself. It is also \emph{likely} to describe the initial stages of
evolution strongly out of equilibrium at least for the relevant
cases of very large occupation numbers.

\bigskip

\textbf{Main results:} The main results of this work are the
following:

\begin{itemize}
\item{We obtain a series of {\bf exact} results in thermal
equilibrium by
    exploiting the equivalence between the classical equilibrium canonical
    averages and the quantum traces of the transfer matrix of the
    anharmonic oscillator\cite{fey}. A body of results on the spectrum and
    matrix elements of the quantum--mechanical anharmonic
    oscillator\cite{ind,hioe} as well as new calculations
    allows us to extract exact results for high, intermediate and low
    temperatures for the observables. We compare these exact results to the
    Hartree approximation for high and low temperatures and we find a
    remarkable agreement. These exact results provide a set of stringent
    criteria for thermalization.}

\item{We implement a light cone method to study the classical
evolution
    which is very accurate and stable, maintains the underlying Lorentz
    symmetries and conserves energy exactly (that is to machine accuracy on
    a computer). }

\item{For a wide range of initial conditions and energy densities
we find
    several distinct stages in the dynamics: the first stage corresponds to
    initial transients with large fluctuations during which energy is
    transferred from low momentum to higher momentum modes. This first
    stage ends at a time scale $ t_i \sim 200 $ in dimensionless
    units. [That is, choosing both the mass and the coupling
    equal one, as one can always do in classical field theory]. This
    value for $ t_i $ turns out to be independent of the
    lattice spacing and the initial conditions provided the energy
    density $ E/L $ is not too small: $ E/L
    \gtrsim 10 $. After a crossover between the $\phi^4$ and the
    $\phi'^2$ terms, a second stage  emerges during which
energy transfer from low to high
    momentum modes becomes very effective and results in an {\bf
      ultraviolet cascade}, namely the power spectrum of the field and its
    canonical momentum acquire support for ever larger values of the
    wavevectors. This ultraviolet cascade leads to a very efficient
    transfer of energy between the interaction term $\phi^4$, the $\phi^2$
    and the $\pi^2$ terms which all diminish and the spatial gradient term
    $\phi'^2$ which grows. After a time scale $ t_0 \sim 50000 $
a third stage emerges during which the {\bf
      ultraviolet cascade} features a {\bf universal behavior} independent
    of the lattice spacing and the initial conditions. During this third
    stage the transfer of energy to higher momentum modes, namely the
    cascade, develops a front $\bk(t)$ which moves towards the ultraviolet
    cutoff and leaves in its wake a state of {\bf local thermodynamic
      equilibrium} characterized by an effective temperature $\Teff$. That
    is, two well separated time scales emerge : a {\bf short} time scale
    leading to a state of local thermodynamic equilibrium with an effective
    temperature that decreases and approaches the equilibrium value on a
    {\bf long} time scale. While $\Teff$ varies slowly with time the
    $k$-modes below the front of the cascade $\bk(t)$ adjust to local
    thermodynamic equilibrium with $\Teff$ on a much shorter time scale.

    The power spectrum of the canonical momentum features a universal
    scaling behavior. This universal scaling stage ends when the front of
    the cascade reaches the ultraviolet cutoff at a time scale $t_1 \sim
    (L/E) \, a^{-1/\alpha}$ with $L$ the size of the system, $E$ the
    conserved energy and $a$ the lattice spacing, and with $\alpha\sim
    0.21-0.25$ a {\bf cutoff--independent exponent} weakly dependent on
    $E/L$. During this stage with $t_1 \gg t \gg t_0$ the front of the
    cascade evolves as $\bk(t) \simeq (E/L)^{1/4} \, t^{\alpha}$ and all
    observables attain their thermal equilibrium values corresponding to
    the effective temperature $\Teff$. At the same time $\Teff$ decreases
    as $\Teff \simeq (\pi/2)\,(E/L)^{3/4} \; t^{-\alpha}$.  For $t > t_1$
    thermalization continues further, however the front of the cascade is
    near the cutoff and universal scaling behavior does no longer hold.
    Exact thermalization is only achieved at infinite time when the power
    spectrum is flat and sharply falls off at the cutoff scale and $ T_{\rm
    eff}(\infty) = T$ coincides with the canonical equilibrium temperature.}

\item{In agreement with most previous studies we find that
    thermalization does occur but strictly speaking only at infinite time.
    Moreover, an {\bf effective local thermal equilibrium} description in
    terms of a (slowly) time dependent temperature $\Teff$ is available
    after the second stage at which the interaction term becomes small as
    compared to the spatio-temporal gradient terms.
It should be stressed that during the stage of universal cascade
we obtain a cutoff-independent expression for $\Teff$ and
therefore for all observables in effective local thermal
equilibrium. This is therefore the quasi-equilibrium state of the
continuum field theory for times
    beyond $t_0$.}

\end{itemize}

In section II we summarize the classical $\phi^4$ theory in
thermal equilibrium, obtain exact results for low, intermediate
and high temperature and compare to the Hartree approximation. In
this section we provide a set of stringent criteria for
thermalization and based on the equilibrium results we discuss the
main features expected from the dynamics.  In section III we
introduce the light cone approach to study the dynamical evolution
in the cutoff theory. In this section we discuss a wide range of
initial conditions and energy densities along with the averaging
procedure and the observables to be studied. In section IV we
study in detail the power spectra and the energy cascade. Here we
discuss the stage during which the power spectrum of the canonical
momentum features universal scaling behavior, which we refer to as
the universal cascade.  We establish the regime of validity of
this universal cascade and show that the wake behind the front of
the cascade describes a state of local thermodynamic equilibrium
at a temperature $\Teff$ which slowly decreases towards the
equilibrium value.

Our conclusions as well as several comments and discussions on the
continuum limit, the quantum theory and the kinetic description
are presented in section V.

\section{Thermal equilibrium}\label{sec:Thereq}

We present here the equilibrium statistical mechanics of the
classical $\phi^4$ theory in $1+1$ dimensions.

\subsection{Generalities}
The Lagrangian density of the $\varphi^4$ field theory reads
\begin{equation*}
   {\cal L}_{m,\lambda}\,(\varphi) = \frac12
    \dm\varphi \, \udm\varphi - \frac{m^2}2\,\varphi^2 -
   \frac\lambda4\,\varphi^4
\end{equation*}
and leads to the  classical equation of motion
\begin{equation}\label{eqnofmotion}
    \dm\udm \varphi + m^2\,\varphi + \lambda\,\varphi^3 = 0 \; .
\end{equation}
The theory is defined in a finite volume with periodic boundary
conditions. In $1+1$ space-time dimensions the coupling $\lambda$
has dimensions of $(\mathrm{mass})^2$.

In the classical theory one can always rescale variables absorbing
the mass $m$ and the coupling $\lambda$ in the field and
coordinates in order to define the following dimensionless
quantities
\begin{equation}\label{adim}
 t \equiv m\,x^0  \quad , \quad x \equiv m\,x^1 \quad ,\quad
 \varphi(x^0,x^1) \equiv \frac{m}{\sqrt{\lambda}} \; \phi(x,t)
\end{equation}
so that the Lagrangian takes the form
\begin{equation}\label{lagra}
 {\cal L}_{m,\lambda}\,(\varphi) = \frac{m^4}{\lambda}\,
 {\cal L}_{1,1}\,(\phi) = \frac{m^4}{2\lambda} \, \left(
 {\dot \phi}^2 - {\phi'}^2 - {\phi}^2 - \frac12 \; {\phi}^4 \right)
\end{equation}
where $ {\dot\phi } \equiv {\partial\phi}/{\partial t } $ and
$\phi' \equiv {\partial\phi}/{\partial x }$. Using the scaled
Lagrangian ${\cal
  L}_{1,1}\,(\phi)$ and the canonical momenta $\Pi=\dot \varphi$ and $\pi =
\dot \phi$, we find for the Hamiltonian
\begin{equation}\label{hamilt}
  H_{m,\lambda}(\Pi,\varphi)=\frac{m^3}{2\lambda}
    H[\pi,\phi]  \; , \quad H[\pi,\phi]=  \int_0^L dx \left(
    {\pi}^2 + {\phi'}^2     + {\phi}^2 + \frac12 \; {\phi}^4 \right) \;.
\end{equation}
We will consider periodic boundary conditions (PBC)  on a ring of
(dimensionless) length $L$, namely $\phi(x+L,t)=\phi(x,t)$.

The equations of motion in terms of the dimensionless field  is
given by
\begin{equation}\label{eqnofmot}
{\ddot \phi} - \phi'' + \phi + \phi^3 = 0 \; ,
\end{equation}
In terms of Fourier mode amplitudes,
\begin{equation}\label{fourier}
  {\tilde\pi}_k = \int_0^L \frac{dx}{\sqrt{L}}\, e^{-ikx} \,\pi(x) =
  {\tilde\pi}_{-k}^{\,\ast} \;, \quad
  {\tilde\phi}_k = \int_0^L \frac{dx}{\sqrt{L}}\, e^{-ikx} \,\phi(x) =
  {\tilde\phi}_{-k}^{\,\ast} \;, \quad
  k=\frac{2\pi}L \; n_k \;,\quad n_k\in\mathbb Z \;,
\end{equation}
the Hamiltonian reads,
\begin{equation*}
  H[\phi,\pi] = \frac12 \sum_k \Big[ |{\tilde\pi}_k|^2 +
    (1+k^2)|{\tilde\phi}_k|^2 + \sum_{q\,q'}
    {\tilde\phi}_q \; {\tilde\phi}_{q'} \; {\tilde\phi}_{k} \;
    {\tilde\phi}_{-q-q'-k}     \Big] \; .
\end{equation*}
The wavenumbers $k$ are dimensionless, the dimensionful momenta
(conjugated to $x^1$) are given by $ m \, k $.

The thermal average of any physical quantity $ \Theta =
\Theta[\phi,\pi]$ in the canonical ensemble is written as
\begin{equation}\label{prog}
  \avg{\Theta [\pi,\phi]} = \frac{\int\!\int D\phi \, D\pi\,
    e^{- \beta H[\pi,\phi]}\,\Theta[\phi,\pi]}
      {\int\!\int D\phi \, D\pi  \; e^{- \beta H[\pi,\phi]} } \; ,
\end{equation}
where $\int\!\int D\phi \, D\pi $ stands for functional
integration over the classical phase space and $ \beta \equiv 1/T
$ is the inverse (dimensionless) temperature in the dimensionless
variables. In terms of the \emph{physical} temperature, here
defined as $T_p$, the effective dimensionless temperature that
will enter in the analysis that follows is given by
\begin{equation}\label{temp}
  T = \frac{1}{\beta}\equiv \frac{\lambda}{m^3} \; T_p \; .
\end{equation}
As a consequence of the field redefinition available in the
classical theory, the relevant variable for equilibrium
thermodynamics is $T$. Therefore for a fixed physical temperature
$T_p$ we see from eq.(\ref{temp}) that the low temperature limit
$T\ll 1$ corresponds to the weak coupling limit and/or $ T_p \ll m
$. This will be relevant in the analysis below.

Translation invariance (which is preserved by PBC) implies that
averages of local observables $\Theta(x)$ which depend on $\pi$
and $\phi$ only at one point $x$, do not depend on $x$, that is
$\avg{\Theta(x)}=\avg{\Theta(0)}\equiv\avg{\Theta}$.

Furthermore, the fact that the Hamiltonian is the sum of a kinetic
and a potential term, namely
\begin{equation*}
H[\pi,\phi]\equiv \mathcal{T}[\pi]+V[\phi] \; ,
\end{equation*}
entails that the average of observables represented by products of
polynomials of the canonical momenta and the field,
$\Theta[\phi,\pi]=\Theta_1[\phi] \; \Theta_2[\pi]$ factorizes in
the form
\begin{equation*}
  \langle \Theta[\phi,\pi] \rangle =
  \langle \Theta_1[\phi]\rangle_{\phi} \;  \langle
  \Theta_2[\pi]\rangle_{\pi}
\end{equation*}
with
\begin{equation*}
  \langle\Theta_1[\phi]\rangle_{\phi}  =  \frac{\int D\phi \;
    e^{- \beta V[\phi]} \; \Theta_{\phi}[\phi]}
  {\int D\phi   \; e^{- \beta V[\phi]} } \quad  ,\quad
  \langle \Theta_2[\pi]\rangle_{\pi}  =  \frac{\int D\pi \;
    e^{- \beta \mathcal{T}[\pi]} \; \Theta_{\pi}[\pi]}
  {\int D\pi   \; e^{- \beta T[\pi]} } \; .
\end{equation*}
Moreover, since the $\pi-$integration in gaussian and ultralocal,
it can be performed quite easily in most cases, leading to the
purely configurational integral over $\phi=\phi(x)$ for $0\le x\le
L$.

\subsection{From quantum to classical field theory at equilibrium}

Before presenting our study of the non--equilibrium dynamics of
the classical field theory, it is important to establish the
equilibrium correspondence with the classical limit of quantum
field theory. This will also allow us to make contact with
non--perturbative treatments that are widely used in quantum field
theory and that have a counterpart in the classical limit.  We
first consider free fields and then the non--perturbative Hartree
approximation.

\subsubsection{Free field theory}

The free-field expansion of the quantum field $\varphi(x)$ and its
canonical momentum $\Pi(x)$ for PBC in the interval $0\leq x^1\leq
L_p$ is given by,
\begin{eqnarray}\label{quanexp}
\varphi(x) & = & \sqrt{\frac{\hbar}{L_p}}\sum_{q}
\frac{1}{\sqrt{2\omega_q}}\left[a_q~ e^{-i(\omega_q \, x^0-q \,
x^1)}+a^{\dagger}_q ~e^{i(\omega_q \,
  x^0-q \,    x^1) }\right]\equiv
\frac{1}{\sqrt{L_p}}\sum_{q}\tilde{\varphi}_q(x^0)\; e^{iqx^1}
\; , \\
\Pi(x) & = & -i\sqrt{\frac{\hbar}{L_p}}\sum_{q}
\sqrt{\frac{\omega_q}{2}} \left[a_q~e^{-i(\omega_q \, x^0-q \,
x^1)} -a^{\dagger}_q ~e^{i(\omega_q \, x^0-q \, x^1)}\right]\equiv
\frac{1}{\sqrt{L_p}}\sum_{q}\tilde{\Pi}_q(x^0) \; e^{iqx^1} \; , \\
\omega_q & = & \sqrt{q^2+m^2} \; .
\end{eqnarray}
The operators $a_q, \; a^{\dagger}_q$ obey canonical commutation
relations. In free field theory in thermal equilibrium at
temperature $T_p$ we have
\begin{equation*}
\langle a^{\dagger}_q \; a_q \rangle = n_q \equiv \frac{1}{
e^{\hbar\omega_q/T_{p}} -1} \quad , \quad \langle a_q  \; a_q
\rangle = \langle a^{\dagger}_q \; a^{\dagger}_q \rangle = 0 \; .
\end{equation*}
The classical limit is obtained for very large occupation number
$n_q \gg1$ which requires that $\hbar\omega_q \ll T_p$. Therefore
in the classical limit
\begin{equation}\label{classlimit}
  n_q \simeq \frac{T_p}{\hbar\omega_q}\; .
\end{equation}
and we obtain the following thermal expectation values
\begin{eqnarray}
\langle \tilde{\varphi}_q\tilde{\varphi}_{-q} \rangle & = &
\frac{\hbar}{2\omega_q}\left[1+2n_q \right] \thickapprox
\frac{T_p}{q^2+m^2} \label{classifi1} \; , \\
\langle \tilde{\Pi}_q\tilde{\Pi}_{-q} \rangle & = & \frac{\hbar
\omega_q}{2}\left[1+2n_q \right] \thickapprox T_p\; .
\label{classifi2}
\end{eqnarray}
In order to avoid the ultraviolet catastrophe in the classical
statistical mechanics of the $\phi^4$ theory we introduce a
ultraviolet momentum cutoff which we write as $\Lambda_p =
\pi\,m/(2a)$, where $2a$ is a dimensionless lattice spacing.

We then get the following results,
\begin{eqnarray}
\langle \varphi^2(x) \rangle_{class} & = & \frac{T_p}{L_p} \sum_q
\frac{1}{\omega^2_q} = T_p \; \int^{+\infty}_{-\infty}
\frac{dq}{2\pi}\frac{1}{q^2+m^2} =
\frac{T_p}{2m} \label{varphi2} \; , \\
\langle \varphi^4(x) \rangle_{class} & = & \frac34 \; \frac
{T_p^2}{m^2}
 \; , \nonumber \\
\langle \Pi^2(x) \rangle_{class} & = & \frac{T_p}{L_p} \sum_q
\rightarrow T_p \; \int^{+\Lambda_p}_{-\Lambda_p} \frac{dq}{2\pi}
=
\frac{T_p}{2a_p} \; ,\nonumber \\
\big\langle \left(\frac{\partial \varphi(x)}{\partial
x^1}\right)^2 \big\rangle_{class} & = & \frac{T_p}{L_p} \sum_q
\frac{q^2}{\omega^2_q} \rightarrow T_p \;
\int^{+\Lambda_p}_{-\Lambda_p} \frac{dq}{2\pi}\frac{q^2}{q^2+m^2}
= \frac{T_p}{2a_p}-\frac{mT_p}{2}\; . \label{varphipri2}
 \end{eqnarray}
\noindent where terms of the order $\mathcal{O}(a)$ were neglected
and we used Wick's theorem to compute $ \langle \varphi^4(x)
\rangle$.  The results (\ref{varphi2})-(\ref{varphipri2}) above
lead to the classical result for the expectation value of the
energy density in free field theory,
\begin{equation*}
\varepsilon_{class}= \frac{1}{L_p} \langle
 H_{m,\lambda}(\Pi,\varphi) \rangle_{class}= \frac{T_p}{2a_p} \; .
\end{equation*}
Furthermore, since the free theory is gaussian we find the ratio
 [ see eqs.(\ref{varphi2})-(\ref{varphipri2})],
\begin{equation}\label{ratio}
\frac{\langle \varphi^2(x)\rangle^2}{\langle \varphi^4(x)
\rangle}= \frac{1}{3}  \; .
\end{equation}

\subsubsection{ The Hartree approximation.}\label{subsec:hartree}

The Hartree approximation is a very useful scheme that has been
implemented as non-perturbative treatment of the equilibrium as
well as non--equilibrium aspects of nonlinear quantum field
theories\cite{hartree,ours}. This approximation is equivalent to a
Gaussian ansatz for the wavefunctionals or the density matrix and
leads to a self-consistent condition. Since it is one of the few
non--perturbative techniques that are available and is a useful
tool to study non--equilibrium phenomena, it is important to test
its validity in the classical theory. Furthermore, we will discuss
below \emph{exact} results for the equilibrium situation in the
$1+1$ dimensional classical theory, which allows us to test the
reliability of the Hartree approximation.

In the Hartree approximation the quartic non--linearity is
replaced by a quadratic form as follows
\begin{equation}\label{hartree}
\frac{\lambda}{4} \; \varphi^4 \rightarrow \frac{3\lambda}{2} \;
\varphi^2 \; \langle\varphi^2\rangle \; ,
\end{equation}
\noindent leading to the following effective mass and frequency
(the momenta in the equations that follow are dimensionful)
\begin{equation}\label{masseff}
M^2=m^2+3\lambda\, \langle \varphi^2 \rangle ~~;~~ \Omega^2_q=
q^2+M^2 \; .
\end{equation}
\noindent Thus the theory becomes Gaussian and the non-linearity
emerges in terms of a self-consistency condition. The equilibrium
occupation number and its classical limit is therefore given by
\begin{equation}\label{harocc}
  \mathcal{N}_q = \frac{1}{e^{\hbar\Omega_q/T_p} -1} \;, \quad
  \mathcal{N}_{q,cl} = \frac{T}{\hbar \Omega_q}\; ,
\end{equation}
\noindent  and the Hartree self-consistency condition in the
quantum theory is given by
\begin{equation*}
\langle \varphi^2(x) \rangle_H = \frac{\hbar}{L_p}\sum_q
\frac{1}{2\Omega_q}[1+2\mathcal{N}_q]\; .
\end{equation*}
Using the classical limit eq.(\ref{harocc}) and the result given
by eq.(\ref{varphi2}) we find the classical limit of the Hartree
self--consistency condition
\begin{equation}\label{selfcons}
  \langle \varphi^2 \rangle_H  = \frac{T_p}{2m}
\frac{1}{\sqrt{1+\frac{3\lambda}{m^2}\langle \varphi^2 \rangle_H
}}
\end{equation}
Assuming that in the high temperature limit $\langle \varphi^2
\rangle_H \gg m^2/\lambda$ (to be confirmed self--consistently
below) we find the solution of eq.(\ref{selfcons}) to be given by
\begin{equation}\label{hartsol}
  \langle \phi^2 \rangle_H =
  \frac{1}{12^{1/3}}\left(\frac{\lambda
      T_p}{m^3}\right)^{2/3} =  0.436790\cdots T^{2/3} \; .
\end{equation}
Since the Hartree approximation is a self--consistent  Gaussian
approximation, the ratio
\begin{equation*}
\frac{\langle \phi^2 \rangle^2_H}{\langle \phi^4 \rangle_H}=
\frac13 \; .
\end{equation*}
takes the same value than for free fields [eq.(\ref{ratio})].

We also find the following results in the classical limit
\begin{eqnarray}
\langle \Pi^2(x)\rangle_{H,cl} & = & \frac{T_p}{2a_p}
\label{pi2H}\\
\big\langle \left(\frac{\partial \varphi(x)}{\partial
x^1}\right)^2\big\rangle_{H,cl} & = & \langle
\Pi^2(x)\rangle_{H,cl} - M^2 \langle \varphi^2 \rangle_{H,cl}
\label{varphipriH}
\end{eqnarray}
\noindent with $M^2$ given in eq.(\ref{masseff}). In order to
compare to the exact results obtained in sec. \ref{qtm} from the
quantum transfer matrix, we summarize the results for the  power
spectra in the classical  limit  in the Hartree approximation in
terms of the dimensionless field and canonical momentum [see eqs.
(\ref{adim}), (\ref{fourier}) and (\ref{temp})] in the high
temperature limit $T\gg 1$,
\begin{eqnarray}
  \langle |\tilde{\pi}_q|^2 \rangle_{H,cl} & = & T   \cr \cr
  \langle |\tilde{\phi}_q|^2 \rangle_{H,cl} & = &
  \frac{T}{k^2+ \left(\frac32 \, T \right)^{2/3}} =
  \frac{T}{k^2+1.310371\ldots T^{2/3}}
\end{eqnarray}
These results will be compared to the exact results obtained from
the transfer matrix in sec.\ref{qtm}  and together with those
provide a yardstick to establish criteria for thermalization and
virialization.

\subsection{Exact results from the quantum transfer matrix}\label{qtm}
In this section we revert to the dimensionless variables given by
eqs. (\ref{adim})-(\ref{temp}).  The expectation value of a
$\pi-$independent observable $\Theta=\Theta[\phi]$ is given by the
configurational integral
\begin{equation} \label{prot}
\avg{\Theta(\phi)} = \frac{\int D\phi \; e^{- \beta \; V[\phi]} \;
\Theta[\phi]  }{\int D\phi \; e^{- \beta \; V[\phi]} } \; ,
\end{equation}
where
\begin{equation*}
 V[\phi] = \frac12 \; \int_0^{L} dx \left[ {\phi'}^2 + {\phi}^2 +
\frac12 \; {\phi}^4 \right] \; .
\end{equation*}
This is just an euclidean path integral for the path
$\phi=\phi(x)$, $0\le x\le L$, with PBC, so that the classical
statistical average in eq. (\ref{prot}) can be interpreted as a
quantum mechanical trace for the one dimensional anharmonic
oscillator in imaginary time $-ix$ with $ 0 \leq x \leq L$
\cite{fey}. Namely,
\begin{equation}\label{msqm}
\frac{\int_{\phi(0) = \phi(L)} D\phi \; e^{- \beta \, V[\phi]}\;
\Theta[\phi] }{\int_{\phi(0) = \phi(L)} D\phi \; e^{-\beta \,
V[\phi]} }
  = \frac{{\rm Tr} \; (e^{-L \, \H} \; {\sf X}\, \Theta[\hphi])}
{{\rm Tr}\; e^{-L \H}}\; .
\end{equation}
where $\hphi$ is the family of (imaginary time) Heisenberg
operators
\begin{equation*}
      \hphi(x) = e^{x\H} \; \hphi(0) \; e^{-x\H}
      \quad , \quad 0 \leq x \leq L \quad,
\end{equation*}
${\sf X}$ stands for operator ordering along $x$. That is,
\begin{equation*}
{\sf X}\, \hphi(x)\hphi(y) = \theta(x-y) \; \hphi(x) \; \hphi(y) +
\theta(y-x) \; \hphi(y) \; \hphi(x)
\end{equation*}
and $\H$ is the quantum Hamiltonian
\begin{equation*}
\H \equiv \frac{1}{2 \,\beta} \; \left[{\hat\Pi}(0)\right]^2 +
\beta\, \left\{ \frac12 \, \left[\hphi(0)\right]^2 + \frac14 \;
  \left[\hphi(0)\right]^4 \right\}\; .
\end{equation*}
Here $\hphi(0)$ and ${\hat\Pi}(0)$ are canonical conjugate
Schr\"odinger operators, that is $[\hphi(0), {\hat\Pi}(0) ] = i $.

It is convenient to make the following canonical change of
variables,
\begin{equation*}
\hphi(0) = \sqrt{T} \; q \quad , \quad {\hat \Pi}(0) =
\frac{1}{\sqrt T} \; p \; .
\end{equation*}
Therefore $[q,p] = i $ and
\begin{equation}\label{hanha}
\H = \frac12 \left( p^2 + q^2 \right) + \frac{T}{4} \; q^4\; .
\end{equation}
Hence, the calculation of classical statistical averages in eq.
(\ref{prot}) reduces to quantum traces of the one--dimensional
anharmonic oscillator in eq.(\ref{hanha}) with anharmonicity equal
to the (dimensionless) temperature [see eq.(\ref{temp})].

The spectrum of $\H$ is an infinite set of discrete energy levels
$E_n(T)$ associated with real eigenfunctions $\psi_n(q\,;T)$ which
solve the Schr\"odinger equation
\begin{equation}\label{sch}
\frac12 \left[
  -\frac{d^2}{dq^2} + q^2 + \frac{T}{2} \; q^4 \right]\psi_n(q\,;T) = E_n(T)
\; \psi_n(q\,;T) \quad , \quad n=0,1,2,\ldots
\end{equation}
and have defined parity:
\begin{equation}\label{par}
\psi_n(-q\,;T) = (-1)^n \; \psi_n(q\,;T) \; .
\end{equation}
since $\H$ is invariant under $ q \to -q $,

In the case of ultralocal observables $\Theta[\phi(x)]$, such as
polynomials in $\phi(x)$, owing to traslational invariance (which
translates into the cyclic property of quantum traces) we can
express the thermal averages as single sums over diagonal matrix
elements,
\begin{equation}\label{sum}
\avg{\Theta(\phi)} = \frac1{Z(T)}\sum_{n=0}^\infty \Theta_{nn}(T)
\; e^{-L \, E_n(T)} \; ,  \; \quad Z(T) \equiv {\rm Tr}\,e^{-L \H}
= {\sum_n\;e^{-L \, E_n(T)}} \; ,
\end{equation}
where
\begin{equation*}
\Theta_{mn}(T) = \int_{-\infty}^{+\infty}\!\!dq\; \Theta( \sqrt{T}
\; q) \; \psi_m(q\,;T) \; \psi_n(q\,;T)
\end{equation*}
with the proper normalization
\begin{equation*}
\int_{-\infty}^{+\infty} \!\!dq\; \left[\psi_n(q;T)\right]^2 = 1
\;.
\end{equation*}
For example, we get for $\Theta=\left[\phi(x)\right]^2$ and
$\Theta=\left[\phi(x)\right]^4$:
\begin{equation}\label{suma}
\avg{\phi^2} = \frac{T}{Z(T)}\sum_n \;
[q^2]_{nn}(T)\,e^{-L\,E_n(T)}\;,\quad \avg{\phi^4} =
\frac{T^2}{Z(T)}\sum_n \;  [q^4]_{nn}(T)\,e^{-L\,E_n(T)} \; .
\end{equation}
where,
\begin{equation*}
  [q^\alpha]_{mn}(T) = \int_{-\infty}^{+\infty}\!\!dq \; q^\alpha \;
  \psi_m(q\,;T) \;  \; \psi_n(q\,;T) \;.
\end{equation*}
The situation is more involved in the case of observables
containing products of fields at different points. For instance
the equal--time two--point correlation function takes the form
\begin{eqnarray}\label{twopoint}
&& \avg{\phi(x) \,\phi(y) } = \frac{\int_{\phi(0) = \phi(L)}
D\phi\, \phi(x) \, \phi(y) \; e^{- \beta \, V[\phi]}
}{\int_{\phi(0) = \phi(L)} D\phi \, e^{-\beta \, V[\phi]} }=\cr
\cr \cr &&= \frac{1}{Z(T)} \; {\rm Tr}\,[e^{-L \H}\,{\sf
X}\,\hphi(x) \, \hphi(y)] \; = \frac{T}{Z(T)} \; \sum_{n,m} \;
(q_{nm})^2 \,e^{-(L-|x-y|)E_n(T) -
  |x-y|E_m(T)}
\end{eqnarray}
Particular care is needed to treat local observables which contain
powers of the field derivative $\phi'(x)$, due to contact terms.
Consider for instance $\left[\phi'(x)\right]^2$; we have
\begin{equation*}
  {\sf X}\,[\hphi(x+dx)-\hphi(x)]^2 = -i[\hphi(x),[\H,\hphi(x)]\,dx
  + ([\H,\hphi(x)]\,dx)^2 + \ldots = T\,dx - T^2\,{\hat\Pi}(x)^2\,
  (dx)^2 + \ldots
\end{equation*}
so that
\begin{equation*}
  \left[\hphi'(x)\right]^2 = T\,(dx)^{-1} - T\, e^{x\H}\,p^2\, e^{-x\H}
\end{equation*}
The $c-$number singularity proportional to $(dx)^{-1}$ reflects
that the relevant configurations the 1D functional integral are
random walks with differentials proportional to $(dx)^{-1/2}$.
Notice that the same singularity appears in the average of
$\pi^2$, since the $\pi-$integration in gaussian and ultralocal,
that is
\begin{equation}\label{corrpi}
  \avg{\pi(x) \, \pi(y)} = T \, \delta(x-y) \; ,
  \quad \avg{\pi^2} = T \, \delta(0) = T \; (dx)^{-1} \; .
\end{equation}
In terms of the lattice cutoff, $\delta(0)=1/dx = 1/(2a)$, therefore
\begin{equation}\label{pi2}
\avg{\pi^2} = \frac{T}{2a} = \frac{T \, \Lambda}{\pi} \;.
\end{equation}
 Since $\pi=\dot\phi$, we obtain
\begin{equation}\label{phipp}
  \avg{\dot\phi^2} - \avg{\phi'^2} = \frac{T}{Z(T)}  \;
  {\rm Tr}\,[e^{-L \H}\,p^2]  = \frac{T}{Z(T)} \; \sum_n \;  [p^2]_{nn}(T)
  \; e^{-L \, E_n(T)}\; ,
\end{equation}
where,
\begin{equation}\label{phipp1}
[p^2]_{nn}(T) =  \int_{-\infty}^{+\infty}\!\!dq \;
[\psi'_n(q;T)]^2 \;.
\end{equation}
It follows from eqs.(\ref{phipp}) and (\ref{corrpi}) that
\begin{equation}\label{fipri2}
\avg{\phi'^2} = \frac{T}{2 \, a} - \frac{T}{Z(T)} \; \sum_n
[p^2]_{nn}(T)
  \; e^{-L \, E_n(T)}\; .
\end{equation}
For long rings ($L \gg 1$) the sums in
eqs.(\ref{sum})-(\ref{suma}) are dominated by the ground state
considerably simplifying  the results and leading to the following
expressions
\begin{eqnarray}
&&\avg{\Theta} \buildrel{L \gg 1}\over=  \Theta_{00}(T) + {\cal
O}\left( e^{-L\,\omega_0(T)} \right) \quad , \quad \avg{\phi^2}
\buildrel{L \gg 1}\over=   T \;[q^2]_{00}(T) + {\cal O}\left(
e^{-L\,\omega_0(T)} \right) \cr \cr &&\avg{\phi^4}  \buildrel{L
\gg 1}\over=   T^2 \;[q^4]_{00}(T) + {\cal O}\left(
e^{-L\,\omega_0(T)} \right) \quad , \quad \avg{\dot\phi^2} -
\avg{\phi'^2}  \buildrel{L \gg 1}\over=   T \;
    [p^2]_{00}(T) + {\cal O}\left(
e^{-L\,\omega_0(T)} \right) \; . \nonumber
\end{eqnarray}
where $\omega_0(T) = E_1(T)- E_0(T)$ is the first gap in the
spectrum of $\tilde{\mathcal{H}}$, namely the effective quantum
Hamiltonian in the transfer matrix. Notice that, since all
eigenvalues $E_n(T)$ as well as all differences $\omega_n(T) =
E_{2n+1}(T)- E_0(T)$ increase monotonically with $T$, these
$L\gg1$ approximations, as well as other to follow, are uniform in
$T$.

\subsection{Classical and quantum virial theorems}\label{secvir}

In equilibrium, the \emph{classical} virial theorem is a
consequence of the vanishing of the Boltzmann weight in the
partition function for large coordinates and momenta. For the
classical field theory under consideration, the classical virial
theorem can be summarized as follows
\begin{equation}\label{classvir}
  \Avg{\eta_i(x)\frac{\delta H}{\delta\eta_j(y)}} =
  T\,\delta_{ij}\,\delta(x-y)
\end{equation}
where $i,j=1,2$ and $\eta_1=\phi,\,\eta_2=\pi$. More explicitly,
\begin{equation} \label{WT}
\Avg{\phi(x)\frac{\delta H}{\delta\phi(y)}} = T\, \,\delta(x-y) =
\Avg{\pi(x)\frac{\delta H}{\delta\pi(y)}} \quad \mbox{and}\quad
\Avg{\pi(x)\frac{\delta H}{\delta\phi(y)}} = 0 =
\Avg{\phi(x)\frac{\delta H}{\delta\pi(y)}} \; .
\end{equation}
For example,
\begin{equation*}
\Avg{\phi(x)\frac{\delta H}{\delta\phi(y)}}= \frac{\int\!\int
D\phi \, D\pi\, \phi(x) \; e^{- \beta H[\pi,\phi]} \; \frac{\delta
H}{\delta\phi(y)} } {\int\!\int D\phi \, D\pi  \; e^{- \beta
H[\pi,\phi]} } = - T \; \frac{\int\!\int D\phi \, D\pi\, \phi(x)
\; \frac{\delta}{\delta\phi(y)} e^{- \beta H[\pi,\phi]}}
{\int\!\int D\phi \, D\pi  \; e^{- \beta H[\pi,\phi]} } = T \,
\,\delta(x-y)
\end{equation*}
and integrating by parts in the last step.
\medskip
For the Hamiltonian \eqref{hamilt} this implies, besides the
already mentioned $\avg{\pi(x)\pi(y)} = T \; \delta(x-y)$ and the
trivial $\avg{\phi(x)\pi(y)} =0$, the nonlinear relation
\begin{equation*}
  \avg{\phi(x)\,[-\phi''(y)+\phi(y)+\phi^3(y)]} = T\,\delta(x-y) \;.
\end{equation*}
Hence upon letting $x\to y$, taking the volume average  and  using
the result (\ref{corrpi}) along with translation invariance, we
find the classical thermal virial theorem
\begin{equation}\label{virial2}
  \avg{\dot\phi^2} = \avg{\phi'^2} + \avg{\phi^2} + \avg{\phi^4} \;.
\end{equation}
When combined with the energy functional $H[\pi,\phi]$ given in
eq.(\ref{hamilt}), it yields
\begin{equation}\label{enerdens}
  \frac{\avg{H[\pi,\phi]}}L = \frac{E}{L}= \avg{\pi^2} - \frac14
  \avg{\phi^4} \;,
\end{equation}
where ${E}$ is the average energy in the canonical ensemble. Since
$N=L/(2a)$ is the total number of degrees of freedom on the
lattice, we find that the  temperature $T$ is related to the
(average) energy per degree of freedom as
\begin{equation}\label{TtoE}
  T = \frac{E}{N} + \frac{a}2 \avg{\phi^4} \;.
\end{equation}
Therefore, \emph{if} $a\,\avg{\phi^4} \rightarrow 0$ then the
temperature is identified with the energy per site. We will study
below the conditions under which the temperature can be directly
identified with the energy per site.
\medskip
Comparing eq.\eqref{virial2} to eq.\eqref{phipp} we obtain
\begin{equation*}
  \avg{\phi^2} + \avg{\phi^4} = T \;
  \frac{{\rm Tr}\,[e^{-L\H}\,p^2]}{{\rm Tr}\,e^{-L \H} } \;.
\end{equation*}
As a consequence of eqs.\eqref{suma}, this entails the identity
\begin{equation*}
  {\rm Tr}\,[e^{-L\H}\,(p^2 - q^2 - T\,q^4)] = 0 \;.
\end{equation*}
This result is  a direct consequence of the quantum virial theorem
\begin{equation*}
    (p^2)_{nn} =  (q\,U'(q)\,)_{nn}
\end{equation*}
for the case of our potential $U(q) = \frac12 q^2+ \frac{T}4 q^4$.

While the virial identity eq.(\ref{virial2}) was obtained in
thermal equilibrium, it is more general and it can be derived out
of equilibrium by focusing on time averages.  To see this,
consider the following quantity:
\begin{equation*}
I(t) \equiv \frac12 \, \int_0^L dx \; \left[\phi(x,t)\right]^2 \;
.
\end{equation*}
Its second derivative with respect to time takes the form,
\begin{equation*}
{\ddot I}(t) = \int_0^L dx \left\{ \phi(x,t) {\ddot \phi} (x,t) +
  \left[ {\dot \phi} (x,t)\right]^2 \right\} \; .
\end{equation*}
Which upon using the equation of motion (\ref{eqnofmot}) and after
integrating by parts  using the PBC yields,

\be \label{devseg} {\ddot I}(t) =  \int_0^L dx \left\{ \left[{\dot
\phi} (x,t)\right]^2 - \left[\phi'(x,t)\right]^2 -
\left[\phi(x,t)\right]^2 - \left[\phi(x,t)\right]^4 \right\} \; .
\ee The time average of a physical quantity is defined by
\begin{equation*}
{\overline \Theta } \equiv \lim_{t \to \infty} \frac{1}{t}
\int_0^{t} dt' \; \Theta(t') \; .
\end{equation*}
For quantities which are total time derivatives of bounded
functions this average obviously vanishes:
\begin{equation*}
{\overline {\frac{d \Theta}{d t}} } = \lim_{t \to \infty} \frac{
  \Theta(t) - \Theta(0)}{t} = 0 \; .
\end{equation*}
In the $\phi^4$ theory every classical trajectory is bounded for
bounded initial conditions; therefore, taking the time average of
eq.(\ref{devseg}) yields
\begin{equation}\label{virial}
  \int_0^L dx \left[ \overline{ {\dot \phi}^2} -
  \overline{\phi'^2} - \overline{\phi^2}  -
  \overline{\phi^4} \right] = 0 \; .
\end{equation}
This is the virial theorem for the $\phi^4$ theory with PBC. If
the time average is translational invariant in space, then the
integral can be omitted and eq.\eqref{virial} takes exactly the
form of eq.\eqref{virial2}, with thermal averages replaced by time
averages. Of course, the two averages should coincide if the
ergodic hypothesis holds, but the validity of eq.\eqref{virial}
does not rely on this fact. This usually entails that the time
scales for virialization are much smaller than those of
thermalization [see sec. \ref{viria} for a discussion on
  this for the present $\phi^4$ model].

Combining eq.(\ref{virial}) along with energy conservation yields
the following  alternative expression for the total energy
\begin{equation*}
  E = \int_0^L dx \left[ \overline{ {\dot \phi}^2} -
    \frac14 \overline{\phi^4} \right]
\end{equation*}

It should also be noticed that an infinite number of relations
similar to eq.\eqref{virial} can be derived by considering the
second derivative of quantities such as $\int_0^L dx \;
[\phi(x,t)]^n $ for $ n > 2 $.

\medskip

These results on virialization, which is different from the
statement of thermalization, also provides a yardstick for the
numerical evolution studied in section \ref{sec:lattice}.

\subsection{Correlation functions and power spectra}

Let us now turn our attention to the equal--time correlation
function \eqref{twopoint}, which by translation invariance is a
function only of $x-y$, that is $\avg{\phi(x)\,\phi(y)} =
G(x-y,T)$.  Its Fourier transform
\begin{equation*}
 {\tilde G}(k,T) \equiv \int_0^L dx\, e^{-ikx}\,G(x,T)
\end{equation*}
coincides with the thermal average of the so--called {\em power
spectrum} of the field $\phi$, that is [see eq.\eqref{fourier}]
\begin{equation}\label{Gtopws}
  {\tilde G}(k,T) = \avg{\,|{\tilde\phi}_k|^2} \;.
\end{equation}
Using completeness, $\sum_n \; (q_{mn})^2 = (q^2)_{mm}$, we verify
that
\begin{equation}\label{sumrule0}
  \frac1{L}\sum_k{\tilde G}(k,T) =  G(0,T) = \avg{\phi^2}(T) \; .
\end{equation}
\medskip
In the  limit, $ L \gg 1 $ with $ x = {\cal O}(L^0) , \; |L-x| \gg
1 $, the correlation function $G(x,T)$ can be approximated as [see
eq.\eqref{twopoint}]:
\begin{equation}\label{twopoint2}
\begin{split}
  G(x,T) &= T \,\sum_{n=0}^\infty e^{-[ E_n(T)- E_0(T)]\; |x|}\;
  ([q]_{0n}(T))^2  + {\cal O}\left(e^{-[ E_1(T)- E_0(T)]\,L}\right) =\\
  & =T \, \sum_{n=0}^{\infty} e^{-\omega_n(T)\, |x|}\;
  \Big([q]_{0,2n+1}(T)\Big)^2+ {\cal O}\left(e^{-\omega_0(T)\,L}\right) \; .
\end{split}
\end{equation}
where $ \omega_n(T) \equiv E_{2n+1}(T)- E_0(T) $ and we used that
$ q_{0,2n} = 0 $ due to parity invariance [see eq.(\ref{par})].
Similarly, in Fourier space we find
\begin{equation}\label{Gtilde}
  {\tilde G}(k,T)  = \sum_{n=0}^{\infty}
  \frac{2\,T\,\omega_n(T)}{k^2 + \omega^2_n(T)}
  \, \Big([q]_{0,2n+1}(T)\Big)^2 +
  {\cal O}\left(e^{-\omega_1(T)\,L}\right) \;,\\
\end{equation}
where $k$ may now be treated as a continuous variable.

Now the normalization eq.\eqref{sumrule0} is written as
\begin{equation*}
  \int_{-\infty}^{+\infty}\dk\; {\tilde G}(k,T) = G(0,T) =
  T\,\sum_{n=0}^{\infty}
  \Big([q]_{0,2n+1}(T)\Big)^2 =  T \,[q^2]_{00}(T) =\avg{\phi^2} \;.
\end{equation*}
This can be supplemented by two another sum rules related to
$\avg{\phi^2}$ and to the (classical or quantum) virial
identities, that is
\begin{equation*}
  \int_{-\Lambda}^\Lambda\dk\;k^2\,{\tilde G}(k,T) =
  T\,\sum_{n=0}^{\infty} \omega_n(T) \left[\frac{2\Lambda}\pi -
  \omega_n(T)\right]
    \Big([q]_{0,2n+1}(T)\Big)^2 = \frac{T}{2a} - T \,[p^2]_{00}(T)
    = \avg{\phi'^2} \;,
\end{equation*}
or
\begin{equation}\label{reglsuma}
  \sum_{n=0}^{\infty} \omega_n(T) \Big([q]_{0,2n+1}(T)\Big)^2 = \frac12
  \quad , \quad \sum_{n=0}^{\infty} \omega_n^2(T) \Big([q]_{0,2n+1}(T)\Big)^2
  = [p^2]_{00}(T) \;.
\end{equation}
Eqs.(\ref{Gtilde})-(\ref{reglsuma}) imply that, for any
temperature, ${\tilde G}(k,T)$ vanishes for large $k$ exactly as
\be \label{Gkinf} {\tilde G}(k,T)\buildrel{k\rightarrow
\infty}\over = \frac{T}{k^2}  \; , \ee just like the free field
case. It should also be noticed that all the sum rules above are
almost saturated by the first term in the sums, for any
temperature. This appears evident from the data plotted in fig.
\ref{equ1} and will be shown in more detail in section
\ref{hightemp}. It follows that ${\tilde G}(k,T)$ is very well
approximated by the first term in its series eq.\eqref{Gtilde}.

\medskip

The two-point correlation function of the conjugate momentum $\pi$
in the classical theory in equilibrium is given by
\begin{equation*}
 \avg{\pi(x)\,\pi(y)} = T \,\delta(x-y) \;.
\end{equation*}
which leads to a flat power spectrum for $\pi$
\begin{equation}\label{Tflat}
  \avg{|{\tilde\pi}_k|^2} = T \;,
\end{equation}
This of course is a consequence of equipartition, and gives a
criterion to identify the temperature: the height of the flat
region in the power spectrum of $\pi$.  This identification will
be very useful for the interpretation of the numerical analysis
presented in sec. \ref{casca}.

\subsection{Low temperature limit}

In the low temperature limit $T \ll 1$ the Hamiltonian
eq.(\ref{hanha}) becomes a harmonic oscillator with eigenvalues $
E_n(0) = n + \frac12 $ and eigenfunctions
\begin{equation*}
\psi_n(q,0) = c_n \; e^{-\frac12 q^2} \, H_n(q) \quad , \quad c_n
\equiv \frac{1}{2^{n/2} \, \pi^{1/4} \, \sqrt{n!}} \; ,
\end{equation*}
where the $ H_n(q) $ are Hermite polynomials. As mentioned above,
in the dimensionless variables the limit $T\ll 1$ is equivalent to
the weak coupling limit of the classical field theory in which the
non-linearity can be neglected.

Using the relation $\phi=\sqrt{T}q$ we find, for $ L \gg 1 $ and $
T \to 0 $, \bea \label{f2f4fp} &&\avg{\phi^2} = T~[q^2]_{00}+
{\cal O}(T^3) + {\cal O}\left(e^{-L} \right) = \frac{T}{2}+{\cal
O}(T^3) + {\cal O}\left(e^{-L} \right) \; , \cr \cr &&\avg{\phi^4}
= T^2 ~[q^4]_{00}+{\cal O}(T^3)+{\cal O}\left( e^{-L}\right) =
\frac34 \; T^2+{\cal O}(T^3) + {\cal O}\left( e^{-L}\right) \;,
\eea and from eqs.(\ref{phipp})-(\ref{phipp1}) we find,
\begin{equation*}
\avg{\dot\phi^2} - \avg{\phi'^2} = \frac{T}{2} +{\cal O}(T^3)+
    {\cal O}\left(e^{-L} \right) \; .
\end{equation*}
Therefore, we recover the free-field theory result [see
eq.(\ref{ratio})]
\begin{equation}\label{fft}
   \lim_{T\to0} \frac{\avg{\phi^2}^2}{\avg{\phi^4}} = \frac13  \quad ,
   \quad
   \lim_{T\to 0}T\,\frac{\avg{\dot\phi^2}-\avg{\phi'^2}}{\avg{\phi^4}}
   = \frac23 \; .
\end{equation}
Similarly, for the matrix elements that define the two--point
correlation function $G(x,T)$ we find  the following low
temperature limit $T\rightarrow 0$,
\begin{equation*}
[q]_{01}(0) = \frac{1}{\sqrt2} \quad , \quad [q]_{0,2l+1}(0) = 0
\quad \mbox{for} \quad l \geq 1 \; ,
\end{equation*}
so that the  free-field results
\begin{equation}\label{Glow}
 G(x,T) = \frac{T}{2} \; e^{-|x|}+{\cal O}(T^3)  \quad , \quad
 {\tilde G}(k,T) = \frac{T}{k^2+1}+{\cal O}(T^3) \; ,
\end{equation}
are obtained. The result above for ${\tilde G}(k,T)$ is the same
as that obtained from the classical limit of the free field theory
eq.(\ref{classifi1}) after the rescaling
(\ref{adim})-(\ref{temp}).

We can evaluate the temperature as a function of the energy
density $E/L$ for  $a \ll 1$ using the low temperature formula
eq.(\ref{f2f4fp}). We obtain from eq.(\ref{TtoE}),
\begin{equation*}
  T =  2\,a\, \frac{E}{L} + \frac{a}{2} \; \avg{\phi^4} \; .
\end{equation*}
Using the results above, in the low temperature limit we find,
\begin{equation}\label{Ta}
\begin{split}
  \frac{E}{L}  &= \frac{T}{2a} -\frac3{16} \; T^2 + {\cal O}(a^3) \;, \\
  T &= 2\,a\,\frac{E}{L} + \frac{3}{2} \; a^3 \;
  \left(\frac{E}{L}\right)^2 + {\cal O}(a^5) \;.
\end{split}
\end{equation}
It should be stressed that the low temperature limit is
asymptotic, in the sense that the perturbation series in $T$ has
zero radius of convergence. So, also the low density limit $E/L\to
0$ at fixed UV cutoff or the continuum limit at finite $E/L$ are
asymptotic expansions.

\subsection{High temperature expansion} \label{hightemp}

For any $T>0$ the quartic terms dominates in the  Hamiltonian
(\ref{hanha}) and the quadratic term can be treated as a
perturbation. In fact, we can make the following change of
variables in eq.(\ref{sch})
\begin{equation*}
  q = \left( \frac{2}{T} \right)^{1/6} \; \xi \; ,
\end{equation*}
so that eq.(\ref{sch}) becomes
\begin{equation}\label{schy}
  \frac12 \left[ -\frac{d^2}{d\xi^2} +  \xi^4 +  \left( \frac{2}{T}
  \right)^{2/3} \; \xi^2 \right]
  \chi_n(\xi;T^{-2/3}) = \epsilon_n(T^{-2/3}) \; \chi_n(\xi;T^{-2/3})
  \; ,\quad n=0,1,2,\ldots
\end{equation}
where
\begin{equation*}
  E_n(T) = \left(\frac{T}{2}\right)^{1/3} \; \epsilon_n(T^{-2/3})
  \quad , \quad
  \psi_n(q;T) =  \left(\frac{T}2\right)^{1/12} \;
  \chi_n(\xi;T^{-2/3}) \; .
\end{equation*}
Since the new eigenvalues $\epsilon_n(T^{-2/3})$ and the new
eigenfunctions $\chi_n(\xi;T^{-2/3})$ are entire functions of
$T^{-2/3}$, all equilibrium quantities multiplied by the
appropriate power of $T$ have convergent high temperature
expansions in powers of $T^{-2/3}$. In particular, for large $T$
we have to leading order\cite{sym},
\begin{equation*}
  E_n(T) \simeq \left(\frac{T}{2}\right)^{1/3} \; \epsilon_n(0) \;,
  \quad   \psi_n(q,T) \simeq  \left(\frac{T}2\right)^{1/12} \;
  \chi_n(y;0) \; .
\end{equation*}
Then the results of refs. \cite{ind,hioe}, such as
\begin{equation*}
\epsilon_0(0) = 0.530181045\ldots \; , \quad \epsilon_1(0) =
1.899836515\ldots \;, \quad \epsilon_3(0) = 3.727848969\ldots \; .
\end{equation*}
prove to be very useful. We thus find for the relevant thermal
averages, when $T \gg 1$,
\begin{eqnarray}\label{sumTg}
&&\avg{\phi^2}\simeq  2^{1/3} \,T^{2/3} \, [\xi^2]_{00} + {\cal
O}\left( e^{-L \;T^{1/3} } \right) \\ \cr &&\avg{\phi^4} \simeq
2^{2/3} \; T^{4/3} \, [\xi^4]_{00} + {\cal O}\left( e^{-L
\;T^{1/3} } \right) \;, \quad \avg{{\dot\phi}^2}-\avg{\phi'^2}
\simeq 2^{-1/3} \, T^{4/3} \, [p_\xi^2]_{00} + {\cal O}\left(
e^{-L \;T^{1/3} } \right)\;.\nonumber
\end{eqnarray}
Solving numerically for the ground state $\chi_0(\xi;0)$ of the
purely quartic oscillator and computing the appropiate integrals
we find
\begin{eqnarray}
&&\lim_{T\to \infty} \frac{\avg{\phi^2}}{T^{{2/3}}} =
0.456119954\ldots,
   \label{y2p} \nonumber \\
&&\lim_{T\to \infty} \frac{\avg{\phi^4}}{T^{4/3}} = \lim_{T\to
\infty} \frac{\avg{{\dot\phi}^2} - \avg{\phi'^2}} {T^{4/3}} =
0.56104\ldots \label{y4p}
\end{eqnarray}
Thus we find following ratios for high temperature,
\begin{equation}\label{raz24}
  \lim_{T\to \infty} \frac{\avg{\phi^2}^2}{\avg{\phi^4}} = 0.37077\ldots \;,
  \quad \lim_{T\to \infty}
  \frac{\avg{\dot\phi^2}-\avg{\phi'^2}}{\avg{\phi^4}} = 1 \; .
\end{equation}
to be the compared with their counterpart in the zero temperature
limit [the second limit in eq.\eqref{raz24} is a restatement of
the virial theorem eq.\eqref{virial2} when $\avg{\phi^2}$ is
subdominant].

\bigskip

The high temperature result for $\avg{\phi^4}$ in eq.(\ref{y4p})
allows to relate the energy density to the temperature using
eq.(\ref{enerdens}) as
\begin{equation}\label{ThiT}
  \frac{E}{L} \simeq  \frac{T}{2a}\left[1- 0.28052\cdots
    \left(Ta^3\right)^{\frac{1}{3}}\right] \;.
\end{equation}
Thus also in the high temperature limit $T \gg 1$ but with $Ta^3
\ll 1$ the energy density is directly proportional to the
temperature, $E/L \simeq T/(2a)$, signalling that the energy is
dominated by the spacetime derivatives of the field.

\bigskip

For high temperatures, the correlation function $G(x,T)$ takes the
scaling form
\begin{equation*}
G(x,T) \simeq  2^{1/3} \; T^{2/3} \; g( x\,T^{1/3})\quad , \quad
{\tilde G}(k,T) \simeq 2^{4/3} \; T^{1/3} \; {\tilde   g}
\left(k\,T^{-1/3}\right) \; ,
\end{equation*}
where \be \label{gaT} g(z) = \sum_{l=0}^{\infty}
e^{-\Omega_{2l+1}\; |z|}\; (y_{0,2l+1})^2 \quad , \quad {\tilde
g}(p) = \sum_{l=0}^{\infty}(y_{0,2l+1})^2 \;
\frac{\Omega_{2l+1}}{p^2 + \Omega^2_{2l+1}} \ee with $
\Omega_{2l+1} \equiv  2^{-1/3} \; [\epsilon_{2l+1} -
  \epsilon_0] $ and
\begin{equation*}
y_{0,2l+1} \equiv \int_{-\infty}^{+\infty} y \; \phi_{2l+1}(y)\;
\phi_0(y)\; dy \; .
\end{equation*}
In particular, $ \Omega_1 = 1.087096267\ldots $ and \cite{ind} $
y_{0,1} = 0.600804942\ldots, \; y_{0,3} = -0.032461289\ldots$.
Since the matrix elements $y_{0,2l+1}$ decrease very fast for $ l
> 0 $ we can approximate eq.(\ref{gaT}) by the first term with the
result, \be \label{aprcor} {\tilde G}(k,T)
=\avg{\,|{\tilde\phi}_k|^2}\simeq T \; \frac{0.988800\ldots}{k^2 +
1.181779\ldots \; T^{2/3}} \quad , \quad G(x,T) \simeq T^{2/3} \;
0.4547894\ldots \; e^{-1.087096\ldots \; T^{1/3} \; |x| }\; .  \ee
Notice that in this approximation $ G(0,T) \simeq T^{2/3} \;
0.45478939\ldots $ while the correct high temperature limit
eq.(\ref{y2p}) is only $0.3\%$ higher. Similarly, at large $k$ the
approximated ${\tilde
  G}(k,T)$ behaves like $0.9888\ldots T/k^2$, while the exact expression is
$T/k^2$ [see eq.(\ref{Gkinf})]. Taking into account the infinite
radius of convergence of the perturbation series in $T^{-2/3}$ and
the fact that $G(0,T=0)$ reduces to the first term only, this
first term approximation must be uniformly good in $T$.

\bigskip

We can now establish the comparison between the exact (and
approximate) results obtained above and those obtained in the
Hartree approximation (in the classical limit
sec.\ref{subsec:hartree}) in the high temperature limit:
\begin{eqnarray}
&& \avg{\phi^2} \simeq 0.45612\ldots T^{2/3} \quad , \quad
\avg{\phi^2}_{H}\simeq 0.437\ldots T^{2/3}
\label{fi2comp}\\
&& \frac{\avg{\phi^2}^2}{\avg{\phi^4}}\simeq 0.37077\ldots \quad ,
\quad \frac{\avg{\phi^2}^2_{H}}{\avg{\phi^4}_{H}}\simeq
0.333\ldots \label{ratiocomp}\\
&&\avg{|\tilde\phi_k|^2} \simeq \frac{0.988800\ldots~T}{k^2+
1.181779\ldots T^{2/3}} \quad , \quad \avg{|\tilde\phi_k|^2}_{H}
\simeq \frac{T}{k^2+ 1.3103707\ldots T^{2/3}} \; .\label{gofkcomp}
\end{eqnarray}
Thus we see that there is agreement between the \emph{exact}
results and those in the Hartree approximation in the high
temperature limit better than $5-10\%$.

An important consequence from these results is that there is a
strong renormalization of the frequencies. In the free field
theory, which is also the low temperature limit, the frequency of
oscillation is $\omega_k = \sqrt{k^2+1}$, but in the high energy
density, or high temperature limit, the effective frequency is
$\omega_k = \sqrt{k^2+1.1818\ldots \; T^{{2/3}}}$. This is an
important observation because in a kinetic description the
particle number or distribution has to be defined with respect to
a definite frequency. It is noteworthy that the Hartree
approximation does indeed capture very efficiently the frequency
renormalization, thus suggesting that a particle number for a
kinetic description can be defined with respect to the Hartree
frequencies.

\subsection{Intermediate temperatures}

For generic temperatures only qualitative properties can be
analytically established. For instance, all eigenvalues $E_n(T)$
as well as all differences $\omega_n(T) = E_{2n+1}(T)- E_0(T)$
increase monotonically with $T$. This implies that the $L\gg1$
approximations of previous sections are uniform in $T$.

\begin{figure}[ht]
\includegraphics[height=160mm,width=150mm]{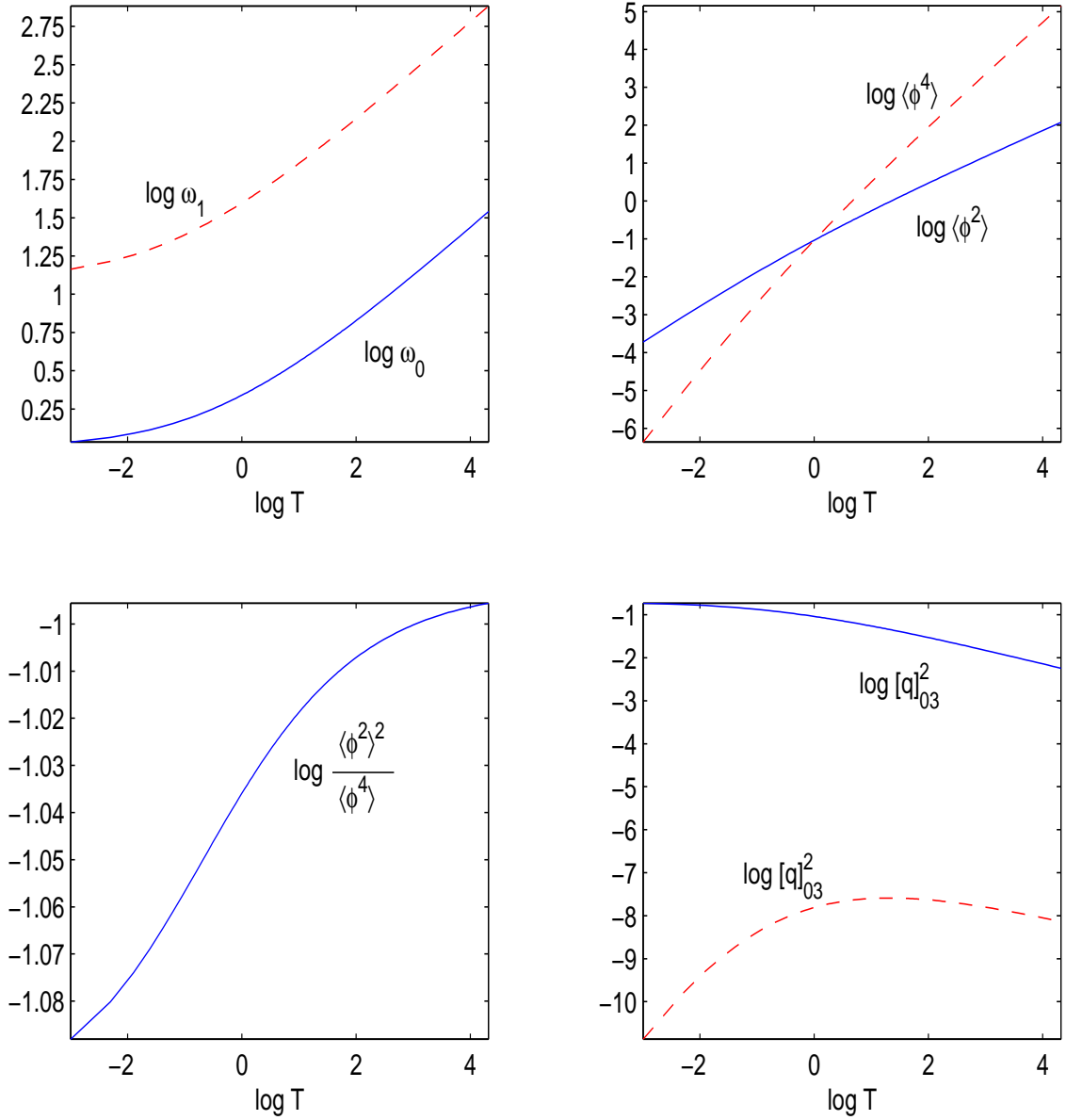}
\caption{\label{equ1} Log-log plot in the temperature of several
quantities relevant for equilibrium observables. }
\end{figure}

To obtain quantitative expressions one has to resort to the
numerical solution of the quantum anharmonic oscillator, which is
nowadays an easy task on modern personal computers. Actually,
accurate determinations of the eigenvalues for few values of the
quartic coupling were obtained already in the late seventies (see
for example refs.\cite{ind,hioe}). For example, we have at $T=2$
\cite{ind},
\begin{equation*}
E_0(2) = 0.69617582\ldots \quad , \quad E_1(2) = 2.324406352\ldots
\quad , \quad E_2(2) = 4.327524979\ldots \; .
\end{equation*}
In fig. \ref{equ1} we plot in log-log scale our numerical
determination of several equilibrium quantities. These plots
nicely interpolate between the low temperature behaviour
eqs.(\ref{f2f4fp})-(\ref{fft}) and the high temperature behaviour
eqs.(\ref{y4p})-(\ref{raz24}). Notice that the ratio
$\avg{\phi^2}^2/\avg{\phi^4} $ is monotonically increasing and
only changes by $11\%$ when $T$ goes from zero to infinity [see
eqs.\eqref{fft},\eqref{ratiocomp}] , with more or less half of the
variation concentrated at $T<2$.

\subsection{Expected dynamical evolution}\label{subsec:expec}

The exact results in the high temperature (large energy density)
regime already suggest a preliminary picture for the dynamics in
the case such that $Ta^3 \ll 1$. In order to present this
preliminary picture, it is convenient to summarize the main exact
results valid in the high temperature limit (and $Ta^3 \ll 1$).
\begin{eqnarray}
&& \langle \pi^2 \rangle = \frac{T}{2a}~~,~~ \langle \phi^2
\rangle = 0.45612\ldots T^{{2/3}} ~~,~~\langle \phi^4 \rangle =
0.56104\ldots T^{{4/3}}~~,~~ \frac{\langle \phi^2
\rangle^2}{\langle \phi^4 \rangle} = 0.37077 \ldots\label{fi2fi4}\\
&&\langle \pi^2 \rangle = \frac{T}{2a}~~,~~\langle \phi'^2
\rangle=\frac{T}{2a}\left[1-1.122\ldots (Ta^3)^{1/3}\right]
\simeq \frac{T}{2a}\label{pifip}\\
&&\langle |\tilde\pi_k|^2 \rangle=T ~~,~~ \langle |\tilde\phi_k|^2
\rangle = \frac{0.98880\ldots~T}{k^2+ 1.1818\ldots
T^{2/3}}\label{powrspec}
\end{eqnarray}
\noindent where we have assumed $Ta^3\ll 1$ in eq.(\ref{pifip}).

The important aspect gleaned from these exact results is that in
equilibrium the nonlinear term is \emph{subdominant} as compared
to the space and temporal derivatives. Namely, while the harmonic
spatial and temporal derivative terms $ \langle \pi^2 \rangle \sim
\langle \phi'^2 \rangle \sim T/a \sim E/L$, the interaction term
$\langle \phi^4 \rangle \sim T^{{4/3}}$, hence $ \langle \phi^4
  \rangle / \langle \pi^2 \rangle \sim \langle \phi^4 \rangle / \langle
\phi'^2 \rangle  \sim (Ta^3)^{1/3} \ll 1$.

\medskip

Consider solving the equation of motion (\ref{eqnofmot}) with
initial conditions $\phi(x,t=0), \; \pi(x,t=0)$ so that the
Fourier transform of $\phi,\pi$ have a power spectrum distributed
on modes $k \ll \Lambda $. Since only small wavevectors are
excited in the initial state, this entails that for large energy
density $E/L \gg 1$ it has to be that $\phi^4 \sim E/L \sim T/a$
and a large fraction of the energy density is initially stored in
the interaction term but a very small fraction is in the
derivative term $\phi'^2$ since the initial power spectrum for
$\phi'^2$ is localized at small wavevectors. Thus initially,
$\phi^4 \gg \phi'^2$. Since the time evolution conserves the
energy, the interaction between modes can only transfer the
energy. If the time evolution leads to a thermal equilibrium
state, then after or near the equilibration time scale it must be
that $\phi'^2 \gg \phi^4$ since in equilibrium the ensemble
averages (which by the ergodic postulate are equivalent to time
averages) are $\langle \phi^4 \rangle/\langle \phi'^2 \rangle \sim
(Ta^3)^{1/3}\ll 1$.

The system will therefore evolve from the initial state to a final
state of thermal equilibrium by transferring energy to modes of
larger $k$, namely the initial power spectrum will necessarily
broaden and energy will flow from the small $k$ values to larger
values in order to increase $\phi'^2$.  Interaction energy will be
then transferred to gradient energy via a cascade of energy
towards larger $k$, namely an \emph{ultraviolet cascade}. As an
equilibrium state is reached the power spectrum of $\pi$, namely
$|\tilde{\pi}_k|^2$ must become flat for all wavevectors up to the
cutoff, since in equilibrium $\langle |\tilde{\pi}_k|^2\rangle = T
$ [see eq.(\ref{pi2})].  Since initially the power spectrum has
been prepared to be localized at small values of the momenta the
flattening of the spectrum must be a direct consequence of the
ultraviolet cascade, namely more and more wavevectors are being
excited by the interaction. Interaction energy is transferred to
spatio-temporal gradient energy at the expense of the interaction
term becoming smaller. We envisage this cascade towards the
ultraviolet as a front in wavector space at a value $\bar{k}(t)$
that moves towards the cutoff as time evolves. For
$k\ll\bar{k}(t), \; |\tilde{\pi}_k|^2$ is approximately
independent of $k$ and eventually must saturate at the value $T$
when the front $\bar{k}(t)$ reaches the cutoff.

As argued above, initially $\phi^4 \gg \phi'^2$ since the initial
power spectrum for the field is concentrated at wavevectors well
below $1/a$ and the energy density (conserved in the evolution) is
$E/L \gg 1$. However if the evolution leads to thermal equilibrium
interaction energy is transferred to gradient energy via the
ultraviolet cascade as described above, resulting in that $\phi^4$
diminishes and $\phi'^2$ increases with time. Therefore there is a
time scale $t_0$ at which both terms are of the same order and the
dynamics crosses over from being dominated by the interaction to
being dominated by spatial derivatives. For $t\ll t_0, \; \phi^4
\gg \phi'^2$ while for $t \gg t_0, \; \phi^4 \ll \phi'^2$.

For $t \gg t_0$ the theory is \emph{weakly coupled} since the
interaction term is much smaller than $\phi'^2$ and $\pi^2$, the
ensuing dynamics becomes slower, the non-equilibrium dynamics
during stage is probably well described in terms of \emph{kinetic}
equations, since the small interaction guarantees a separation of
time scales.

These arguments that describe the expected dynamics are fairly
robust and hinge upon very general features: i) energy
conservation, ii) large energy density, iii) the initial condition
determined by power spectra localized at small values of
wavevectors, iv) the assumption that the dynamical evolution leads
to a state of thermal equilibrium, v) the {\bf exact} results
obtained above in thermal equilibrium, which describe the final
state of the dynamics.

A detailed numerical analysis of the evolution presented below
confirms the robust features of these arguments (and more).

\subsection{Criteria for thermalization}\label{subsec:criteria}

In the following section we undertake an exhaustive numerical
study of the non-equilibrium evolution with the goal of
understanding the dynamics that leads to thermalization and to
assess the validity of the description presented above. We solve
numerically the equation of motion (\ref{eqnofmot}) with given
initial conditions. The initial conditions determine the energy
density from which we can extract the equilibrium temperature
either via eq.(\ref{Ta}) in the low temperature limit $T\ll 1$ or
via eq.(\ref{ThiT}) in the high temperature limit $T \gtrsim 1$.
Since the classical field theory must always be understood with a
fixed lattice (or UV) cutoff $2a$, the equilibrium temperature is
obtained by providing the initial conditions, which fix the
(conserved) energy density $E/L$ and the value of the cutoff $a$.
Furthermore our primary interest is in studying the high
temperature limit $T \gtrsim 1$, hence large energy density $E/L$,
and will always work with values of the temperature and cutoff
such that $Ta^3 \ll 1 $ in which case eq.(\ref{ThiT}) entails that
the energy density $E/L \sim T/2a$.

In order to recognize when thermalization occurs we must define a
consistent and stringent set of criteria. These are the following

\begin{itemize}
\item{The ensemble averages should give the same result as the
  temporal averages over a macroscopically long time plus spatial
  volume averages.  That is, we check the validity of ergodicity.}

\item{The spatial (volume) and temporal average of $\pi^2(x,t)$
should
    approach the canonical ensemble average $\langle \pi^2 \rangle=
    T/(2a)=E/L$.  }

\item{The spatial (volume) and temporal average of
    $\phi(x,t)~, \; \phi^2(x,t),  \; [\phi'(x,t)]^2,
    ~[\dot{\phi}(x,t)]^2$ and $[\phi(x,t)]^4$
    should approach their  {\bf exact} high temperature result
    eqs.(\ref{y2p})-(\ref{y4p}) for high temperature and the behaviour
    plotted in figs. \ref{equ1} for all temperatures. }

\item{ The spatial (volume) and temporal average of the canonical
  momentum correlator must tend to $T \delta(x-x')$, which on the lattice
    would translate to $T / (2a) $ [see eqs.(\ref{corrpi})-(\ref{pi2})].}

\item{The temporal average of $|\tilde\pi_k(t)|^2$ (Fourier
transform of the
    $\pi$-correlator) , which in equilibrium is given
    by $\langle |\tilde\pi_k|^2\rangle = T$. The temporal average of
    $|\tilde\phi_k(t)|^2$ (Fourier transform of the $\phi$-correlator) must
    reach their thermal values.}

\end{itemize}

\section{Dynamics of thermalization. }\label{sec:lattice}

Having provided an analysis of exact as well as approximate
results of the cutoff classical field theory in equilibrium, we
now pass on to the study of the dynamical evolution. In this
section we study numerically the solution of the equation of
motion (\ref{eqnofmot}) with different initial conditions, of a
given large energy density. By changing the initial conditions for
a fixed energy density we are studying the evolution of different
members of a microcanonical ensemble on a fixed energy (density)
shell.

Before embarking on the numerical study, we detail below our
approach to solving the equations of motion by discretized
dynamics in light--cone coordinates.

\subsection{Discretized dynamics in light--cone coordinates}\label{ddlc}
In order to solve numerically the evolution of the $\phi^4$ theory
it is necessary to discretize space and time. We choose to do that
on a light--cone lattice\cite{cono}. In this approach space and
time are simultaneously discretized in light--cone coordinates
with the same lattice spacing $a\sqrt2$, thus preserving as much
as possible of the original relativistic invariance of the field
equation
\begin{equation} \label{evo}
  {\ddot \phi} - \phi'' + \phi + \phi^3 = 0 \; .
\end{equation}
 Moreover, we choose a scheme where the discretized dynamics
possesses an exactly conserved energy on the lattice.

 Given a space--time field configuration $\phi(x,t)$, consider
the two quantities
\begin{equation*}
\begin{split}
  \E_\pm(x,t) = &\frac1{2a} \big[\phi(x+a,t)-\phi(x,t\pm a) \big]^2 +
  \frac1{2a}\big[\phi(x-a,t)-\phi(x,t\pm a) \big]^2 + \\ &+
  \frac{a}4 \big[1 + \phi^2(x,t\pm a) \big] \big[ 2 +
  \phi^2(x+a,t) + \phi^2(x-a,t) \big] - \frac{a}2 \;.
\end{split}
\end{equation*}
In the limit $a\to0$, assuming $\phi$ to be smooth enough, we
obtain
\begin{equation*}
        \E_\pm = a \left(\dot\phi^2 + |\nabla\phi|^2
        + \phi^2 + \tfrac12 \phi^4 \right) +{\cal O}(a^2) \;,
\end{equation*}
identifying the leading term in $a$ of both $\E_+$ and $\E_-$ with
the energy density of the $\phi^4$ theory. However, to higher
orders in $a$ they do differ; in fact
\begin{equation*}
  \E_+(x,t) - \E_-(x,t) = \frac{\phi(x,t+a) - \phi(x,t-a)}{a}\; Q(x,t)\;,
\end{equation*}
where
\begin{equation*}
  Q(x,t) =\big[\phi(x,t+a) + \phi(x,t-a)\big]  \Big\{
  1 + \tfrac{a^2}4 \big[ 2 + \phi^2(x+a,t) + \phi^2(x-a,t) \big]\Big\}
  - \phi(x+a,t) - \phi(x-a,t) \;.
\end{equation*}
Hence, if $Q(x,t) = 0$, then $\E_+(x,t) = \E_-(x,t)$ also on the
lattice. In this case the total lattice energy
\begin{equation}\label{emas}
  E = \sum_n \E_+(2n\,a,t)
\end{equation}
is exactly conserved in time, since it can also be written
\begin{equation}\label{emenos}
  E = \sum_n \E_-(2n\,a+a,t+a) \;.
\end{equation}
This holds exactly on infinite space. If space is restricted to
the segment $[0,L]$, suitable boundary conditions on $\phi(x,t)$
are necessary; PBC are of this type if $L=2Na$ with $N$ an
integer.

In conclusion, we may regard $Q(x,t) = 0$ as a discrete field
equation which conserves the total energy $E$. More explicitly,
$Q(x,t) = 0$ can be written as the recursion rule
\begin{equation}\label{recursion}
  \phi(x,t+a) = -\phi(x,t-a) + \frac{\phi(x+a,t) + \phi(x-a,t)}
  {1 + \tfrac{a^2}4 \big[ 2 + \phi^2(x+a,t) + \phi^2(x-a,t) \big]}
\end{equation}
which evidently allows to propagate in time any configuration
known in a time interval of width $a$.

It is easy to check that $Q(x,t)=0$ indeed becomes eq.(\ref{evo})
in the continuum $ a \to 0 $ limit. The order $a^0$ is trivially
satisfied, odd powers of $a$ vanish identically as a consequence
of the symmetry of eq.(\ref{recursion}) under $a\to-a$, while the
order $a^2$ produces eq.(\ref{evo}).

Keeping up to $\mathcal{O}(a^4)$ in eq.(\ref{recursion}) yields,
\bea\label{correca2} {\ddot \phi}-\phi'' + \phi + \phi^3 = a^2
\left\{-\frac{1}{2}\left[ \phi'' - \phi - \phi^3 \right] (1 +
\phi^2) +
  \frac{1}{12}(\phi'''' - {\ddddot \phi} ) - \; \phi (\phi'^2 + \phi \;
  \phi'') \right\}+ {\cal O}(a^4) \; .  \eea

\bigskip

To cast eq.\eqref{recursion} in a form suitable for numerical
simulations, we define the lattice fields $ F(n,s) $ and $ G(n,s)
$ as
\begin{equation*}
 F(n,s) \equiv \phi(2n \, a,2s \, a) \; , \quad
 G(n,s)\equiv \phi([2n-1]\,a,[2s + 1] \, a)\;,  \quad n,s \in {\mathbb Z}\; .
\end{equation*}
We then obtain the iterative system
\begin{eqnarray}\label{evod}
&& F(n,s+1) = -  F(n,s)+ \frac{G(n,s)+G(n+1,s)}{1 + \frac{a^2}{4}
  \left[ 2 + G^2(n,s) + G^2(n+1,s)\right]} \cr \cr
&& G(n,s+1) = - G(n,s)+ \frac{F(n-1,s+1)+F(n,s+1)}{1 +
\frac{a^2}{4}
  \left[ 2 + F^2(n-1,s+1) + F^2(n,s+1)\right]} \; .
\end{eqnarray}
with the PBC $F(n+N,s)=F(n,s)$ and $G(n+N,s)=G(n,s)$. As initial
conditions we have to specify $ F(n,0) $ and $ G(n,0) $ for $0
\leq n \leq N-1 $. Once these values of the fields are specified,
the iteration rules (\ref{evod}) uniquely define $ F(n,s) $ and $
G(n,s) $ for all $s\neq0$. A comparison of this discretized
dynamics with other more traditional numerical treatments of
hyperbolic partial differential equations was performed in
\cite{zanlungo}. Here we only notice that this approach is
particularly efficient, stable and accurate, specially when the
continuum limit $a\to0$ and very long evolution times are of
interest.

All observables of the continuum are rewritten on the lattice in
terms of the basic fields $F(n,s)$ and $G(n,s)$. In particular,
time and space derivatives are replaced by finite diferences. We
always choose symmetric discretization rules such that lattice
observables differ from their continuum limit by ${\cal O}(a^2)$.
In the sequel, while referring to properly discretized
observables, we shall keep using the continuum notation for
simplicity,

\subsection{Initial Conditions}\label{incond}

We studied a variety of initial conditions with a fixed energy
density in our calculations. In these studies the power is
concentrated in the infrared; that is, $|{\tilde\phi}_{k}(0)|^2$
and $|{\tilde\pi}_{k}(0)|^2$ are large for wavenumbers well below
the cutoff $\Lambda = \frac{\pi}{2a} $. We considered the
following sets of initial conditions:

\begin{itemize}
\item{ \textbf{Superpositions of plane waves}:  The initial fields
    have the form
    \begin{equation}\label{ondp}
      \phi(x,0) = A\sum_i c_i \; \cos(k_i \, x+2 \, \pi \, \gamma_i) \quad ,
      \quad    \pi(x,0) = A\sum_i d_i \;  \cos(k_i \, x+2 \, \pi \,
      \delta_i) \; .
    \end{equation}
    The wavenumbers $k_i=2\pi n_i/L$ are chosen in the interval $[0,k_{\rm
      max}]$ with $k_{\rm max}=50\pi/L = 25\pi/(N \, a) \ll \pi/(2a) $.  We
    will refer to these initial conditions as {\bf hard} because these
    plane waves have sharp values for the wavenumbers. The power spectra
    $|\tilde\phi_k|^2~;~|\tilde\pi_k|^2$ are sharply peaked at
      discrete (and small )
    values of the momenta small compared with the cutoff $\Lambda =
      \frac{\pi}{2a} $.}

\item{\textbf{Superpositions of localized wave packets}: These are
    configurations of the form
    \begin{equation} \label{campa}
      \phi(x,0) = A\sum_i c_i \; w(x-x_i) \quad , \quad
      \pi(x,0) = A\sum_i d_i  \; w(x-y_i) \;.
    \end{equation}
    where we enforce the PBC by choosing,
    \begin{equation*}
      w(x) = \sum_{n \in {\mathbb Z}} w_0(k_{max}[x-n\,L])
    \end{equation*}
    with $ w_0(x) $ either the gaussian, $ w_0(x)=e^{-x^2}$, or the
    lorentzian, $w_0(x)=\frac{1}{1+x^2}$. In practice, with our choice
    $L\sim 20$, only few terms in the sum are needed. These fields have
    support throughout Fourier space, but peaked as gaussians or simple
    exponentials at low wavenumbers $ k \lesssim k_{\rm max}$.  We will
    refer to these initial conditions as {\bf soft} since these correspond
    to continuous and slowly varying power spectra.}

\item{\textbf{Random on a fixed energy shell}: A systematic way to
decrease
    fluctuations is to average over initial conditions corresponding to a
    given value of the total energy. A microcanonical description gives
    equal a priori probability to all the configurations on the same energy
    shell.

    Thus, we choose smooth initial conditions as defined by
    eqs.(\ref{ondp})-(\ref{campa}) and we average over them with unit
    weight choosing random values for the various coefficients. That is, we
    choose in eq.(\ref{ondp}) the wavenumbers $k_i=2\pi n_i/L$ at random
    both in number and in location, within the interval $[0,k_{\rm max}]$
    with $k_{\rm max}=50\pi/L = 25\pi/(N \, a) \ll \pi/(2a) $. The
    positions $x_i\,,y_i$ in eq.(\ref{campa}) are chosen at random in
    $[0,L]$. The phases in eq.(\ref{ondp}), $\gamma_i\,,\delta_i$, and the
    relative amplitudes $c_i\,,d_i$ are also chosen at random in the
    interval $[0,1]$. Finally, for any given realization of
    $k_i\,,\gamma_i\,,\delta_i\,,c_i\,,d_i$, the overall amplitude $A$ is
    fixed through the energy density $E/L$ so all the configurations chosen
    are on the same energy shell.

    Typically, we performed averages over 30 initial conditions. }
\end{itemize}

\noindent For a fixed energy (density) the overall amplitude $A$
in eqs.\eqref{ondp} and \eqref{campa} is fixed for any choice of
$c_i,\,d_i,\,\gamma_i,\,\delta_i$ by the requirement that the
initial configurations had always the same energy $E$.  We
considered several values of the energy density $E/L$ ranging from
$2.6$ to $ \sim 10^4 $.

We notice also that all our initial configurations have vanishing
total momentum $\int dx\,\pi\phi'$, which is a conserved quantity
for PBC.

\subsection{Averaged observables}

The key observables in our investigation are the basic quantities
\begin{equation} \label{lista}
  \phi, \,\phi^2, \,\phi^4, \,{\dot \phi}^2, \,{\phi'}^2 \; ,
\end{equation}
as well as the power spectra of $\phi$ and $\pi$, that is
$|{\tilde\phi}_{k}(t)|^2$ and $|{\tilde\pi}_{k}(t)|^2$, where,
\begin{equation*}
  {\tilde\phi}_{k}(t) \equiv \int_0^L \frac{dx}{\sqrt{L}} \; e^{ikx} \;
  \phi(x,t) \quad , \quad {\tilde\pi}_{k}(t) \equiv \int_0^L
  \frac{dx}{\sqrt{L}} \; e^{ikx} \; \pi(x,t) \; ,
\end{equation*}
as in eq.(\ref{fourier}).

In reference\cite{wett} the correlation function of the canonical
momenta was studied which is particularly important since in
equilibrium in the classical theory it is Gaussian. We here study
\emph{a large number} of independent observables and correlation
functions since the criteria established above for thermalization
based on the \emph{exact} results apply to many different
correlation functions.

In accordance with the anticipation at the end of sec. \ref{ddlc},
we are using here the continuum notation also for the Fourier
transforms, although they actually are discrete Fourier
transforms.

The fluctuations of all these observables do not vanish upon time
evolution. Hence for generic initial conditions they do not have a
limit as $t\to\infty$. These are fine--grained or microscopic
observables. Typically there are several spatio-temporal scales,
the microscopic scales correspond to very fast oscillations and
short distance variations that are of no relevance to a
thermodynamic description. We are interested in longer,
macroscopic scales that describe the relaxation of observables
towards a state of equilibrium.

In particular the ergodic postulate states that ensemble averages
must be identified with \emph{long} time averages as well as
spatial averages over macroscopic-sized regions.

To make contact between the time evolution and the thermal
averages we need to properly average the microscopic fluctuations.

First of all, for local quantities such as those in
eq.\eqref{lista} we take the spatial average. Secondly, we take
suitable time averages of all key observables in the following way
\begin{equation}\label{promT}
  \overline{\phi^2}(t) \equiv
  \frac1\tau\int_{t-\tau}^t dt' \; \;\frac1{L}\; \int_0^L dx\,
                [\phi(x,t')]^2 \;.
\end{equation}
where $\tau\gg a $ stands for the length of the time interval
where we average. We find that the period of the fast (microscopic
oscillations) themselves vary in time suggesting a sliding
averaging in which $ \tau $ grows in time to compensate for the
growth in the period of the fast time variation. We find that a
practical and efficient manner to implement this averaging is to
use
\begin{equation} \label{taut}
  \tau(t) = \tau(0) + C \; t
\end{equation}
where $\tau(0);C$ small and positive with typical values $ \sim
0.1 $. In this way $t \gg \tau(t)$ and the dependence on the
initial values becomes negligible for practically accessible
times. This method is quite effective in revealing general
features of the (logarithmic) time evolution such as the presence
of distinct stages characterized by well separated macroscopic
time scales.

\begin{figure}[htbp]
\includegraphics[height=100mm]{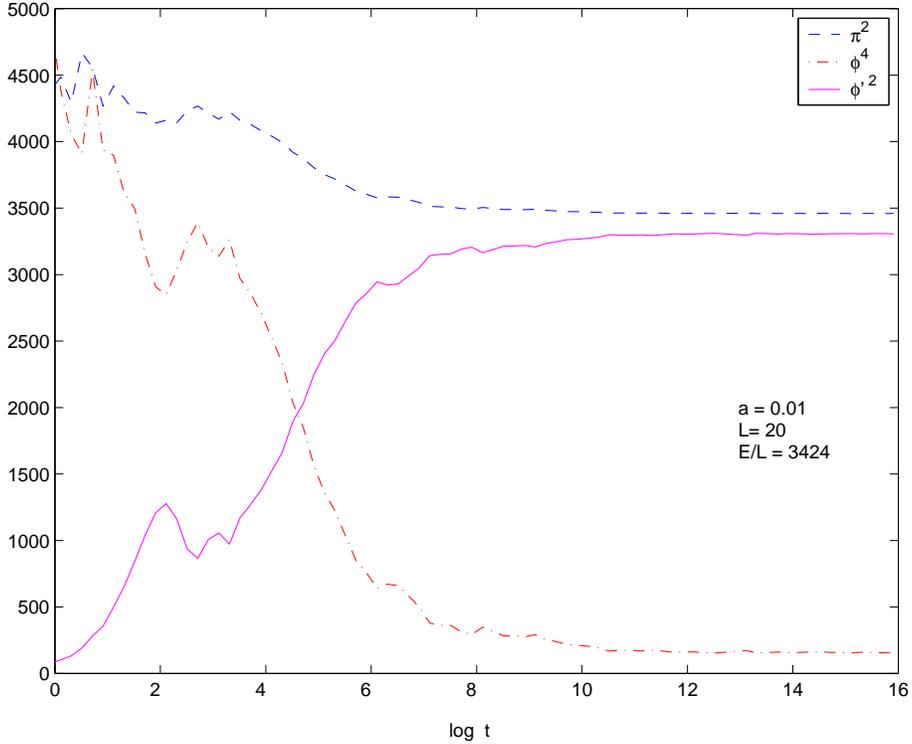}
\caption{${\overline {\phi'^2}}(t), \; {\overline {\phi^4}}(t) $
  and $ {\overline {\pi^2}}(t)$ as a function of the logarithm of the
  physical time for $E/L = 3424 , \; L = 20 $ and $ a = 0.01 $.}
\label{gfiT69NA001}
\end{figure}

\begin{figure}[htbp]
\includegraphics[height=100mm]{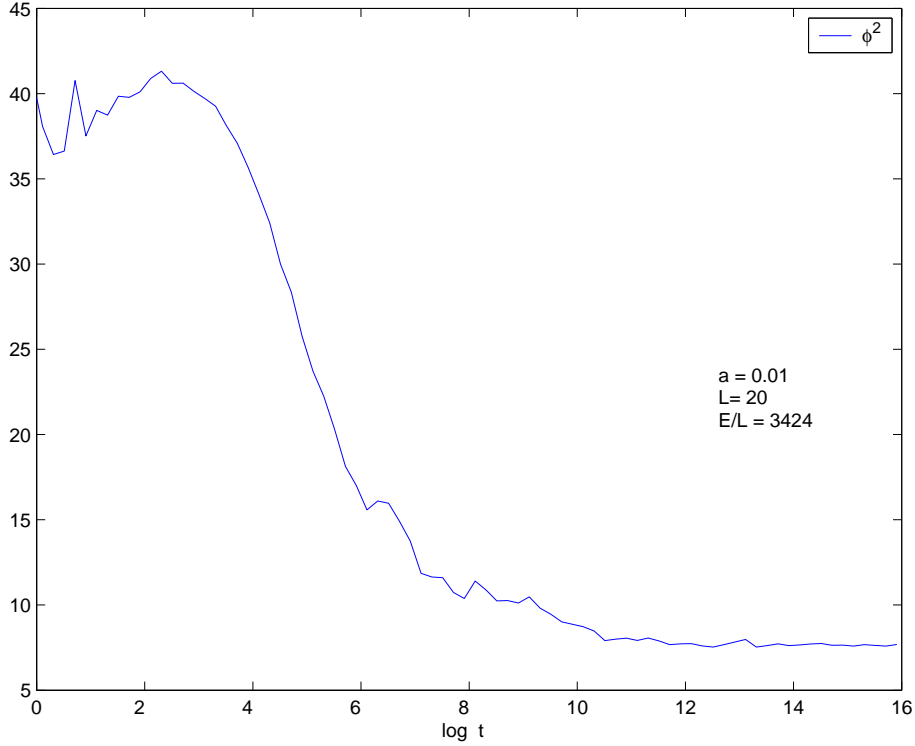}
\caption{${\overline {\phi^2} }(t)$ as a function of the
  logarithm of time for $T = 69.2 , \; E/L = 3424 , \; L = 20 $ and $ a =
  0.01 $.}
\label{fi2T69NA001}
\end{figure}

Altogether, we denote the results of all coarse-grainings simply
with an overbar (not to be confused with the complete time average
of section \ref{secvir}) to avoid cluttering of notation. For
example we have in the case of random initial conditions on the
energy shell
\begin{equation*}
  \overline{\phi^4}(t) = \frac{1}{M} \sum_{i=1}^M \;
  \frac1\tau \; \int_{t-\tau}^t dt' \; \frac1{L} \; \int_0^L dx \;
                [\phi^{(i)}(x,t')]^4 \;.
\end{equation*}
where the superscript $^{(i)}$ labels the $M$ different choices of
smooth initial conditions, all with the same energy density $E/L$,
we have taken $M\sim 30$ in our study. Likewise,
\be\label{promcondi}
  \overline{|{\tilde\pi}_k|^2}(t) =  \frac{1}{M} \sum_{i=1}^M \;
  \frac1\tau \; \int_{t-\tau}^t dt' \; |{\tilde\pi^{(i)}}_{k}(t')|^2
\ee where we used the reality condition ${\tilde\pi}_k =
{\tilde\pi}_{-k}^{\,\ast}$. Analogous expressions hold for $\phi,
\; \phi^2, \;  \phi^4,  \; \pi^2,  \; \phi'^2$ and
$|{\tilde\phi}_{k}|^2$.

In particular, due to the linearity of these averages, we have the
sum rules:
\begin{equation}\label{sumrules}
  \int_{-\Lambda}^{+\Lambda} \dk \; \overline{|{\tilde\phi}_k|^2}(t) =
  \overline{\phi^2}(t)
  \quad ,\quad
  \int_{-\Lambda}^{+\Lambda} \dk \; \overline{|{\tilde\pi}_k|^2}(t) =
  \overline{\pi^2}(t) \quad , \quad
  \int_{-\Lambda}^{+\Lambda} \dk \; k^2 \; \overline{|{\tilde\phi}_k|^2}(t) =
  \overline{\phi'^2}(t)\;.
\end{equation}
Where UV cutoff in the light--cone lattice is $\Lambda=\pi/(2a)$.

The power spectra $\overline{|{\tilde\phi_k}|^2}(t)$ and
$\overline{|{\tilde\pi_k}|^2}(t)$ are connected to equal--time
correlation functions of $\phi(x,t)$ and $\pi(x,t)$, respectively,
just as it happens at thermal equilibrium [see {\em e.g.}
eq.\eqref{Gtopws}]. We have, for instance
\begin{equation}\label{phiphi}
\begin{split}
  \overline{|{\tilde\phi_k}|^2}(t) &= \int_0^L dx\; e^{-ikx} \;
  \overline{\phi\phi}(x,t) \;, \\
  \overline{\phi\phi}(x,t) &=
  \frac1M \sum_{i=1}^M \frac1\tau \int_{t-\tau}^t dt' \;
  \frac1L\int_0^L dy \; \phi^{(i)}(y,t') \; \phi^{(i)}(x+y,t') \; .
\end{split}
\end{equation}
with similar relations between $\overline{|{\tilde\pi_k}|^2}(t)$
and or $\overline{\pi\pi}(x,t)$ and between
$k^2\overline{|{\tilde\phi_k}|^2}(t)$ and
$\overline{\phi'\phi'}(x,t)$.

\bigskip

\begin{figure}[htbp]
\includegraphics[height=100mm]{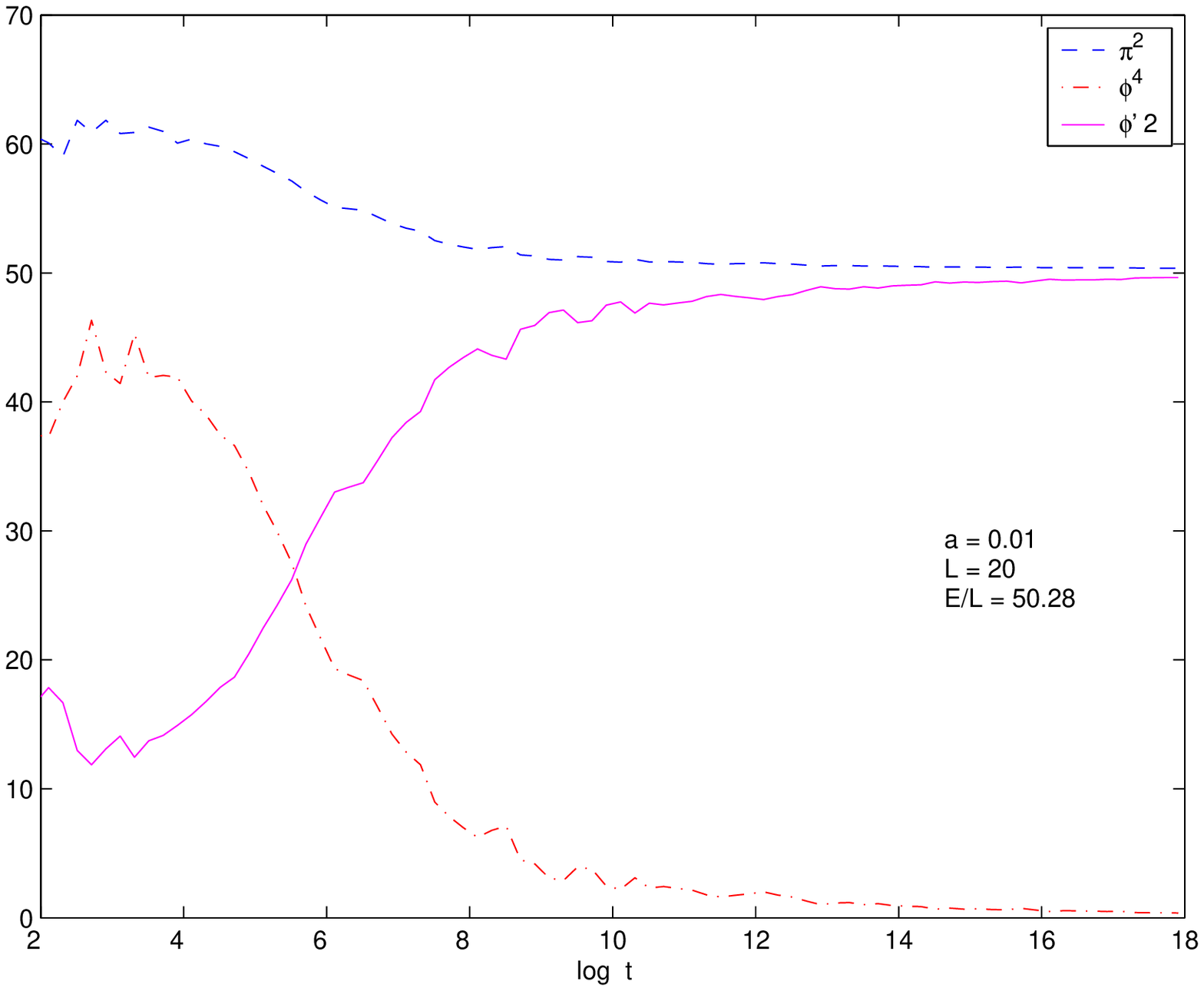}
\caption{${\overline {\phi'^2}}(t), \; {\overline {\phi^4}}(t) $
and $ {\overline {\pi^2}}(t)$ as a function of the
  logarithm of the physical  time   for $ E/L = 50.28,
\;   L = 20 $ and $ a = 0.01 $.} \label{gfiT1NA001}
\end{figure}

\begin{figure}[htbp]
\includegraphics[height=100mm]{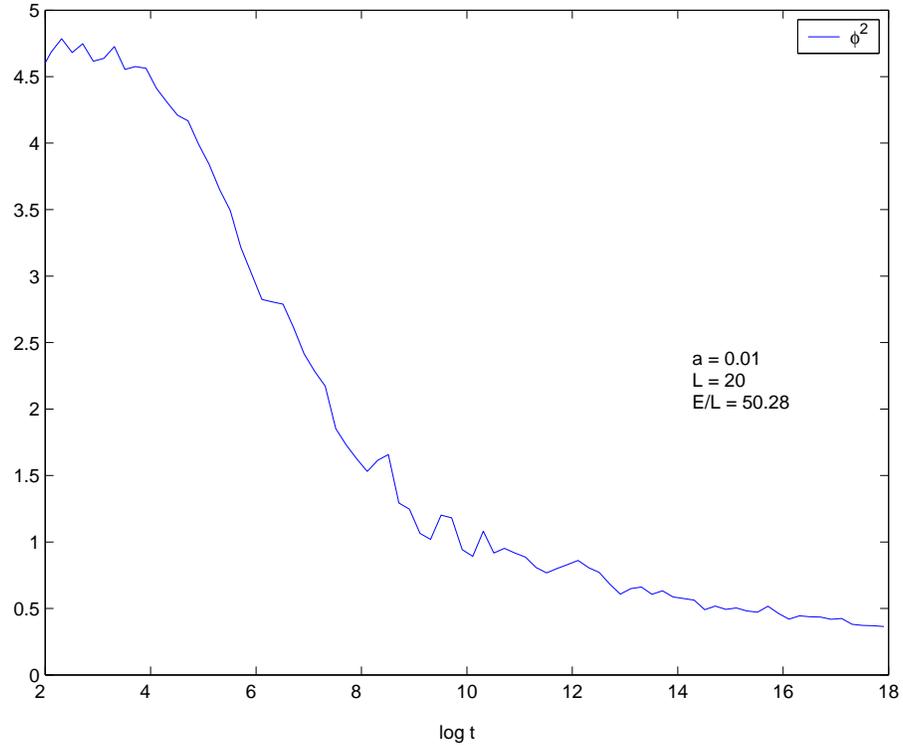}
\caption{${\overline {\phi^2} }(t)$ as a function of the
  logarithm of time    for $ E/L = 50.28,
\;   L = 20 $ and $ a = 0.01 $.} \label{fi2T1NA001}
\end{figure}

It is useful to define  the normalized power spectrum of $\pi$,
\begin{equation}\label{npws}
   P(k,t) \equiv \frac{\overline{|{\tilde\pi}_k|^2}(t)}{\overline{\pi^2}(t)}\;,
\end{equation}
which describes the distribution of power over the wavenumbers, it
is  normalized so that, \be\label{norP} \int_{-\Lambda}^{+\Lambda}
\dk \; P(k,t) = 1 \ee

Another important quantity of paramount importance to describe the
cascade described in section \ref{subsec:expec} above is the
average wavenumber
\begin{equation}\label{defkbar}
   \bk(t) =  \int^{+\Lambda}_{-\Lambda} \dk \; |k| \; P(k,t) \;.
\end{equation}
In thermal equilibrium $\langle |\tilde{\pi}_k|^2 \rangle = T$,
$\avg{\pi^2} = T/(2a) = T\Lambda/\pi$, therefore in equilibrium
$\bk_{eq} = \Lambda/2$.

The physical significance of this quantity becomes obvious by
considering the situation in which the power spectrum
$|\tilde{\pi}_k|^2$ is approximately flat in a region of
wavevectors $|k|\leq k_M$ and negligible elsewhere, namely
$|\tilde{\pi}_k|^2 \propto \Theta(k^2_M-k^2)$. In this case
$\bk(t) \equiv \frac{k_M}{2}$. The relevance of this effective
quantity will become clear below when we study in detail the
process of cascade of energy towards the UV, described in section
\ref{subsec:expec}, where it will become clear that $2 \bk(t)$
determines the front of the ultraviolet cascade as described in
section \ref{subsec:expec}.

To summarize, we perform spatio-temporal averages in the isolated
system with fixed energy (density) which is equivalent to
microcanonical ensemble averages at long time by the ergodic
postulate. In the thermodynamic limit, it is expected that
microcanonical and canonical ensembles will yield the same
equilibrium results provided the equilibrium temperature $T$ is
identified with the energy density as per eqs.(\ref{Ta}) or
(\ref{ThiT}) in the low or high temperature limit respectively.

While in refs. \cite{wett} and \cite{zanlungo} only the variance
of $\pi^2$ was studied as a measure of gaussianity and
thermalization, we study \emph{many different} correlators, since
as described above the exact solution in thermal equilibrium
furnishes a stringent set of criteria for thermalization.

\subsection{Time evolution of basic observables}\label{basico}

We use the lattice field equations, eq.\eqref{evod}, to evolve the
initial configurations in time and compute the time and space
average of the basic quantities (\ref{lista}) as  a function of
time for $ 10 < t < 10^8 $ and $ 0.1 < a < 0.0001 $.

We confirmed  that the lattice energy
eqs.(\ref{emas})-(\ref{emenos}) is indeed conserved with large
accuracy.

Figs. \ref{gfiT69NA001}, \ref{fi2T69NA001}, display ${\overline
  {{\phi'}^2}}(t), \; {\overline{\phi^4}}(t) ,\; {\overline{ \pi^2}}(t) $
and $ {{\overline {\phi^2}}}(t) $, respectively, as functions of
time for $ E/L = 3424, \; L = 20 $ and $ a = 0.01 $ corresponding
to an equilibrium temperature $T\sim 68.5$.  Figs.
\ref{gfiT1NA001} and \ref{fi2T1NA001}, display the same quantities
as functions of time for $ E/L = 50.28, \; L = 20 $ and $ a = 0.01
$

In figs. \ref{gfiT69NA001}-\ref{fi2T1NA001} the initial conditions
are the plane waves eq.(\ref{ondp}) with parameters that fix
$\pi^2 \sim \phi^4 \sim E/L$ initially. No average over initial
conditions is performed and we used the sliding time average with
linearly growing time intervals as in eq.(\ref{taut}).

These figures clearly reveal the expected dynamics as described in
section \ref{subsec:expec} above. Initially $\bar{\phi'^2}(t)$ is
small reflecting the fact that the initial conditions determine a
power spectrum localized at wavectors $ k \ll 1/a $. The mode
mixing entailed by the interaction is transferring power to larger
wavevectors, thus effectively transferring energy from the
interaction term, which diminishes, to the spatial gradient term
which increases. As is clear from these figures, all magnitudes
tend to a limit for late times. The late time limits are the
thermal equilibrium values, as we discuss below in detail,

The growth of $ {\overline {{\phi'}^2}}(t) $ at the expense of the
interaction term $ {\overline {\phi^4}}(t)$ shows that
thermalization is a result of the flow of energy towards larger
$k$ modes, namely the ultraviolet cascade ultimately leads to the
thermal equilibrium state.

\bigskip

We find {\bf three} distinct stages of evolution for a  wide
choice of initial conditions.

\begin{figure}[htbp]
\includegraphics[height=100mm]{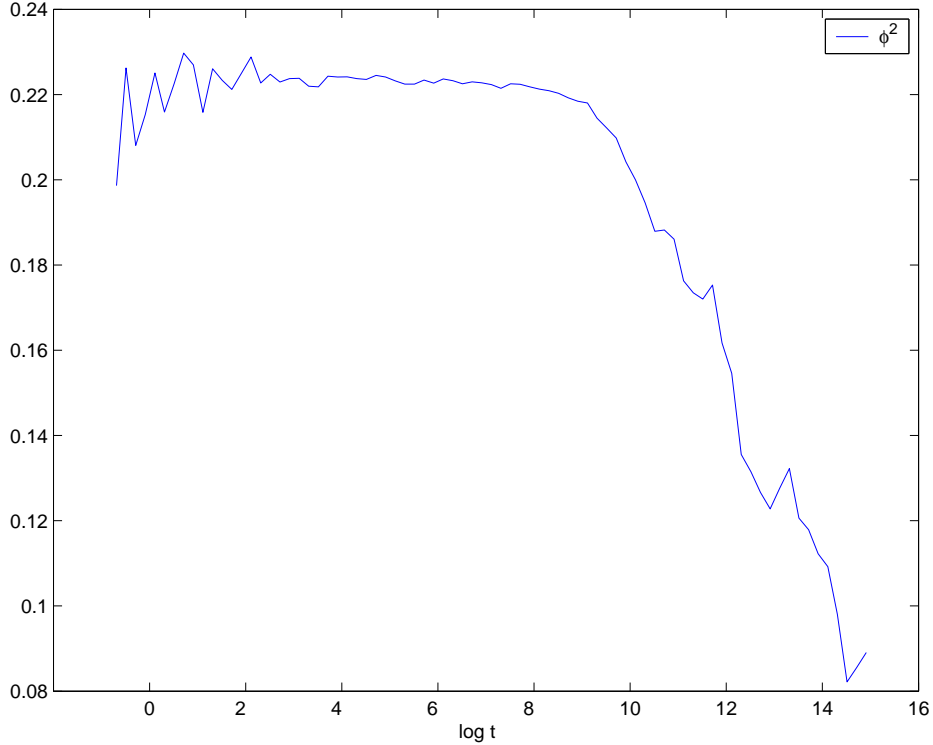}
\caption{${\overline {\phi^2} }(t)$ as a function of the logarithm
   of the  time $ t $ for $ E/L = 3.67  , \;  L = 20 $ and $ a
   = 0.01 $. For such a low value of $ E/L $ the cascade
   starts much later for $ \log t_0 \sim 17$}
\label{Tbaja}
\end{figure}

\begin{itemize}

\item
  A first stage with relatively important fluctuations and whose precise
  structure depends on the initial conditions. Such stage can be seen in
  figs. \ref{gfiT69NA001}-\ref{fi2T1NA001} for $ \ln t < 5 $, that is $ t
  < 200 $. The value $ t_i \sim 200 $ turns to be independent of both
  $a$ and $E/L$ as long $E/L$ is not very small. Namely, we
  find $ t_i \sim 200 $ for $ 0.1 > a > 0.0002 $ and $ E/L
  \gtrsim 10 $. For lower values of $E/L, \; t_i $ increases
  sharply as shown in figs. \ref{Tbaja} and \ref{ratio24}. In
  addition, for such low values of
  $E/L, \; t_i $ becomes dependent on the details of the initial
  conditions but is not relevant for our study which focused on the high
  density case. During this first stage the transfer of energies via the
  cascade begins to be operative and becomes most effective at the time
  scale $t_i$. We see that $t_i$ can be identified with the time scale at
  which the interaction and gradient terms are of the same order and the
  crossover from a strongly interacting $\phi^4\gg\phi'^2$ to a weakly
  interacting theory occurs. Thus the first stage corresponds to $t \leq
  t_i \sim 200$ during which the cascade is the result of large
  interaction energy which is redistributed to larger wavevectors via mode
  mixing and the dynamics is dominated by the interaction.

\item
  There is a second stage where ${\overline {{\phi'}^2}}(t), \; {\overline
    {{\dot \phi}^2}}(t), \; {\overline{ \phi^2}}(t) $ and $
  {\overline{\phi^4}}(t)$ are about the same order $\sim E/L$.  During this
  stage there is a crossover from a strongly to a weakly interacting theory
  since at the end of this second stage the spatio-temporal gradient terms
  are much larger than the non-linear term. During this second stage the
  cascade is very efficient in redistributing the energy.

  In figs. \ref{gfiT69NA001}-\ref{fi2T1NA001} this second stage corresponds
  to $ 5 < \ln t < 10 $ for $ a=0.01 $.

  We find that the behaviour of the observables during this second
  transient stage depends to some extent on the initial conditions.  For
  hard initial conditions [as eq.(\ref{ondp})] we find much {\bf steeper}
  curves for ${\overline {{\phi'}^2}}(t), \; {\overline {{\dot
        \phi}^2}}(t), \; {\overline{ \phi^2}}(t) $ and $
  {\overline{\phi^4}}(t)$ than for soft initial conditions [as
  eqs.(\ref{campa})]. Thus this second stage corresponds to an
  interval after  the time at which the spatial gradient and
  interaction terms cross.

  While the details of the dynamics during this stage depend on the initial
  conditions, the \emph{presence} of this stage during a time interval $5
  \lesssim \ln t \lesssim 10 \sim \ln t_0 $ is fairly robust. We found
  such intermediate
  stage for all types of initial conditions, and a wide range of lattice
  spacings and energy density.

During this second stage $\overline{\pi^2}(t)$ varies with time
more slowly than ${\overline {{\phi'}^2}}(t), \;  {\overline{
\phi^2}}(t) $ and $
  {\overline{\phi^4}}(t)$.

\item
  After this transient second stage there is a third stage where these
  physical quantities approach their thermal equilibrium values. The
  third stage starts at a time scale $ t_0 \sim 50000 $. The value of
  $ t_0 $ turns to be independent  of both
  $a$ and $E/L$ as long $E/L$ is not very small. That is, for $ E/L
  \gtrsim 10 $.

In particular, ${\overline{\pi^2}(t)}$ slowly decreases to its
asymptotic
  value which agrees with the thermal equilibrium value $T/(2a)$, that is,
  using eq.\eqref{Ta}:
  \begin{equation}\label{pi2t}
    \lim_{t\to\infty}\overline{\pi^2}(t) = \frac{T}{2a} =   \frac{E}{L} +
    {\cal O}(a^2) \;.
  \end{equation}
  The dynamics during this third stage is also driven by the ultraviolet
  cascade but unlike the earlier two stages wherein the details depend on
  the initial conditions, we find that this third stage is
  described by a {\bf universal cascade} independent of the initial
  conditions as we shall see in the next section. In particular, the
  results obtained during the third stage for sharp initial conditions
  [eq.(\ref{ondp})] are independent of the chosen wavenumbres $ k_i $
  provided they do not approach the lattice cutoff. The third stage
  ends at a time $ \sim t_1 $ when the lattice size effects start to
  play a role. For the cases depicted in
  figs. \ref{gfiT69NA001}-\ref{fi2T69NA001} and
  figs. \ref{gfiT1NA001}-\ref{fi2T1NA001}
we have $ \ln t_1 \sim 12 $ and  $ \ln t_1 \sim 17 $,
respectively.

\item The fourth  and final stage extends for times beyond
  $t_1$. Thermalization is reached here, strictly speaking, only for
  infinite times. Since the dynamics feels here the details of the
  discretization adopted, we shall not study this stage in detail.

\end{itemize}

\begin{figure}[htbp]
\includegraphics[height=95mm]{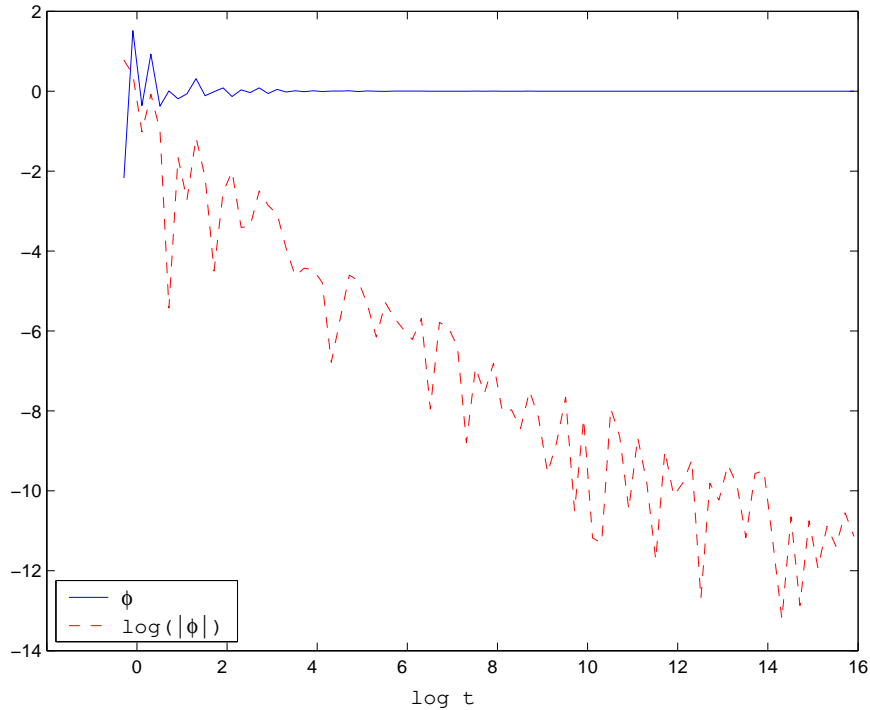}
\caption{${\overline {\phi} }(t)$ and $\log|{\overline {\phi}
}(t)|$ as a function of the   logarithm of the  time $ t $
   for $ E/L = 3424  , \;  L = 20 $ and $ a = 0.01 $.}
\label{fimT69NA001}
\end{figure}

\begin{figure}[htbp]
\includegraphics[height=95mm]{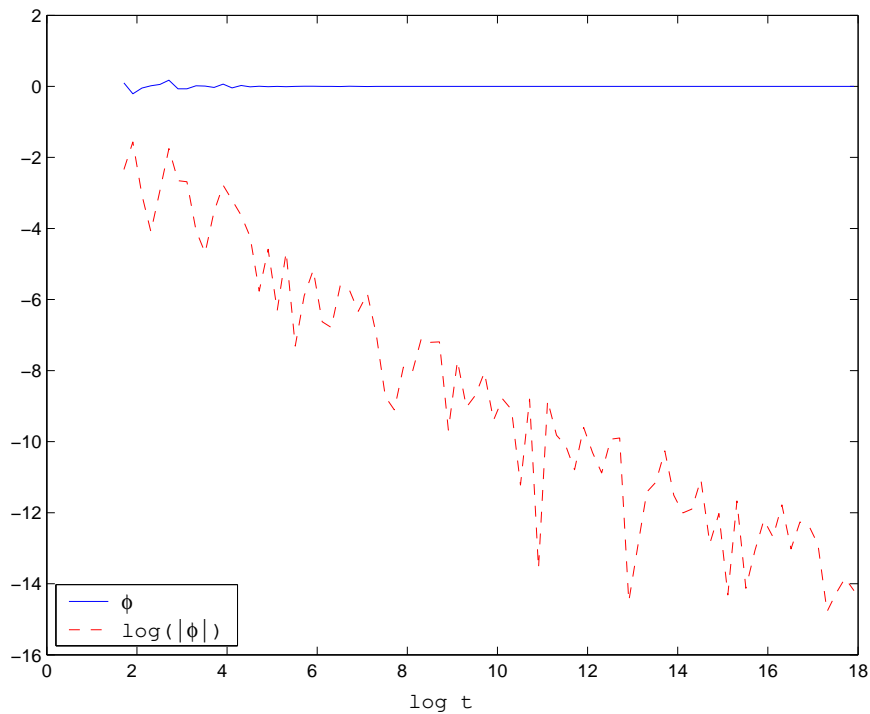}
\caption{${\overline {\phi} }(t)$ and $\log|{\overline {\phi}
}(t)|$ as a function of the   logarithm of the  time $ t $
   for $ E/L = 50.28 , \;  L = 20 $ and $ a = 0.01 $.}
\label{fimT1NA001}
\end{figure}

Figs. \ref{fimT69NA001}-\ref{fimT1NA001} display ${\overline
{\phi}}(t)$ and $\log|{\overline {\phi}}(t)|$ as a function of the
   logarithm of the   time for  $ E/L = 3424 $ and $
   E/L = 50.28 $ for $  L = 20 $ and $ a = 0.01 $, respectively.

   We see that the relaxation of $ \phi $ towards its thermal equilibrium
value ($\phi=0$) is different from the other physical quantities
previously discussed. This is due to the fact that the vanishing
of ${\overline   {\phi}}$ is connected to a symmetry of the model.
We find that ${\overline {\phi}}(t)$ relaxes as  $1/t$ for $ t
\lesssim t_0 $ (during the first two stages of thermalization) and
as $ \sim 1/\sqrt{t}$ for $ t \gtrsim t_0 $ (during the two
subsequent stages).

\subsection{Time Evolution of the Correlation Functions}\label{correti}

We now turn our attention to the study of correlation functions.

According to the Fourier transform relationship eq.\eqref{phiphi}
between the power spectrum $\overline{|{\tilde\phi}|^2}(k,t)$ and
the equal--time correlation function $\overline{\phi\phi}(x,t)$,
there are two approaches to the numerical evolution of such
quantities (we specialize here on $\phi$ but the discussions
applies equally well to $\pi$). We extract the field $\phi(x,t)$
from the lattice fields $F(n,s)$ and $G(n,s)$, Fourier--transform
it to ${\tilde\phi}_k(t)$ and then perform all needed averages on
$|{\tilde\phi}_k(t)|^2$. Or we directly compute averages of the
correlations of $F(n,s)$ and $G(n,s)$ and extract from them the
correlations $\overline{\phi\phi}(x,t)$ and
$\overline{\pi\pi}(x,t)$. We found that both methods yield the
same numerical results (see Appendix).

Moreover, when using the approach with growing time averages as in
eq.\eqref{taut} with a unique initial condition, one realizes that
the simply time--averaged correlation
\begin{equation}\label{corfi}
  \frac1\tau \; \int_{t-\tau}^t dt' \; \phi(x,t') \; \phi(x',t') \; .
\end{equation}
very soon (in the logarithm of time) becomes translation
invariant, that is a function only on the distance $|x - x'|$,
making the time consuming space average unnecessary.

\begin{figure}[ht]
\includegraphics[height=100mm]{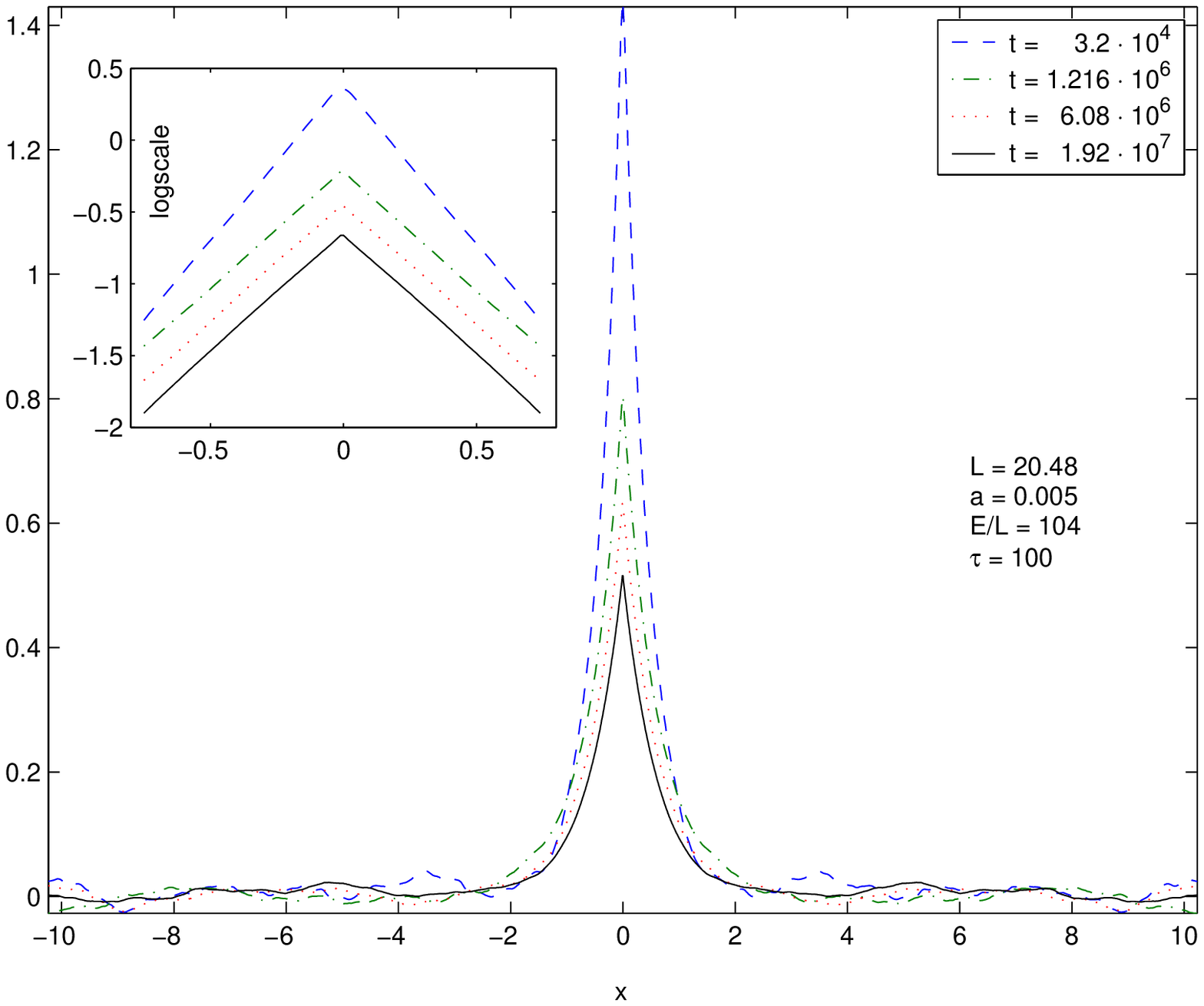}
\caption{\label{Gcorr} The correlation $\overline{\phi\phi}(x,t)$
as a
  function of $x$ at four different times. It takes the thermal form
  eq.(\ref{aprcor}) for an effective temperature $\Teff$.}
\end{figure}

\begin{figure}[ht]
\includegraphics[height=100mm]{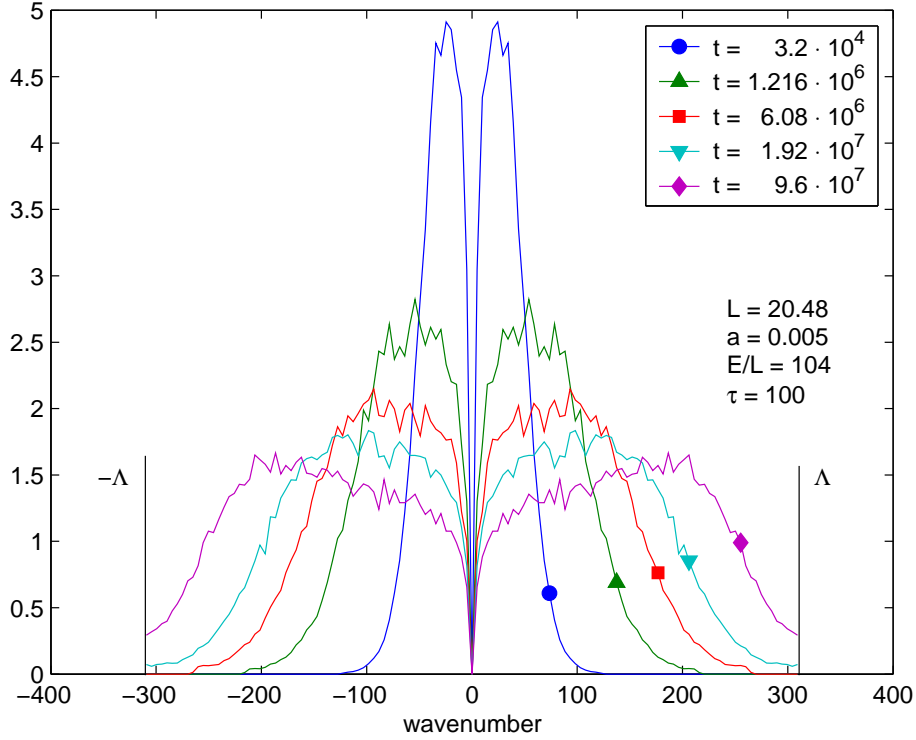}
\caption{\label{k2pw1} $k^2\overline{|{\tilde\phi_k}|^2}(t)$ vs.
$k$
  at the same times of fig. \ref{Gcorr}. It is described in the
  central plateau by the thermal spectrum  eq.(\ref{aprcor}) for an
  effective temperature $\Teff$. $\Teff$ can be read here
  approximately from the height of the plateau.}
\end{figure}

\begin{figure}[ht]
\includegraphics[height=100mm]{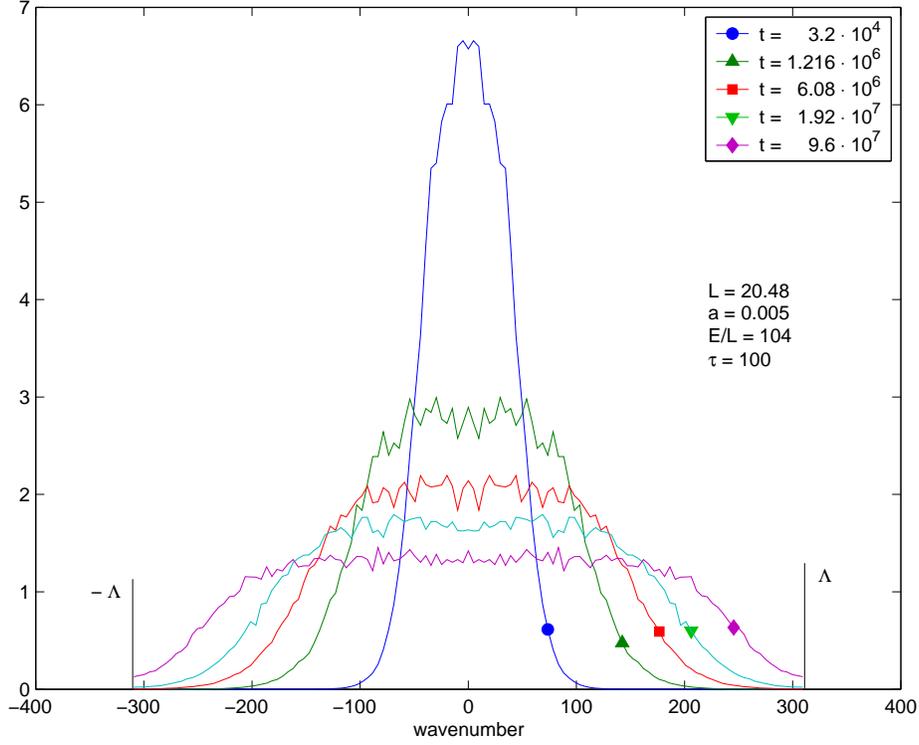}
\caption{\label{pw2} $\overline{|{\tilde\pi_k}|^2}(t)$ vs. $k$ at
the
  same times of figs. \ref{Gcorr} and \ref{k2pw1}. It is described in the
  central plateau by the flat thermal spectrum  eq.(\ref{Tflat})for an
  effective temperature $\Teff$. The height of the plateau is
  identical to fig.\ref{k2pw1} as expected. }
\end{figure}

We plot in figs. \ref{Gcorr}, \ref{k2pw1} and \ref{pw2} thee
example of the typical profiles of $\overline{\phi\phi}(x,t)$,
$k^2\,\overline{|{\tilde\phi}_k|^2}(t)$ and
$\overline{|{\tilde\pi}_k|^2}(t)$, respectively, for one given
choice of parameters. We see from fig \ref{Gcorr}. that the angle
point at $x=0$, characteristic of the equilibrium correlation
function $G(x)$ [see
  eqs.\eqref{twopoint} and \eqref{twopoint2}] is developping. This
appears more evident in Fourier space, fig \ref{k2pw1}, since the
region where $k^2\,\overline{|{\tilde\phi}_k|^2}(t)$ is large and
almost constant is spreading towards the UV cutoff. Likewise we
see the same UV cascade for $\overline{|{\tilde\pi}_k|^2}(t)$ in
fig \ref{pw2}.  The power spectrum of the canonical momentum $\pi$
will be discussed in detail in the next section.


\begin{figure}[ht]
\begin{turn}{-90}
\includegraphics[height=120mm]{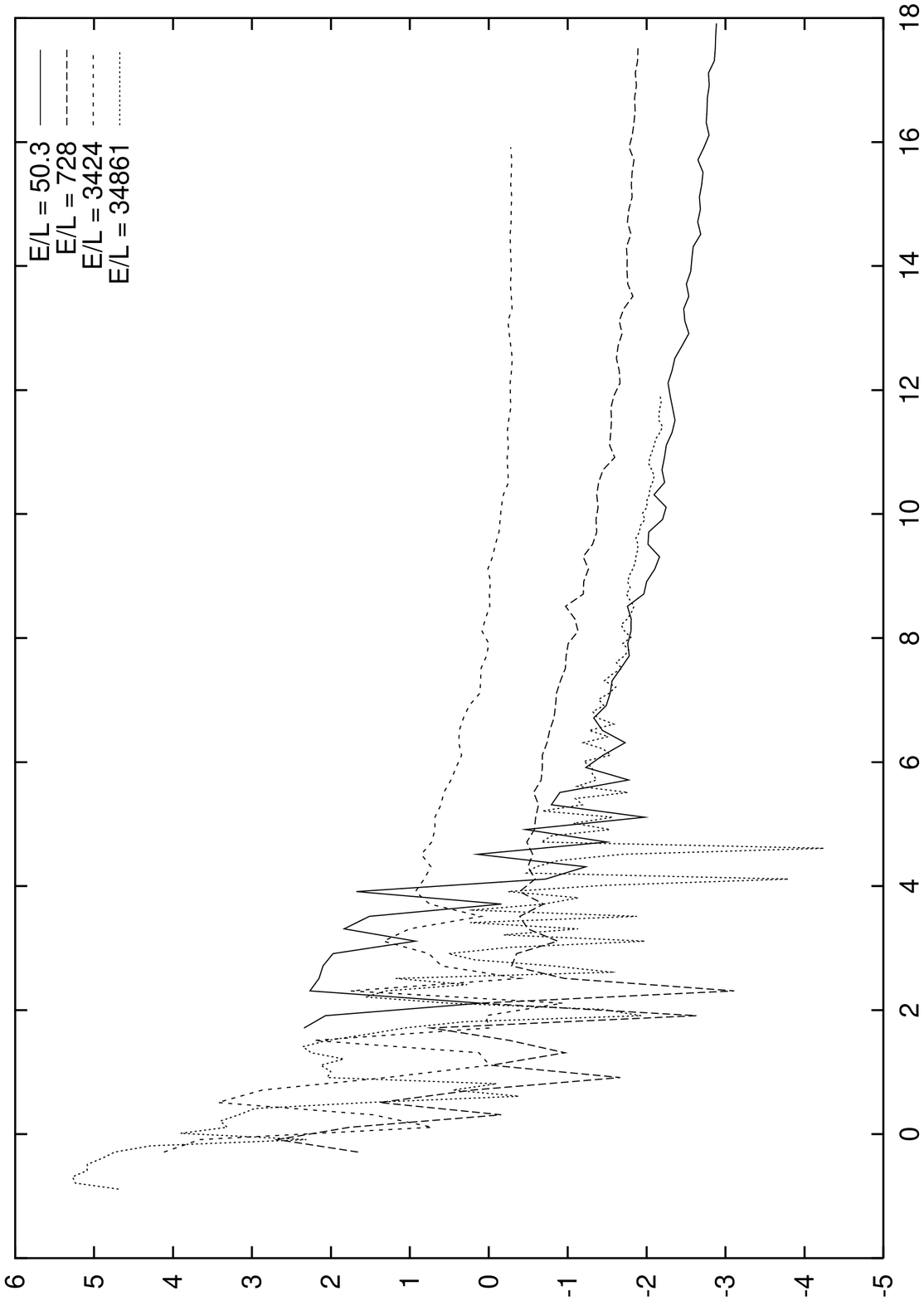}
\end{turn}
\caption{\label{vir} Virialization: the logarithm of
$\frac{L^2}{E^2
    \; a^3} \times   |\Delta(t)|$ vs.
  the logarithm of the time for $ E/L = 50.3, \;
  728$ and $3424$ with $ a = 0.01 $ and for $ E/L = 34861 $
  with $ a = 0.001 $.}
\end{figure}

\subsection{Early Virialization}\label{viria}

We depict in fig. \ref{vir} the quantity,
\begin{equation*}
\Delta(t) \equiv\avg{\dot\phi^2}(t) - \avg{\phi'^2}(t) -
\avg{\phi^2}(t) - \avg{\phi^4}(t) \; .
\end{equation*}
This quantity vanishes when the virial theorem is fulfilled [see
  eq.(\ref{virial2})]. It turns out to be negative for finite times
and nonzero $a$. We see from fig. \ref{vir} that $|\Delta(t)|$
starts to decrease at times earlier than $ t_0 $. Therefore, the
model starts to virialize {\bf before} it starts to thermalize.
$|\Delta(t)|$ keeps decreasing with time and tends  to a nonzero
value which is of the order $ {\cal O} (a^3) $ for $ t \to \infty
$. This is to be expected since eq.(\ref{virial2}) only holds in
the continuum limit and receives corrections in the lattice.

\section{The Energy Cascade} \label{casca}

We describe here the flow of energy towards higher frequencies
leading towards thermalization. Such cascade turns to be universal
(independent of the lattice spacing and of $E/L $) and exhibits
scaling properties within a wide range  of time.

\subsection{Power spectrum of $\pi$ and the universal cascade}

The chosen initial conditions eqs.(\ref{ondp})--(\ref{campa}) are
such that the power is concentrated in long wavelength modes $k$
well below the ultraviolet cutoff $\Lambda = \pi/(2a)$. Therefore,
$\overline{|{\tilde\pi_k}|^2}(0)\,;\,\overline{|{\tilde\phi_k}|^2}(0)$
are concentrated on small $k$. During the time evolution the
non-linearity  gradually transfers energy off to higher $k-$modes
leading to the ultraviolet cascade as discussed above.

\begin{figure}[ht]
\includegraphics[height=100mm]{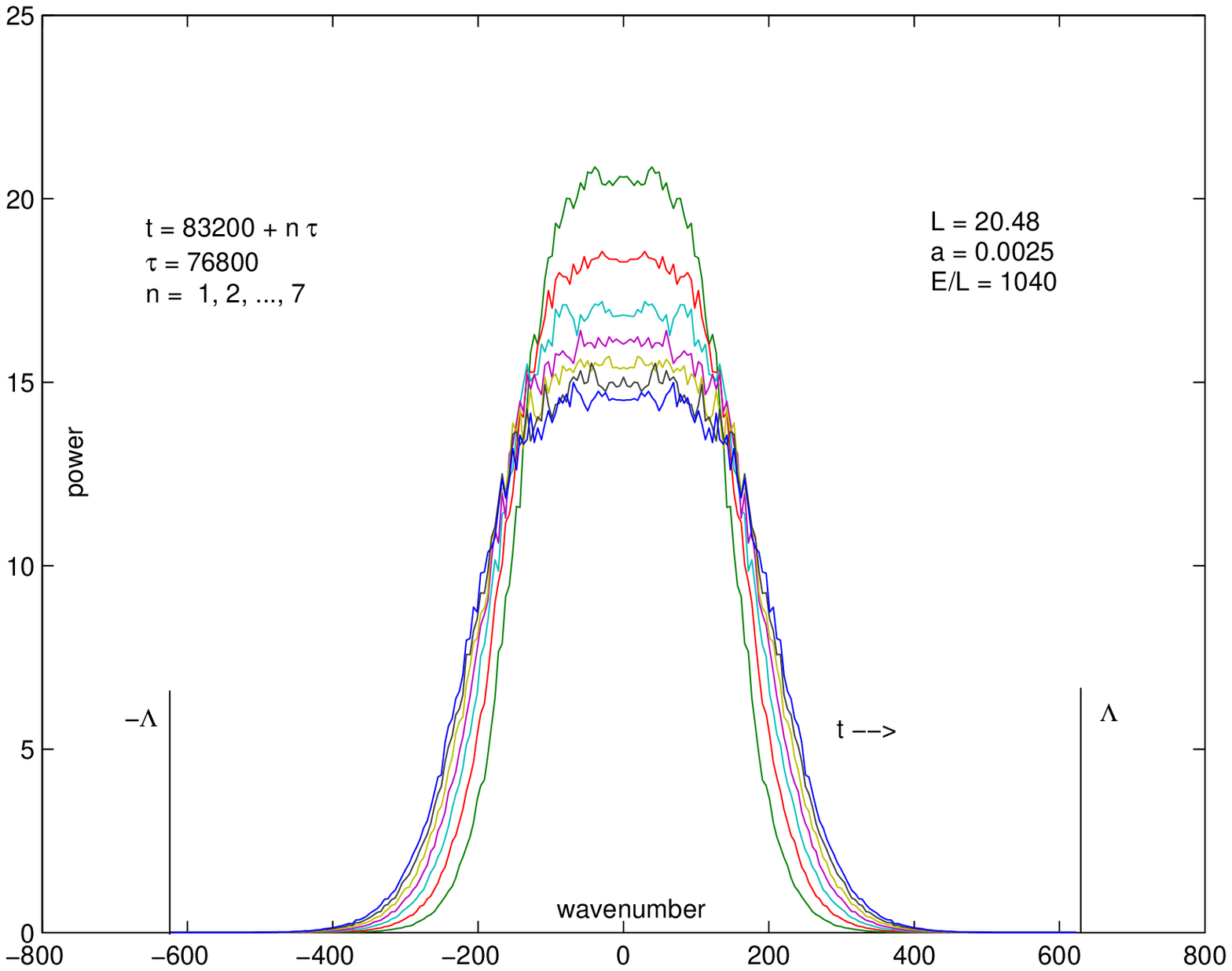}
\caption{\label{cascade1} $\overline{|{\tilde\pi_k}|^2}(t)$
cascade still
 far from the UV cutoff.}
\end{figure}

\begin{figure}[ht]
\includegraphics[height=100mm]{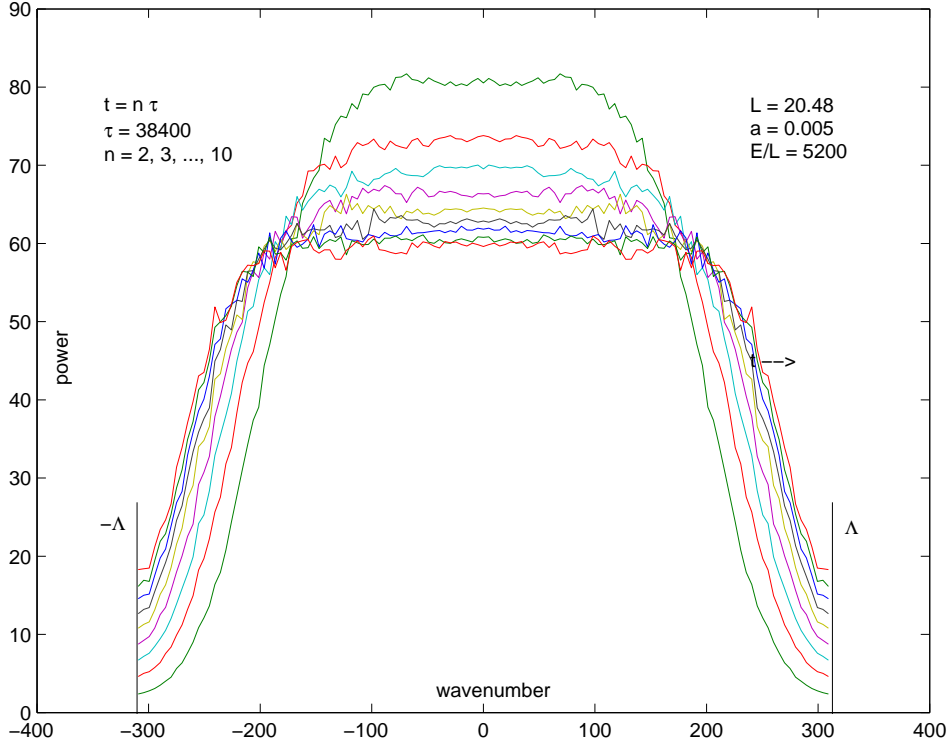}
\caption{\label{cascade0} $\overline{|{\tilde\pi_k}|^2}(t)$
cascade reaching the UV cutoff.}
\end{figure}

\begin{figure}[ht]
\includegraphics[height=100mm]{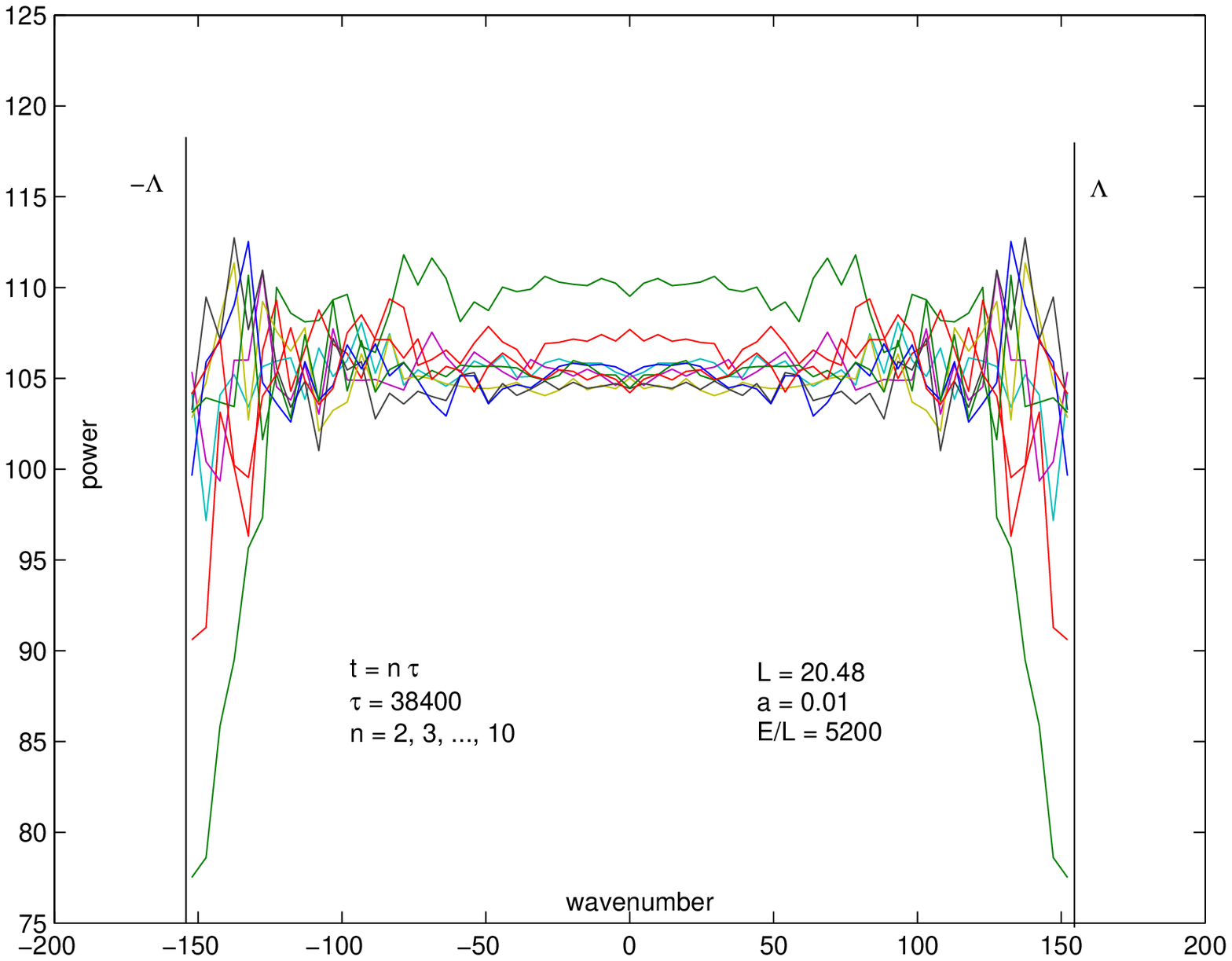}
\caption{\label{cascade2} $\overline{|{\tilde\pi_k}|^2}(t)$
cascade close to complete thermalization.}
\end{figure}

The typical behaviour of $\overline{|{\tilde\pi_k}|^2}(t)$ is
shown in figs. \ref{cascade1}, \ref{cascade0} and \ref{cascade2}.
In these figures we have averaged over time with constant
intervals $ \tau $, and on the initial conditions as defined by
eq.(\ref{promcondi}). This plots have also been smoothed by a
moving average over wavenumbers with averaging intervals of size
$32\pi/L$ [see the Appendix for details].


The time evolution of the power spectrum
$\overline{|{\tilde\pi_k}|^2}(t)$ features the ultraviolet cascade
in  a very clear manner. The gradual transfer of energy to larger
wavevectors results in a the formation of a central plateau which
spreads over higher wavenumbers as time grows decreasing its
height. This plateau ends abruptly at a value of the wavector
which determines the front of the cascade. From the definition of
$\bk(t)$ given by eq.(\ref{defkbar}) and the discussion that
follows it,  it is clear that this cascade front is given by $2 \;
\bk(t)$.

\begin{figure}[ht]
\includegraphics[height=100mm]{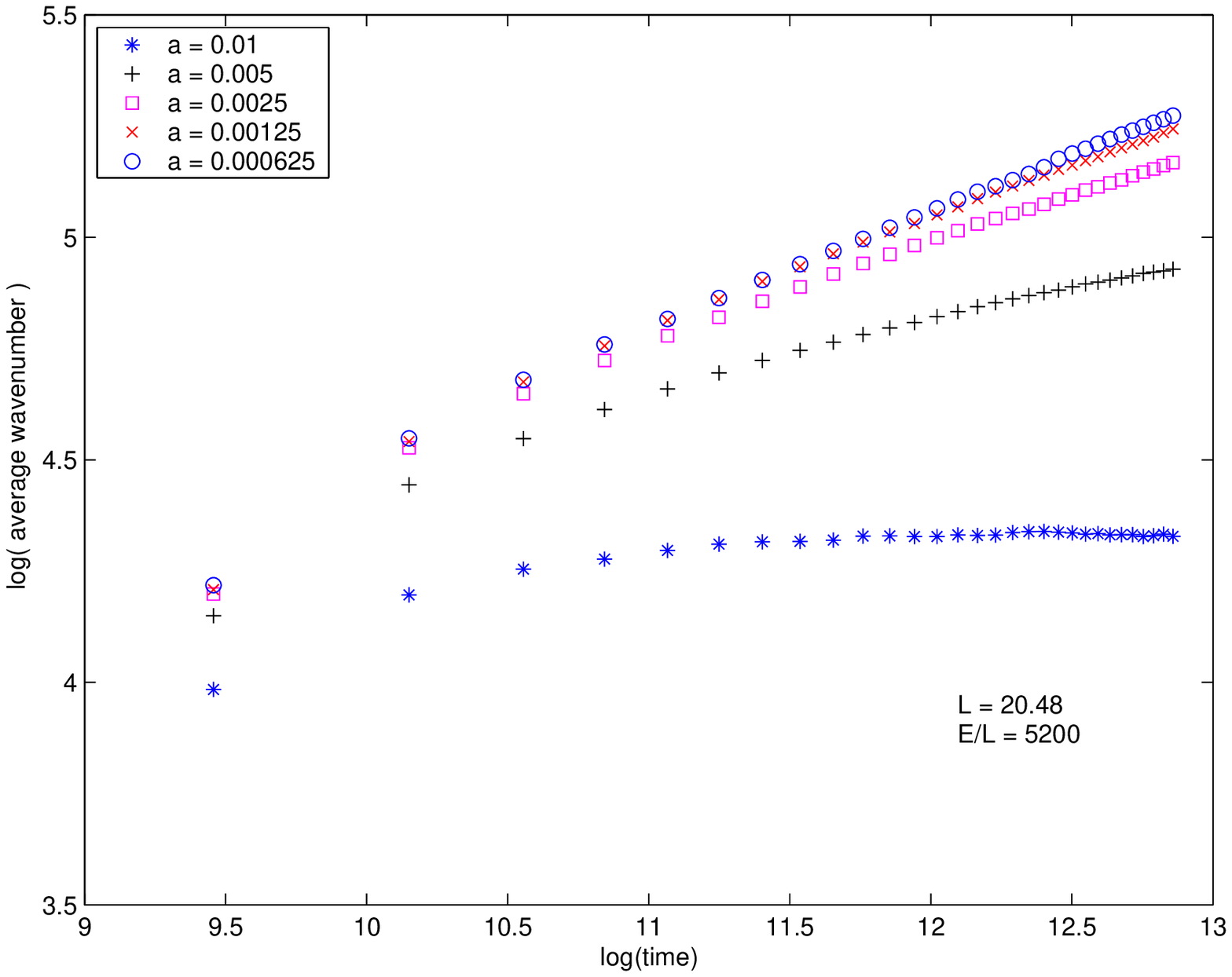}
\caption{\label{bark} Numerical evaluation of the average
wavenumber
  $\bk(t)$ for different values of the UV cutoff, at fixed energy density.
  When $a=0.01$ the system almost thermalizes [see fig. \ref{cascade2}] and
  $\bk(t)$ saturates to its equilibrium value $\Lambda/2$.
  When $a=0.005$ the bulk of the cascade reaches the UV cutoff, causing a
  significant bending. For smaller $a$ the bulk of the cascade is still
  far away from the cutoff at the times depicted here.}
\end{figure}

The limiting form  of $\overline{|{\tilde\pi_k}|^2}(t)$ for
$t\to\infty$ is flat, as expected from thermal equilibrium,
eq.\eqref{Tflat} [see fig. \ref{cascade2}]:
\begin{equation}\label{limpower}
  \lim_{t\to\infty}\overline{|{\tilde\pi_k}|^2}(t) = T \;.
\end{equation}
The second and third stages described in sec. III.D for $t\gg t_0$
correspond to the steady flow of energy towards higher wavenumbers
with a steady increase of the average wavenumber $\bk(t)$  towards
its asymptotic value of thermal equilibrium [see figs. \ref{bark}]
and decrease of the height of the plateau. Thus the front of the
cascade $2 \; \bk(t)$ advances towards its limiting value, given
by the cutoff $2 \; \bk(\infty)=\Lambda=\frac{\pi}{2a} $ leaving
behind a wake in local thermal equilibrium at a temperature
$\Teff$ corresponding to the height of the plateau.

Figs. \ref{cascade1}-\ref{cascade2} clearly show that there are
only two relevant wavevector scales in the power spectrum
$\overline{|{\tilde\pi_k}|^2}(t)$: the front of the cascade $2 \;
\bk(t)$ and the momentum cutoff $\Lambda$. Neglecting fluctuations
in the plateau, it is clear from these figures that for $k
\lesssim 2 \; \bk(t)$ the power spectrum is flat just as in the
equilibrium case, and for $2 \; \bk(t) \ll \Lambda$ the shape of
the power spectrum is very simple and insensitive to the cutoff.

\begin{figure}[ht]
\includegraphics[height=140mm]{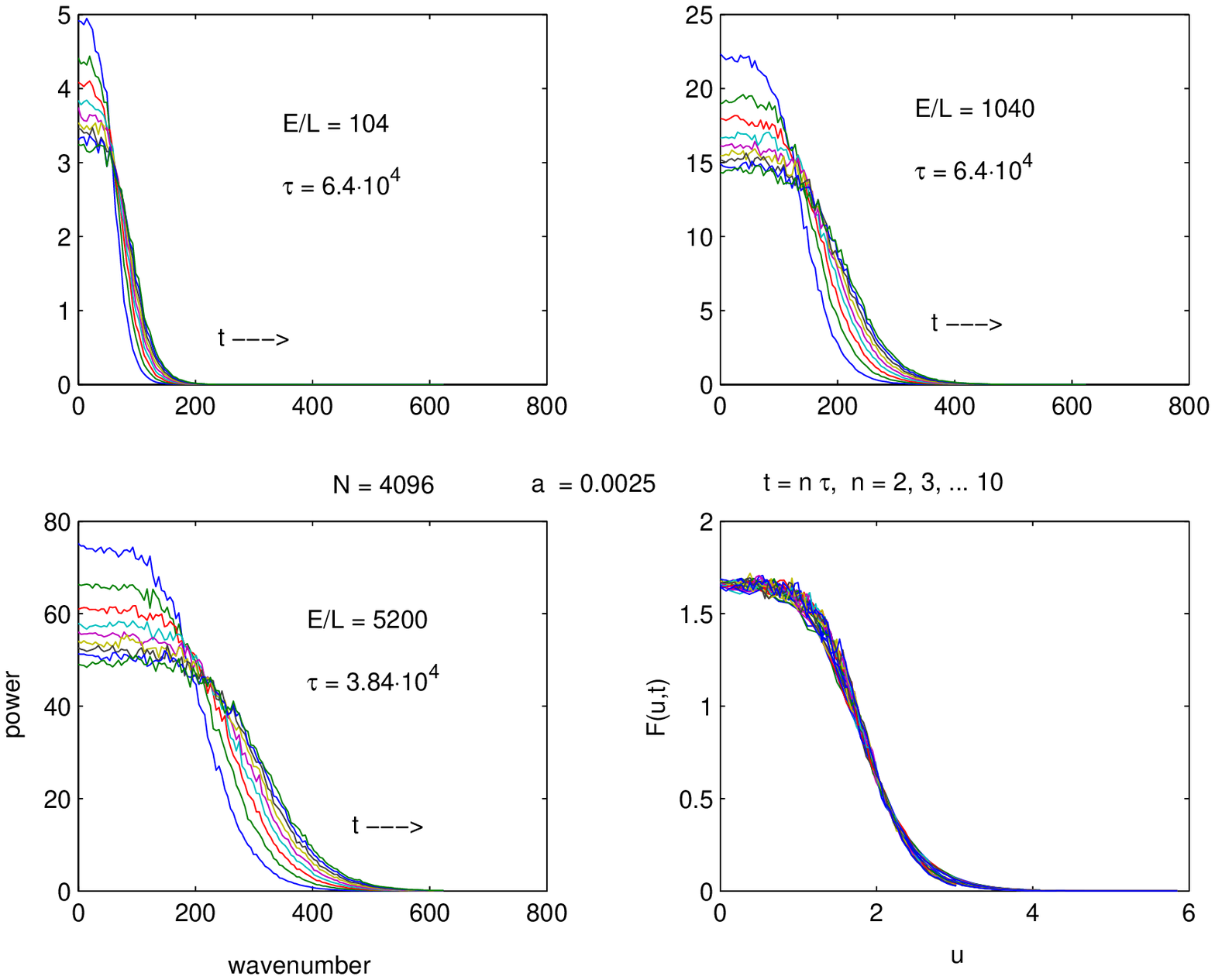}
\caption{\label{cascade} The ultraviolet cascade for $k\ge0$ and
three
  different values of the energy density (top and bottom-left). In all
  three cases and for all different times, a very good data collapse is
  obtained, upon rescaling, on the universal scaling function
  $F(u,0)$ (bottom--right).}
\end{figure}

\begin{figure}[ht]
\includegraphics[height=100mm]{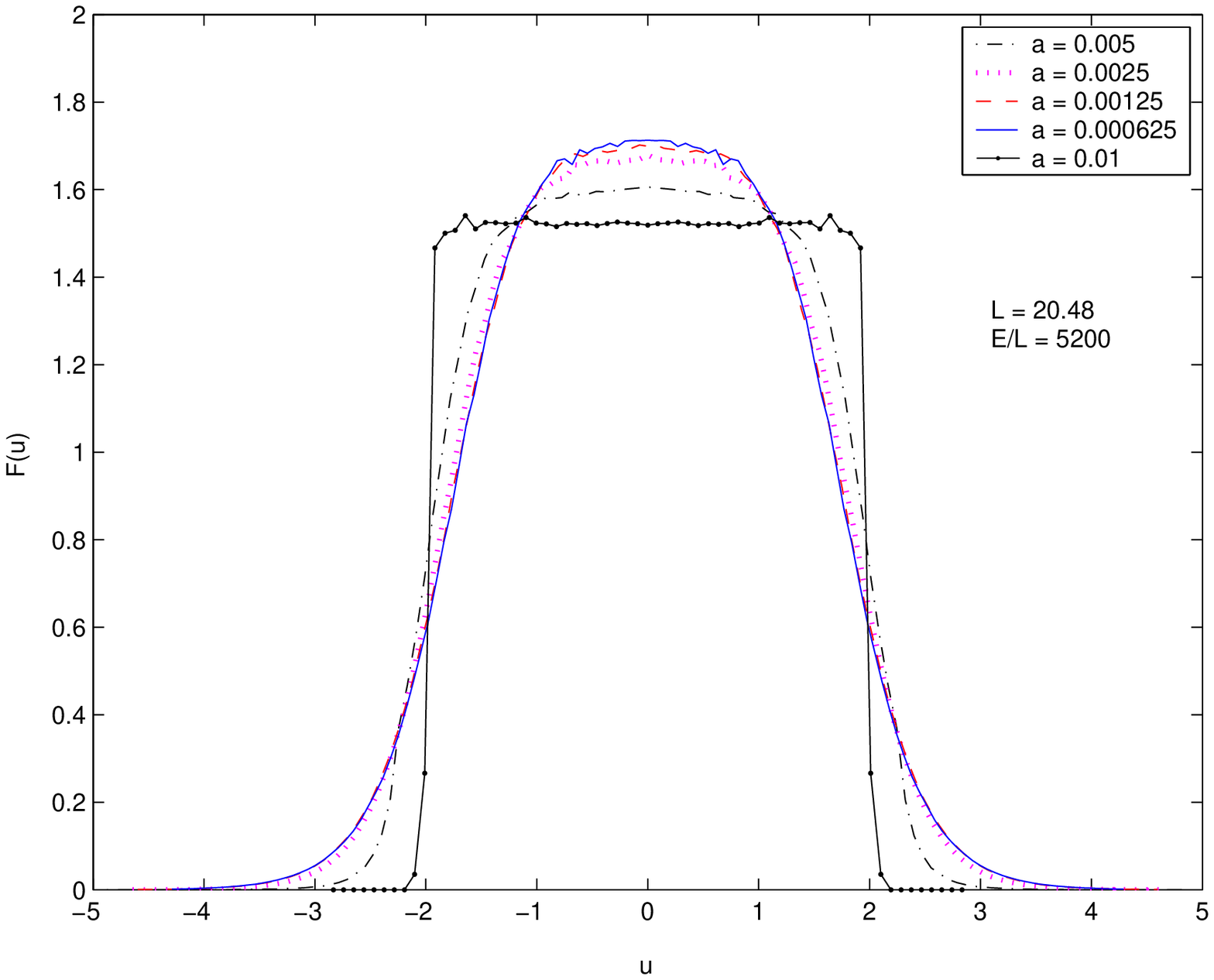}
\caption{\label{sfun} Numerical results for the universal function
  $F(u,v)$ obtained through time averages of $P(k,t)$ from $t=51200$ to
  $t=384000$, for several values of the UV cutoff, at fixed energy
  density. When $a=0.01$ the cascade is almost over and we obtain a profile
  very close to $F(u,1)$ [see eq.(\ref{limsfun})]. When $a=0.005$ the
  bulk of the
  cascade reaches the UV cutoff, causing significant deformations. In the
  other cases the bulk of the cascade is still far away from the cutoff and
  $F(u,0)$ is approached. $F(u,0)$ is obtained in the continuum limit
  $a\to0$. }
\end{figure}

We therefore reexpress the distribution $P(k,t)$ defined by
 eq.(\ref{npws}) in terms of dimensionless ratios as follows,
\begin{equation}\label{sfun0}
    P(k,t) =   \frac{1}{\bk(t)} \,F (u,v) \;,\quad
    u \equiv \frac{k}{\bk(t)}\;, \quad
    v(t) \equiv \frac{2\,\bk(t)}{\Lambda}=\frac{4}{\pi} \; \bk(t) \; a  \;.
\end{equation}
\noindent with the asymptotic behavior of $v(t)$ given by
\begin{equation*}
v(t) \buildrel{t\rightarrow \infty}\over = 1 \label{asyv}
\end{equation*}
Our numerical analysis shows that during a fairly large time
interval $t_1\gg t\gg t_0$, during which $2 \; \bk(t) \ll
\Lambda$, the shape function $F(u,v)$ becomes {\bf universal},
namely independent on the initial conditions, including the energy
density $E/L$ [see fig. \ref{cascade}], and independent on the UV
cutoff as well [see fig \ref{sfun}]. The new time scale $t_1$ is
determined by when the front of the cascade begins to reach the
cutoff, namely $2 \; \bk(t_1) \sim \Lambda$. At this point the
wake in the power spectrum behind the front of the cascade
corresponds to the plateau at the temperature $T$, namely $T_{\rm
eff}(t_1) \simeq T$. The time scale $t_1$ marks the end of the
third stage. Beyond it, nonuniversal ($a$-dependent) effects
become relevant.

Our exhaustive numerical analysis leads us to conclude the
following picture for the cascade and the process of
thermalization. For a wide range of initial conditions, lattice
cutoff and energy density, there exists a {\bf scaling window} for
the cascade, namely a time interval $t_1\gg t \gg t_0$
characterized by the following properties:
\begin{itemize}
\item
  For $t_1\gg t \gg t_0$ so that $2 \; \bk(t)\ll \Lambda$ we find that
  $\bk(t)$ is described by power--like behaviour in time, as shown for
  instance in fig.\ref{bark}, namely
  \begin{equation}\label{bkh}
    \bk(t) \simeq h_1 \; t^\alpha\;;
  \end{equation}
  where $\alpha$ is slowly growing with $E/L$, typically $0.21\le \alpha \le
  0.25$ for $ E/L \gtrsim 10 $ , while $ h_1\sim (E/L)^\gamma$, with
  $\gamma\sim 0.25$, for $E/L \gtrsim 10 $. The numerical evidence
  suggests that both $\alpha$ and $\gamma$ depend on the initial conditions only
  through $E/L$.

\item
  During the interval $t_1\gg t\gg t_0$ the front of the cascade is far
  away from the cutoff, namely $v\ll 1$ and $F(u,v)$ depends very weakly on
  $v$ therefore $F(u,2 \, \frac{\bk(t)}{\Lambda})$ is almost time
  independent. Hence,
  during this interval we can make the approximation $F(u,v) \simeq F(u,0)$
  so that $P(k,t)$ satisfies the following {\bf scaling law } with great
  accuracy [see fig. \ref{cascade}]
  \begin{equation}\label{sclaw}
    P(k,t) \simeq  \frac1{\bk(t)} \, F\Big(\frac{k}{\bk(t)}\,, 0\Big)
    \quad \mbox{for} \quad a \; \bk(t) \ll 1 \; .
  \end{equation}
  Hence $P(k,t)$ satisfies a renormalization group-like relation for
  its time dependence, namely
  \begin{equation}\label{ley}
    P(k,t)\simeq  \frac{\bk(t')}{\bk(t)} \; P\Big(
    \frac{\bk(t')}{\bk(t)} \; k,\,t' \Big)\;.
  \end{equation}
  and $F(u,0)$ plays the role of {\bf scaling function}.

\end{itemize}

The behavior of $\bk(t)$ as a function of the logarithm of time
for several values of the lattice cutoff $a$ displayed in
fig.\ref{bark} shows a power-law behavior given by eq.(\ref{bkh})
in the window  $t_1\gg t \gg t_0$ and a saturation for times
larger than $t_1$ (which  depends on the lattice cutoff).

\begin{figure}[ht]
  \includegraphics[height=100mm]{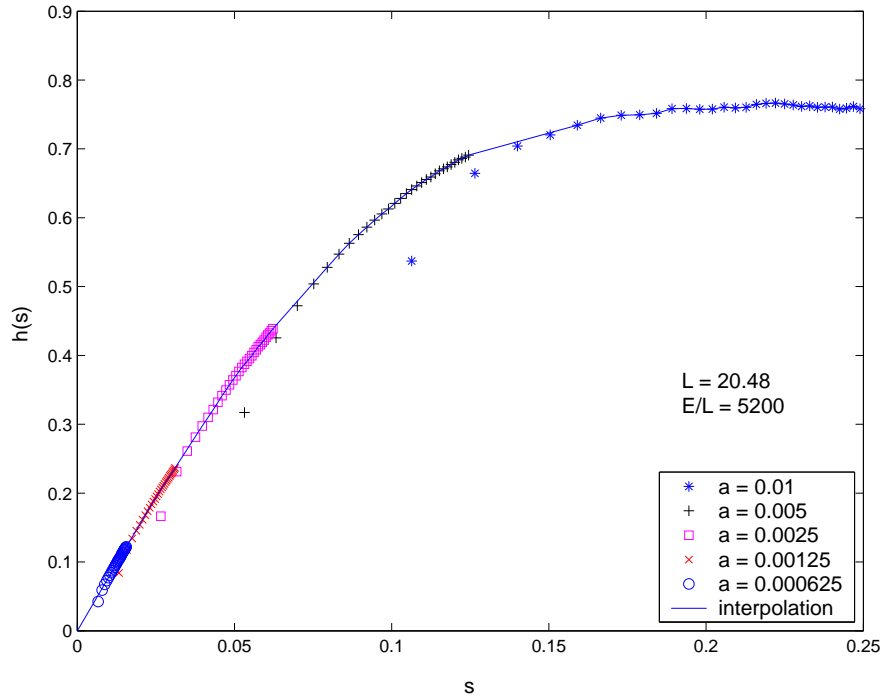}
  \caption{\label{hfun} Numerical estimate of the function $h(s)$
    obtained by interpolating the data of $a \; \bk(t)$ vs. $s=a \; t^{1/4}$.
    Deviations occur here when the time $t$ is not large enough compared to
    the averaging interval and are not included in the interpolation.  }
\end{figure}

\begin{figure}[ht]
  \includegraphics[height=100mm]{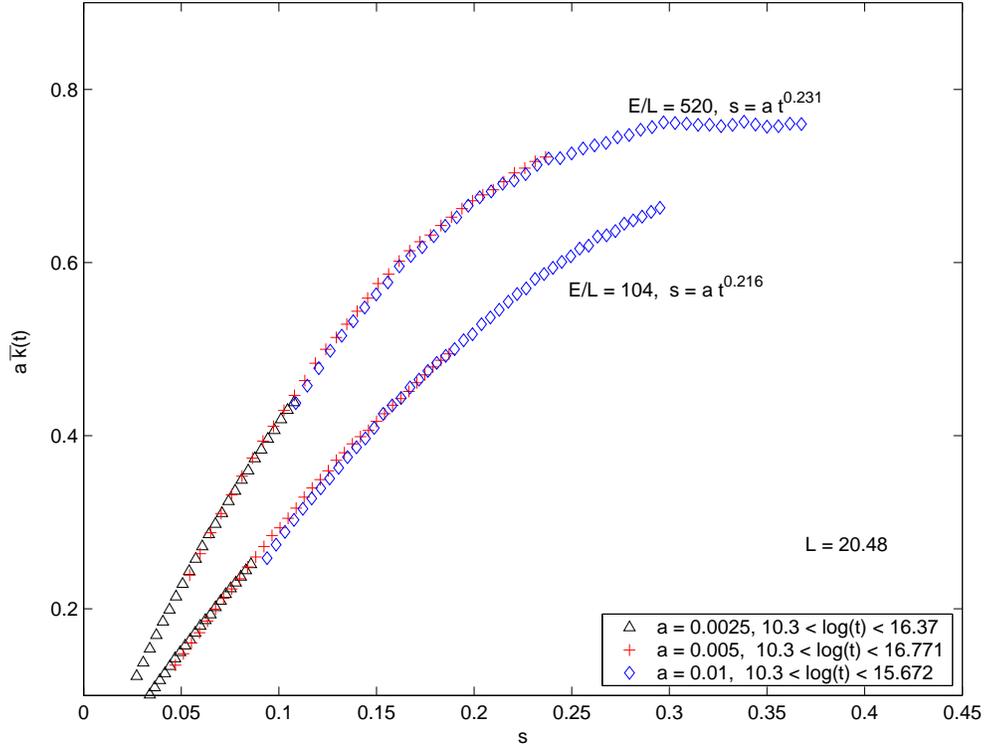}
  \caption{\label{hfun5} Data collapse over the scaling function $h(s)$ of
    $a\bk(t)$ for $E/L=520$ and $E/L=104$.}
\end{figure}

Proposing that, for very small lattice spacings and late times, $a
\; \bk(t)$ is solely a function of the combination
$a\,t^{\alpha}$, namely
\begin{equation}\label{bkfunc}
  \begin{split}
    a\,\bk(t) &= h(s) \;, \; s = a \; t^{\alpha} \;,\; a\to 0\;,\;
    t\to\infty\;, \\
    h(s) &\simeq h_1\, s \;,\quad s \to 0 \;;
  \end{split}
\end{equation}
we find numerically the function $h(s)$ displayed in fig.
\ref{hfun} for $E/L=5200$ and in fig. \ref{hfun5} for $E/L=520$
and $E/L=104$. These results show quite clearly that $\alpha$ does
depend, although quite weakly, on $E/L$. Actually, as explained in
the Appendix, this scaling--based approach is more effective in
the determination of $\alpha$ than direct fitting. Moreover, these
results show that $h(s)$ is not universal, but depends on the
initial conditions at least (and most likely only) through $E/L$.

\begin{figure}[ht]
  \includegraphics[height=80mm, width=120mm]{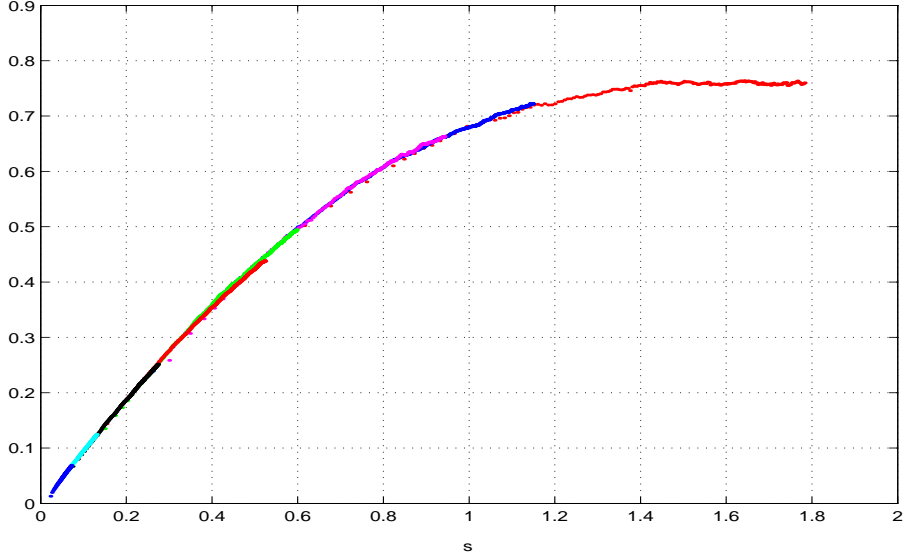}
  \caption{\label{htilfun} The double scaling function ${\tilde h}(s)$
    reconstructed from several set of data at different $E/L$ and $a$.}
\end{figure}

In fact, it turns out that the data for $h(s)$ all collapse very
well on a unique profile upon rescaling $s\to (E/L)^\gamma\,s$, if
$\gamma=0.25$. In other words, for the average wavenumber the
following double scaling form holds true when $E/L$ is large
enough
\begin{equation}\label{bkh1}
  \bk(t;\,a;\,E/L) \simeq a^{-1} \,{\tilde
  h}\Big(a\,t^\alpha\,\left(\frac{E}{L}\right)^\gamma\Big)\;,
\end{equation}
where ${\tilde h(s)}$ is now a universal function of order 1 with
${\tilde h(s)}\sim {\tilde h}_1 s$ as $s\to0$ and ${\tilde h}_1
\sim 1.1$ [our numerical reconstruction of ${\tilde h}(s)$ is
reported in fig. \ref{htilfun}]. Therefore, a fairly good
approximation for $ \bk(t) $ during the universal cascade ($t_1\gg
t\gg t_0$) is given by,
\begin{equation}\label{kbarap}
\bk(t) \simeq 1.1 \; \left(\frac{E}{L}\right)^{\frac14} \;
t^\alpha \; .
\end{equation}
with $ 0.21 < \alpha < 0.25 $ according to the value of $ E/L $
[see Table I].

\bigskip
The function ${\tilde h(s)}$ saturates for infinite time to the
value $\pi/4$ which translates into the maximum value for the
front wavevector $\bk_{max}=\pi/4a=\Lambda/2$. Actually ${\tilde
h}(s)$ in practice saturates already when $\gtrsim \pi/2$ (see
fig. \ref{htilfun}). Therefore we extract the \emph{thermalization
time scale}
\begin{equation}\label{t1}
  t_1 \simeq  \Lambda^{1/\alpha} \,(E/L)^{-\gamma/\alpha} \; ,
\end{equation}
at which $2\,\bk(t_1) \lesssim \Lambda$. For $t\gg t_1$ the power
spectrum $|\tilde{\pi}_k|^2$ features a plateau for all
wavevectors up to the cutoff, thus describing the thermal
equilibrium state.

As far as the shape function $F(u,v)$ is concerned, we see from
fig. \ref{sfun} that for $v\ll 1$ it features a bell-shaped
profile with exponentially small tails for large $u$,
\begin{equation*}
   F(u,v) \simeq  {\cal C} \; \exp[-\gamma(v)|u|] \;,\quad |u|\gg1 \;,
\end{equation*}
with $\gamma(0)\sim 2$ and ${\cal C}$ a constant. As $v$
increases, the tails decrease even faster, the lateral walls
steepen and the central region flattens. Finally, as $v\to1$,
which corresponds to complete thermalization, we have [see
eqs.\eqref{limpower} and \eqref{asyv}]
\begin{equation}\label{limsfun}
  \lim_{t\to\infty}P(k,t) = 2a \;, \quad
  F(u,1) =
\begin{cases}
  \pi/2 & |u| < 2 \\ 0 & |u| > 2
\end{cases} \;,
\end{equation}
to be compared to the approximation $F(u,v)\simeq F(u,0)$ valid in
the scaling window. Strictly speaking, the function $F(u,0)$ is
truly universal  in the continuous limit where $ v = \frac{4}{\pi}
\; \bk(t) \; a \to 0 $. However, one can see a very good
approximation to $F(u,0)$ in fig. \ref{sfun} for the plot at the
smallest $ a = 0.000625 $.

Finally, by construction, $F(u,v)$ is even in $u$ and must
satisfy, following eqs.(\ref{norP}) and (\ref{defkbar}),
\begin{equation*}
  \int_0^{2/v} du\, F(u,v) = \int_0^{2/v} du\, u\,F(u,v) =  \pi \;,
\end{equation*}
for any value of $v$.

Summarizing, after the transient stage, the effects of the initial
conditions on $P(k,t)$ are entirely accounted for by the average
wavenumber $\bk(t)$, which fixes the scale of the universal
ultraviolet cascade with shape described by $F(u,v)$. For $v\ll
1$, namely when the front of the cascade is far away from the
cutoff, the function $F(u,0)$ is {\bf universal} as shown
explicitly by the collapse of the data for several values of the
initial energy density onto one function in fig. \ref{cascade}.

\subsection{Time dependent effective temperature and local equilibrium}

The shape and universality of the power spectrum
$\overline{|\tilde{\pi}_k|^2(t)}$ featuring a flat plateau behind
the wake of the cascade leads to the definition of the effective
temperature $\Teff$ as the height of the plateau. The
interpretation of this temperature is that for $t\gg t_0$ but well
before the true thermalization time $t_1$, there is an description
in terms of \emph{local
  equilibrium} (local in Fourier space) at the effective temperature
$\Teff$.

In order to firmly state this conclusion, however, we must
understand if the other criteria for thermalization established in
section \ref{subsec:criteria} are fulfilled.

Combining eqs.\eqref{npws} and \eqref{sfun0} we can write the
$\pi$ power spectrum as
\begin{equation*}
  \overline{|{\tilde\pi}|^2}(k,t) = \frac{\overline{\pi^2(t)}}{\bk(t)} \;
   F\Big( \frac{k}{\bk(t)},\,\frac{2 \; \bk(t)}\Lambda \Big) \;.
\end{equation*}
For $t_0\ll t\ll t_1$, within the scaling window, this is well
approximated using eq.(\ref{bkh}) as
\begin{equation*}
  \overline{|{\tilde\pi}|^2}(k,t) \simeq
  \frac{\overline{\pi^2(t)}}{h_1 \; t^\alpha} \;
   F\Big( \frac{k}{h_1 \; t^\alpha},\,0 \Big) \;.
\end{equation*}
Now, for any fixed $k$ in the bulk of the cascade and large enough
$t$ (but still $t\ll t_1$), taking into account the flat profile
of $F(u,0)$ around $u=0$, we can write
\begin{equation*}
  \overline{|{\tilde\pi}|^2}(k,t) \simeq
  \frac{\overline{\pi^2(t)}}{h_1 \; t^\alpha} \,F(0,\,0) \equiv\Teff \;,
\end{equation*}
which is indeed $k-$independent. Moreover, the observed change in
time of the integrated power spectrum ${\overline{\pi^2}(t)}$ is
not as important [see figs. \ref{gfiT69NA001} and
\ref{gfiT1NA001}]. It slowly decreases to its asymptotic value
which agrees with the thermal equilibrium value $T/(2a)$ [see
eq.\eqref{pi2t}]. Hence, to leading order in $t$ (but always with
$t\ll t_1$) we may write:
\begin{equation}\label{Teff}
  \Teff \simeq \frac{E}L \; \frac{F(0,0)}{h_1\,t^\alpha} \;.
\end{equation}
Finally, using the large $E/L$ behavior of $h_1$ [see
eq.\eqref{bkh1}], we arrive at
\begin{equation}\label{Teff1}
  \Teff \simeq T_1\,\Big(\frac{E}L\Big)^{1-\gamma}t^{-\alpha} \;,
\end{equation}
The constant $T_1=F(0,0)/{\tilde h}_1$ is estimated to be
$T_1\simeq 1.55$, very close to $\pi/2$. We recall also that
$\alpha$ is weakly dependent on $E/L$ for $E/L \gtrsim 10$,
saturating approximately at $0.25$ for large $E/L$, and that
$1-\gamma\simeq 3/4$ quite precisely from the scaling argument of
eq.\eqref{bkh1}. Therefore, fairly good approximations for $ t_1 $
and $\Teff$ are given by,
\begin{equation}\label{t1yTef}
  t_1 \simeq \frac{L\,\Lambda ^4}{E} \; , \quad \Teff\simeq \frac\pi2
  \left(\frac{E}{L}\right)^{3/4} \; t^{-\alpha} \; .
\end{equation}

\bigskip

\begin{figure}[ht]
\includegraphics[height=100mm]{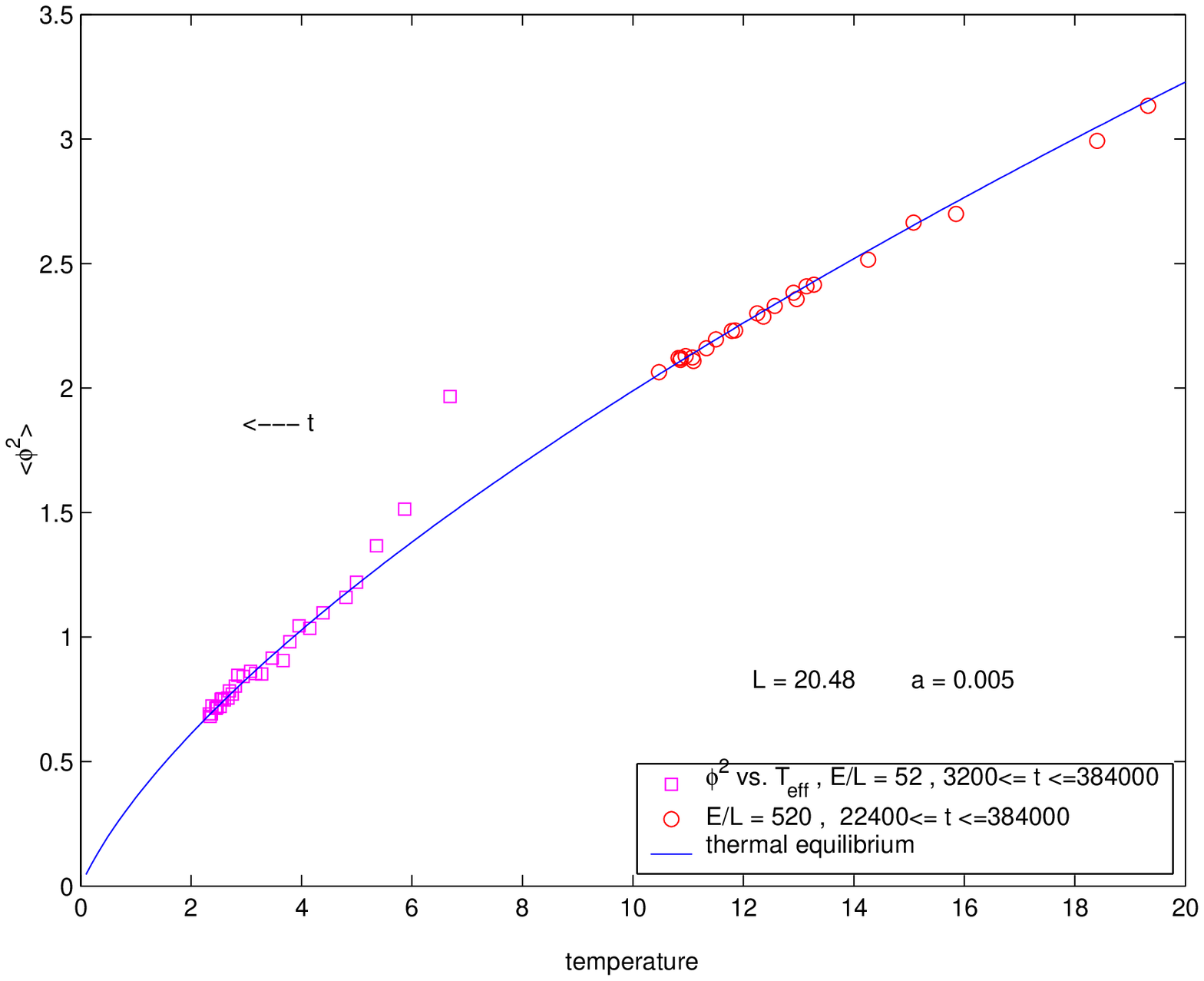}
\caption{\label{Teffphi2} Effective thermalization of
$\overline{\phi^2}(t)$. }
\end{figure}

\begin{figure}[ht]
\includegraphics[height=100mm]{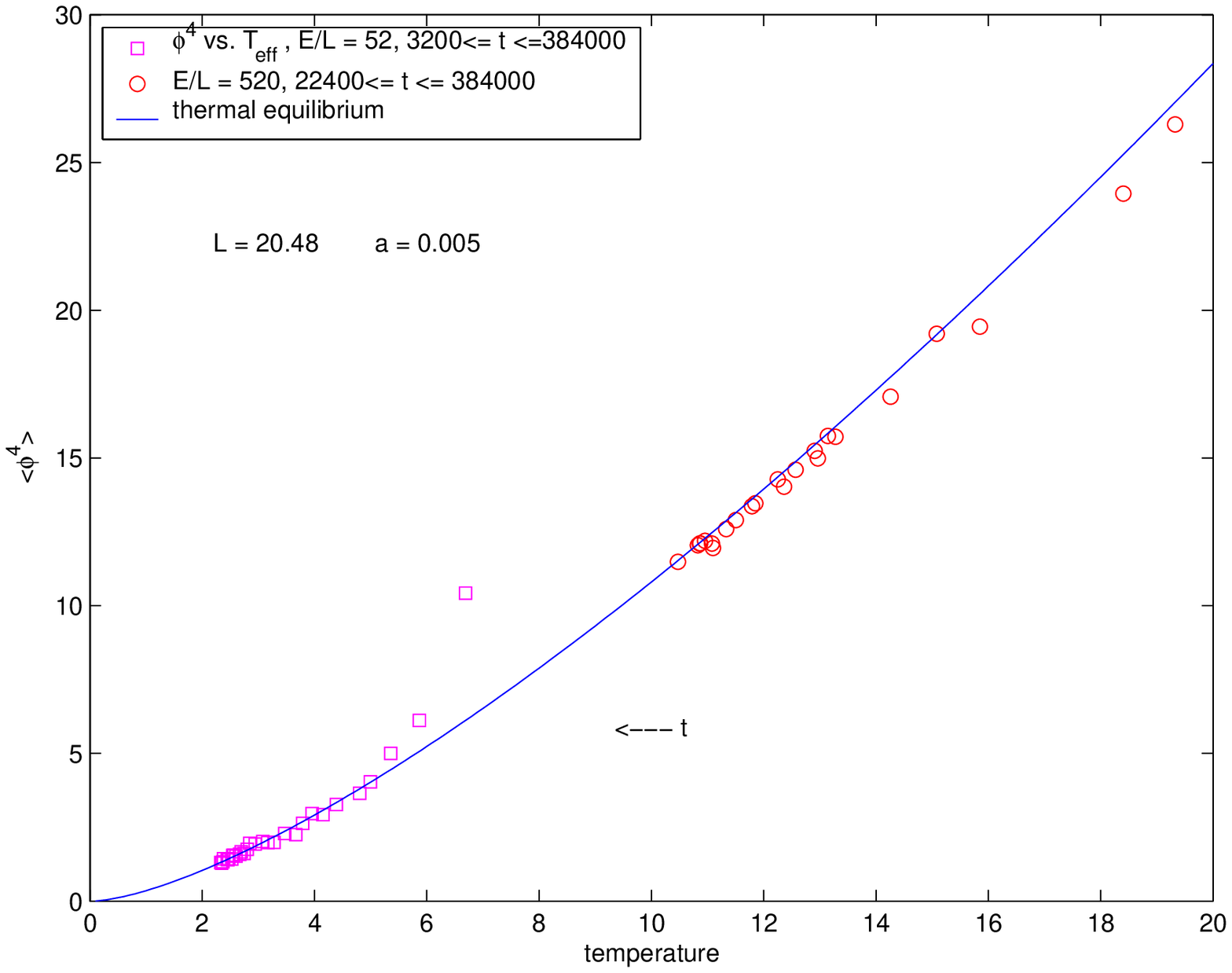}
\caption{\label{Teffphi4} Effective thermalization of
$\overline{\phi^4}(t)$.}
\end{figure}

The identification of the effective temperature allows us to study
whether the other observables, such as $\overline{\phi^2(t)}$,
$\overline{\phi^4(t)}$ or the correlation function
$\overline{\phi\phi}(x,t)$, depend on this effective temperature
in a manner consistent with local equilibrium. A numerical
analysis reveals that indeed to high accuracy we have [see figs.
\ref{Teffphi2} and \ref{Teffphi4}]
\begin{equation}\label{FTeff}
  \overline{\phi^2}(t) = \avg{\phi^2}(\Teff) \;, \quad
  \overline{\phi^4}(t) = \avg{\phi^4}(\Teff) \;,
\end{equation}
in terms of the equilibrium functions $\avg{\phi^2}(T)$ and
$\avg{\phi^4}(T)$ plotted in fig. \ref{equ1} early times (but
$t\gg t_0$), that is high effective temperature these reproduce
the thermal equilibrium results given by eq.(\ref{fi2fi4}), while
for late times and large UV cutoffs, that is low effective
temperature, these agree with the free field results
(\ref{f2f4fp})-(\ref{fft}).

\begin{figure}[ht]
\includegraphics[height=100mm]{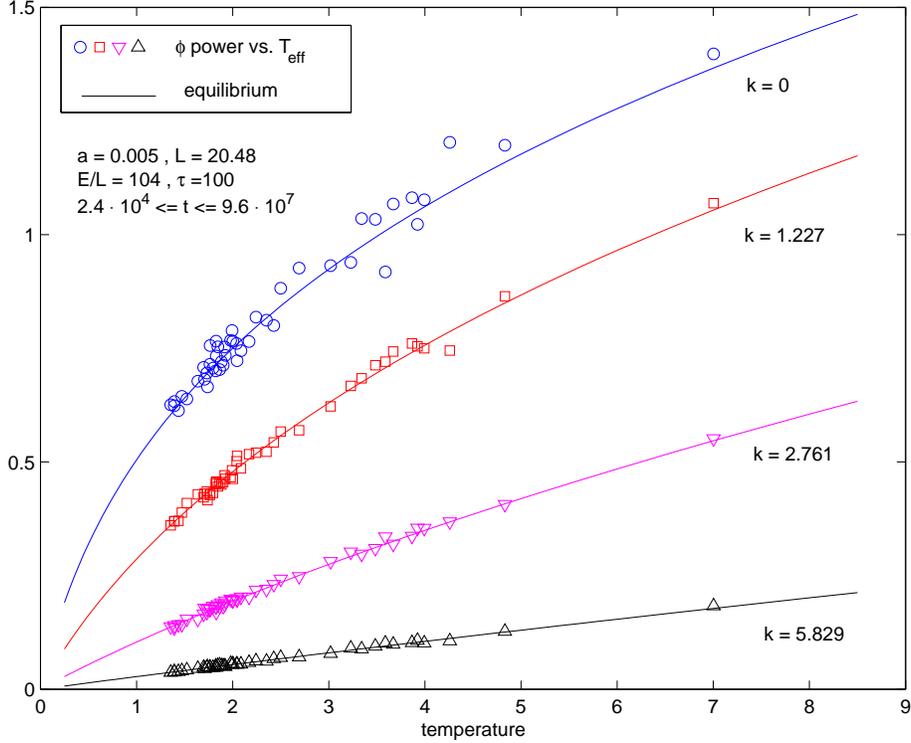}
 \caption{ \label{pw1fit} Effective thermalization of
   $\overline{|{\tilde\phi}|^2}(k,t)$ for different values of $k\ll
   2\,\bk(t)$ .}
\end{figure}

A more precise analysis can be performed on
$\overline{\phi\phi}(x,t)$, or better on its Fourier transform
$\overline{|{\tilde\phi}|^2}(k,t)$. Indeed, as evident from fig.
\ref{pw1fit}, we find that for $k \ll 2\,\bk(t) \ll \Lambda$ and
$t \gtrsim t_0$
\begin{equation}\label{GTeff}
   \overline{|{\tilde\phi}|^2}(k,t) \simeq {\tilde G}(k,\Teff) \;,
\end{equation}
in terms of the equilibrium distribution ${\tilde G}(k,T)$. For
$k$ too close to the front of the cascade $k < 2\,\bk(t)$
effective thermalization has not occurred yet, but this does not
affect significantly quantities like $\overline{\phi^2(t)}$ in
which all wavenumbers are summed over, since the equilibrium
${\tilde G}(k,T)$ vanishes as $T/k^2$ for large $k$. It instead
affects significantly quantities like $\overline{\phi'^2(t)}$,
since the $\phi$ equilibrium power spectrum $k^2 \; {\tilde
G}(k,T)$ goes to $T$ for large $k$.

The physical meaning and interpretation of the effective
temperature and the description in terms of local equilibrium at
this temperature behind the front of the ultraviolet cascade
applies solely to long wavelength physics. Namely the wavevectors
behind the front of the cascade $k<2 \; \bk(t)$ can be considered
to be thermalized by the mode mixing entailed by the interaction.
Physical observables that do not depend on the cutoff are
described in terms of this local thermal equilibrium concept.

This description in terms of a universal cascade in local thermal
equilibrium at an effective temperature $\Teff$ applies for $t\gg
t_0$ up to the   time $t_1$ at which the front of the cascade $2
\; \bk(t)$ reaches the cutoff. From this time onwards (during the
fourth stage, see sec. \ref{basico}) the power spectrum $\langle
|\tilde{\pi}_k|^2 \rangle$ is no longer universal and is sensitive
to the cutoff.

There is a clear separation between the time scale $ t_0 \sim
50000 $ where the cascade forms with the effective time-dependent
temperature $\Teff$ and the much longer time scale $ t_1 \propto
a^{-4} $ which signals that the front of the cascade is near the
cutoff and the end of the universal cascade.

Thermalization does continue beyond $t_1$ during the fourth stage
[see sec. \ref{basico}] and  the power spectrum $\langle
|\tilde{\pi}_k|^2 \rangle$ eventually becomes $T \; \Theta(\Lambda
-k)$ with $T$ the true equilibrium value of the temperature
(determined by the energy density) at infinite time. Thus true
thermalization takes an infinitely long time.

Hence the dynamics for long-wavelength phenomena can be described
to be in local thermodynamic equilibrium for $t>t_0$ at a time
dependent temperature $\Teff$ while short wavelength phenomena on
the scale of the cutoff will reach thermalization only at a much
later time scale $ t \gg t_1 $.

\bigskip

\begin{figure}[htbp]
\begin{turn}{-90}
\includegraphics[height=120mm]{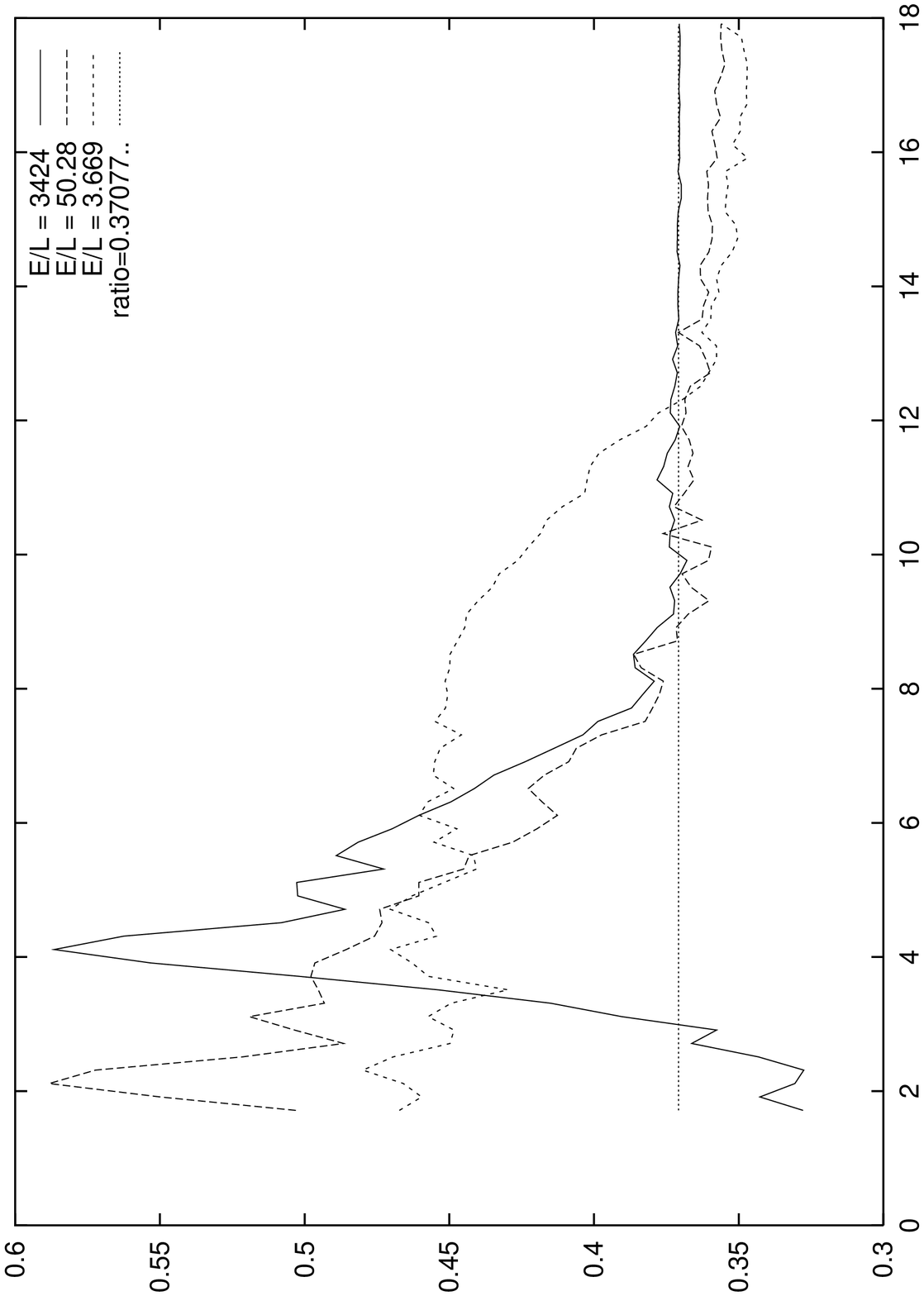}
\end{turn}
\caption{The ratio  $\frac{\avg{\phi^2}^2}{\avg{\phi^4}}(t)$ as a
   function of the   logarithm    of the  time $ t $ for $ E/L
   = 3424, \; 50.28 $ and $3.669$ for $L = 20 $ and $ a = 0.01 $. This
   ratio in thermal equilibrium takes values between the low
   temperature limit $1/3$ and the high temperature limit $
   0.37077\ldots$ depicted as a dotted line. Hence, effective
   thermalization  may  start only when the ratio falls within such
   strip [see eq.(\ref{cotar})]. For  $E/L= 3.669$  effective
   thermalization only starts for $ \ln t > 20 $.}
\label{ratio24}
\end{figure}

The ratio $ \frac{\avg{\phi^2}^2}{\avg{\phi^4}} $ plotted in
thermal equilibrium as a function of $T$ in fig. \ref{equ1}
provides a simple test of effective thermalization. This ratio
increases monotonically from its zero temperature value $ \frac 13
$ up to its infinite temperature limit  $0.37077\ldots $
[eq.(\ref{raz24})]. Therefore, a {\bf necessary} condition for
effective thermalization is that \be\label{cotar} \frac 13 \leq
\frac{[\overline{\phi^2}(t)]^2}{\overline{\phi^4}(t)} \leq
0.37077\ldots \ee We depict in fig.\ref{ratio24} the ratio
   $\frac{\avg{\phi^2}^2}{\avg{\phi^4}}(t)$ as a function of the
   logarithm of the time $ t $ for three values of $ E/L
   $. Effective thermalization  may  start only when the ratio falls within
the inequality (\ref{cotar}). We see from fig. \ref{ratio24} that
for $E/L \gtrsim 10 $ this happens around  $ \ln t \sim 9 $ which
is about the begining of the universal cascade [see sec.
\ref{basico}]. For  $E/L= 3.669 $  effective
   thermalization starts much {\bf later } around $ \ln t \sim 12
   $. More generally, we find for $E/L \lesssim 10 $
that thermalization is significantly delayed.

\section{ Conclusions and discussions}

In this article we have studied the approach to equilibrium in the
classical $\phi^4$ theory in $1+1$ dimensions with the goal to
understand the physical process that lead to thermalization and to
establish criteria for the identification of a thermalized state
in a strongly interacting theory. After discussing the classical
theory as the limit of the quantum field theory, we exploited the
equivalence of the classical partition function to the transfer
matrix for the quantum anharmonic oscillator. A body of
established results on the spectrum of the quantum anharmonic
oscillator allows us to obtain \emph{exact} results for low, high
and intermediate temperatures which furnish a yardstick and a set
of criteria to recognize thermalization from the physical
observables. We compared these exact results to those obtained for
the same quantities in the Hartree approximation and found that
this simple approximation describes the equilibrium properties of
the theory remarkably accurately, to within $10\%$ for most
observables. We point out that in the high temperature, or
equivalently the large energy density regime there is a {\bf
strong} renormalization of the single particle frequencies.

After studying the equilibrium properties, we presented a method
to solve the equations of motion based on the dynamics on a
light--cone lattice. This method is particularly suitable to study
the evolution in field theories within the setting of heavy ion
collisions where the initial dynamics is mainly along the light
cone, it is very stable  and conserves energy to  high accuracy.
The equations of motion for the classical field where solved with
a broad range of initial conditions and energy densities,
corresponding to microcanonical evolution. In all of these initial
conditions the initial energy density was stored in few (or a
narrow band) of long wavelength modes.

Our main results reveal several distinct stages of evolution with
different dynamical features: a first transient stage during which
there are strong fluctuations is dominated by the interaction
term, mode mixing begins to transfer energy from small to larger
wavevectors. A second stage is distinguished by the onset of a
very effective transfer of energy to larger wavevectors, this is
an {\bf energy cascade towards the ultraviolet}. During this
second stage the interaction term $\phi^4$ diminishes and
eventually becomes smaller than the spatio-temporal gradient
terms. At the end of this stage $\phi^4 \ll \phi'^2;\pi^2$. The
third stage is dominated by the ultraviolet cascade and the power
spectrum of the canonical momentum $|\tilde\pi_k|^2$ features a
{\bf universal scaling} form. The cascade is characterized by a
front $\bk(t)$ which slowly moves towards the ultraviolet cutoff
as $\bk(t) \simeq (E/L)^{\frac14} \; t^{\alpha}$ with $\alpha \sim
0.25$ an almost universal exponent independent of the lattice
spacing and the details of initial conditions but weakly dependent
on the energy density.  During this stage and while the front of
the cascade is far away from the cutoff, the power spectrum is
universal. Behind the front of the cascade the power spectrum is
that of thermal equilibrium with an effective temperature $\Teff$
which slowly decreases towards the equilibrium value. During this
stage we find that {\bf all observables} have the same functional
form as in thermal equilibrium but with the time dependent
effective temperature.  Namely the wake behind the front of the
ultraviolet cascade is a state of {\bf local thermodynamic
equilibrium}. This stage of universal cascade ends when the front
is near the cutoff, at a time scale $t_1 \simeq \frac{L}{E}
\;a^{-\frac{1}{\alpha}}$ with $a$ the lattice spacing.

The dynamics continues for $t>t_1$ but is no longer universal,
true thermalization is actually achieved in the infinite time
limit and occurs much slower than in any previous stage.

Thus we find that thermalization is a result of an energy cascade
with several distinct dynamical stages. Universality and scaling
of power spectra emerges in one of the later stages and local
thermodynamic equilibrium is associated with this stage. While we
find scaling and anomalous dynamical exponents we did  not find
any turbulent cascade.

We concur with previous studies that thermalization is indeed
achieved but on extremely long time scales, however our study
reveals that virialization starts to set  earlier than local
thermodynamic equilibrium.

It is important to  emphasize that the universal cascade described
above is different from a turbulent regime  described in
ref.\cite{tkachev} where the classical massless $\phi^4$ model was
studied in $3+1$ dimensions. In that reference, the spectrum for
the distribution of particles defined with respect to the Hartree
frequencies is studied and the numerical results are interpreted
in terms of wave turbulence and fit to a  Kolmogorov spectrum in
\cite{tkachev}.

We do not find a turbulent spectrum in the power spectra of the
field or its canonical momentum, instead we find local
thermodynamic equilibrium with an effective time dependent
temperature which diminishes slowly. While our study in $1+1$
dimensions reveals local thermal equilibrium in contrast to the
results of ref.\cite{tkachev} in the $3+1$ dimensional case, there
are several features in common: a)extremely long thermalization
time scales, as compared to the two natural time scales $m^{-1}$
or $T^{-1}$, b) a cascade of energy towards the ultraviolet and a
cascade front that evolves towards the cutoff as a function of
time. This latter feature can be gleaned in fig. (2) in
ref.\cite{tkachev}.

The theory of weak wave turbulence which describes the dynamical
evolution in terms of kinetic equations leads to a turbulent
spectrum for nonrelativistic and nonlinear wave equations as the
three wave equation\cite{lvov}. It was argued in ref.\cite{son}
that the non-equilibrium evolution in $\phi^4$ in $3+1$ dimensions
but in the small amplitude regime  features a scaling behavior and
a power law albeit different from the one with $\alpha \sim
0.21-0.25$ that we find in $1+1$ dimensions.

More recently\cite{newell}, a study of local wave turbulence for a
system dominated by four-wave interactions, again via wave kinetic
equations and following the methods of ref.\cite{lvov}, reports
the formation of cascades that feature a front that moves forward
in time, the wake behind the front features a Kolmogorov-Zakharov
spectrum of weak turbulence. The behavior of a moving front found
in ref.\cite{newell} is similar to the front of the cascade that
we find numerically, but the wake is different, we find local
thermodynamic equilibrium, while in ref.\cite{newell} the
distribution is of the Kolmogorov type.

While  a kinetic description may  not reliable during the first
stage of the dynamics, a suitable kinetic description in terms of
a distribution function for particles of the (strongly)
renormalized frequency may be available during the second and
third stages. We comment on these and other issues below.

\bigskip

\textbf{Discussions and comments: }

\bigskip

As discussed in section II, the classical limit must be understood
with a lattice (or ultraviolet) cutoff to avoid the Rayleigh-Jeans
divergence. For a finite energy density the naive continuum limit
leads to a vanishing temperature in equilibrium. Furthermore the
classical approximation is valid when occupation numbers are
large. In the cases under study for large energy density and
initial conditions for which the energy is stored in
long-wavelength modes, the classical approximation is valid and
reliable for small wavevectors. Our study reveals that the cascade
of energy proceeds very slowly after the first stage, thus
suggesting that the dynamics after the first stage could be
studied within a kinetic approach. However, such kinetic approach
should include the strong renormalization of the single particle
frequencies in order to define the slowly varying distribution
functions.

Another result of our study is that the simple Hartree
approximation is in  remarkably good agreement with the
\emph{exact} results in equilibrium.

Thus we conjecture that the following scenario for studying
thermalization in a \emph{quantum field theory} may prove suitable
to understanding the dynamics from initial conditions that entail
large occupations for long-wavevectors: study the numerical
evolution in the quantum Hartree approximation up to the time
scale at which the interaction and consequently the evolution
becomes on long time scales. At that stage read out the occupation
numbers for suitably defined single quasiparticles. Obtain the
kinetic equations for the evolution of these quasiparticle
occupations for a \emph{weakly} coupled theory, and continue the
evolution via these kinetic equations towards final equilibration
using the occupation numbers at the end of the Hartree evolution.

 This program, if proven suitable, is certainly much
more economical and efficient than trying to implement  more
sophisticated and calculational intensive approximation schemes
that lead to non-local update equations. This suggestion  is
analogous to that advocated in\cite{son2} for the evolution after
nucleus-nucleus collisions.

We highlight that we do not find a thermalization threshold as
found in the study of FPU chains\cite{FPU,varios}. Namely, even
for low values of the energy density the $\phi^4$ theory reaches
thermal equilibrium for very long times. Indeed, the
thermalization scale $ t_1 $ diverges in the $ a \to 0 $ limit
[see eq.(\ref{t1yTef})]. The fact that the thermalization slows
down when $\bk(t)$ approaches the UV cutoff (Brillouin zone) can
be understood as follows. The UV cascade can be understood as
follows. The equation of motion (\ref{eqnofmot}) for the Fourier
component of the classical field with wavevector $k$ features the
non-linearity $ \sum_{k_1} \sum_{k_2} \tilde{\phi}_{k_1} \;
\tilde{\phi}_{k_2} \; \tilde{\phi}_{k-k_1-k_2} $. Thus, if
initially the amplitude corresponding to a wavevector large $k$ is
small, the modes with small $k$ and large amplitude lead to a
large non-linearity that acts as a source for the amplitude with
$k$. In this manner, if the initial power spectrum is localized at
small wavevectors the amplitude of the higher  modes increases via
the non-linearities and the transfer of energy by mode mixing. By
periodicity the sum $k-k_1-k_2$ is constrained to be in the first
Brillouin zone. Therefore when $k-k_1-k_2$ is larger than the
cutoff, periodicity implies that this sum must be folded back
inside the first Brillouin zone by adding a vector in the
reciprocal lattice, namely $\pi/2a$. Hence once the amplitudes for
wavevectors near the cutoff are beginning to be large, the
non-linearity transfer the energy to low momentum modes.

Hence, the cascade mechanism is less efficient for $ k \lesssim
\Lambda $ than for $ k \ll \Lambda $.

While we have focused on the study of the dynamics in a classical
field theory with a cutoff as a timely  and interesting problem
all by itself, the applicability of our results to the realm of
quantum field theory descriptions of early Universe cosmology and
ultrarelativistic heavy ion collisions must be discussed.

Classical field theory is \emph{likely} to describe the QFT
non-equilibrium dynamics at least in the cases in which the
occupation numbers are very large. Strongly out of
equilibrium when a temperature cannot yet be defined, the
classical description is valid for the range of momenta
corresponding to large occupation numbers.
 Near equilibrium this would be the case for frequencies much smaller
 than the temperature. Our study reveals that effective thermalization
 is reached for times $ t > t_0 $  with a time
dependent effective temperature. Hence,
our classical results should apply to QFT for modes with
frequencies much smaller than  the effective temperature.

In order to extract a criterion for the validity of the classical
approximation we must revert to the physical variables
$T_{p}, \; a_{p}$ related to the rescaled variables used in this study
as in eqs. (\ref{adim})-(\ref{temp}), namely $T_p = m^3 \;
T/\lambda, \;  a_p= a/m, \; k_p = m \; k$. $T_p$ is the
 temperature in physical units and the physical momenta in the lattice
are restricted to be  $-\pi/a_p \leq k_p \leq \pi/a_p$ (first
Brillouin zone).

The exact equilibrium results obtained above indicate that the
exact equilibrium frequencies are of the form $\omega(k)=
\sqrt{k^2+ b \, T^{2/3}}$ with $b$ of order 1, where $T$ is the
dimensionless rescaled temperature $ T= \lambda \;  T_p/m^3 $. While in
the classical theory the mass and coupling constant can be
absorbed in an overall rescaling of the dimensionful parameters
and the amplitude of the fields, that is not the case in the
quantum theory where the amplitude of the fields and their
canonical momenta are fixed by the commutation relations. Thus a
comparison of the classical and quantum theory requires reverting
to the physical variables. Moreover, the dimensionless ratio
$\lambda/m^2$ plays a r\^ole in the discussion as we show below.

By following the numerical evolution of single modes it is easy to see that
for all modes in the bulk of the cascade, where local equilibrium holds,
the time-dependent dimensionless frequencies have the expected form
\begin{equation*}
  \omega_k(t)= \sqrt{k^2+ b \; \Teff^{2/3}}
\end{equation*}
We emphasize that the cascade corresponds to a
state of \emph{local equilibrium} with the effective temperature
$T_{eff}(t)$ which is \emph{much larger} than the final
thermodynamic equilibrium temperature.

A simple criterion for the validity of the classical
approximation in QFT is that
\begin{equation}\label{condQ}
\Tpeff \gg \omega_p(\bk_p(t))
\end{equation}
where $\omega_p(k_p)= \sqrt{k_p^2+ b \, [\lambda \; \Tpeff]^{2/3}}$
are the dimensionful frequencies.  In dimensionless variables
eq.\eqref{condQ} takes the form,
\begin{equation}\label{conQd}
  \frac{m^3}\lambda \Teff \gg  m\,\sqrt{\bk(t)^2+ b \; \Teff^{2/3}}
\end{equation}
It is enough to discuss the modes at the wake to see the applicability of
the classical approximation since $ k < \bk(t) $ for the modes in the
cascade. At times early enough, in regimes such that the plateau is already
formed but $\bk(t)\ll \Teff^{1/3}$, eq.(\ref{condQ}) holds provided $ \Teff
\gg (\lambda/m^2)^{3/2}$, which is certainly true for $\lambda$ small
enough.

On the other hand if $\bk(t)\gg \Teff^{1/3}$ the classical approximation
will be valid provided
\begin{equation}\label{condQ2}
  \frac{m^2}\lambda \; \Teff \gg \bk(t)
\end{equation}
Using our late time estimates eqs. \eqref{kbarap}, \eqref{t1yTef}
for $\alpha=0.25$ one obtains
the condition
\begin{equation}\label{condQ3}
  t \ll t_Q \;,\quad t_Q \sim \frac{m^4\,E}{\lambda^2\; L} =
  \frac{E_p}{\lambda \, L_p}
\end{equation}
This shows that for sufficiently small coupling constant and fixed UV
cutoff $ \Lambda = \pi/a_p$ the classical approximation could be valid
up to full
thermalization, that is $t_Q\gg t_1$, which using the first of eq.
\eqref{t1yTef} is equivalent to the condition
\begin{equation*}
  \frac{E}L \gg \frac{\lambda}{m^2 \; a^2}
\end{equation*}
Recalling that $E/L\simeq T/(2a)$, this is nothing but $T_p\, a_p \gg 1$.
In fact, as the front of the cascade moves towards the cutoff, for a fixed
cutoff, the classical approximation will continue to be valid for
all modes behind the wake of the cascade if $ T_p \; a_p \gg 1$.

The validity of the classical approximation for $T_p\, a_p \gg 1$
is well known in the elementary theory of harmonic solids where
phonons are the elementary excitations. In the Debye model of
solids the classical limit is valid for $T_p>T_D$ where $T_D
\propto \pi/a_p$ is the Debye Temperature (and $\pi/a$ is the edge
of the first Brillouin zone, namely the maximum
momentum)\cite{solid}.

As per the discussion in section \ref{subsec:expec} the
non-linearities are \emph{subdominant} for $Ta^3 \ll 1$, condition
which all but guarantees the cascade [see discussion in section
\ref{subsec:expec}]. Therefore, the validity of the classical
approximation is warranted provided,
\begin{equation}\label{condQ4}
T_p\,a_p \gg 1 \quad \mbox{or ~equivalently} \quad T \, a  \gg
\frac{\lambda}{m^2} \; ,
\end{equation}
which is the same condition for the validity of the classical
limit in elastic solids.

Since our results are valid for $T \; a^3\ll 1$ the classical results
obtained here apply in the quantum theory  for a \emph{fixed cutoff}
in the whole dynamical range where eq.(\ref{condQ2}) is valid.

We have studied the dynamical evolution in a broad range of parameters from
$0.001 \leq Ta \lesssim 1$ (with $Ta^3 \ll 1$) and found a similar behavior
for the dynamics during the stage of the universal cascade. Therefore, our
classical results applies in QFT {\bf provided} one chooses values of $
\lambda/m^2 $ fulfilling eq.(\ref{condQ4}) for the values of $ T a $
indicated in the respective plots. Indeed, our \emph{classical} results
turn to be relevant for the weakly coupled \emph{quantum field theory}.
Otherwise, quantum effects certainly appear in the process of
thermalization\cite{berges,serreau}. Nonetheless, even in the continuum
limit at fixed coupling constant, when inevitably $T_p\,a_p\to0$, it is
conceivable that, as long as $t\ll t_Q$, the classical approximation is
quite accurate for all modes in the bulk of the cascade.  One should expect
that the shape of the cascade at the forefront indeed depends sensibly on
quantum effects, since occupation numbers fall off rapidly across it. But
the rate at which the plateau behind decreases is most likely almost purely
classical, at least until occupation numbers are high, that is for $t\ll
t_Q$, when the effective temperature is high enough.

Hence, in the quantum field theory with initial conditions
strongly out of equilibrium with a power spectrum featuring large
occupations for long wavelengths we expect that the initial stages
of the dynamics will \emph{likely} be well described by the
classical field theory. In this case we expect a cascade to form,
the modes behind the front will be well described by the classical
field theory, whereas in the theory with a large cutoff, the modes
with small occupation in front of the cascade all the way to the
cutoff will require a quantum description. The backreaction of
these quantum modes onto the long-wavelength modes will modify the
time evolution of the cascade.

In the quantum $\phi^4$ theory we expect that the
time scale for complete thermalization to be finite even in the continuum
limit. While in the classical limit the equilibrium power spectrum
for the field falls off with a power law and that of the canonical
momentum is flat, hence sensitive to the cutoff, in the quantum
theory the exponential suppression will make the theory
insensitive to the cutoff in the limit $ T\; a \ll 1$.   Hence,
we conjecture that time scale for complete thermalization diverges in the
continuum quantum theory when $ \hbar \to 0 $. Furthermore, we
expect faster thermalization in the quantum theory since zero
point fluctuations imply finite amplitude for all modes in the
initial state.

The numerical studies previously reported on the dynamics of
classical and quantum field theories \cite{patkos}-\cite{marcelo}
have not yet focused on studying the mechanism of energy transfer
from long to short wavelengths as a function of time, although
ref.\cite{marcelo} reported the emergence of spatio-temporal
structures. It would indeed be interesting to study if the
universal cascade found in the classical theory remains at least
during some early and intermediate stages in the quantum theory.

Our study in conjunction with a previous body of work clearly
suggests that thermalization, at least in this scalar theory,
occurs on extremely long time scales, much longer than the natural
scales $m^{-1}\,,\,T^{-1}$. However,  local thermodynamic
equilibrium  with a slowly decreasing effective temperature $
\Teff$ emerges relatively soon. This result all by itself raises
important questions on the current assumptions on thermalization
both in cosmology as well as in heavy ion collisions. If
thermalization involves a time dependent effective temperature,
both in cosmology as well as in the formation and evolution of the
quark gluon plasma, there are immediate phenomenological
consequences and a reassessment of the current assumptions and
models is called for.

An intense program to study these issues and their impact on both
fields is therefore warranted.

\section{Appendix}

We provide here more details on our numerical calculations. They
were performed on a cluster of personal computers at LPTHE-Paris
and at the Physics Department of Milano--Bicocca. The computers
were used in parallel with different values of relevant parameters
on each one of them.

\begin{figure}[ht]
\includegraphics[height=100mm]{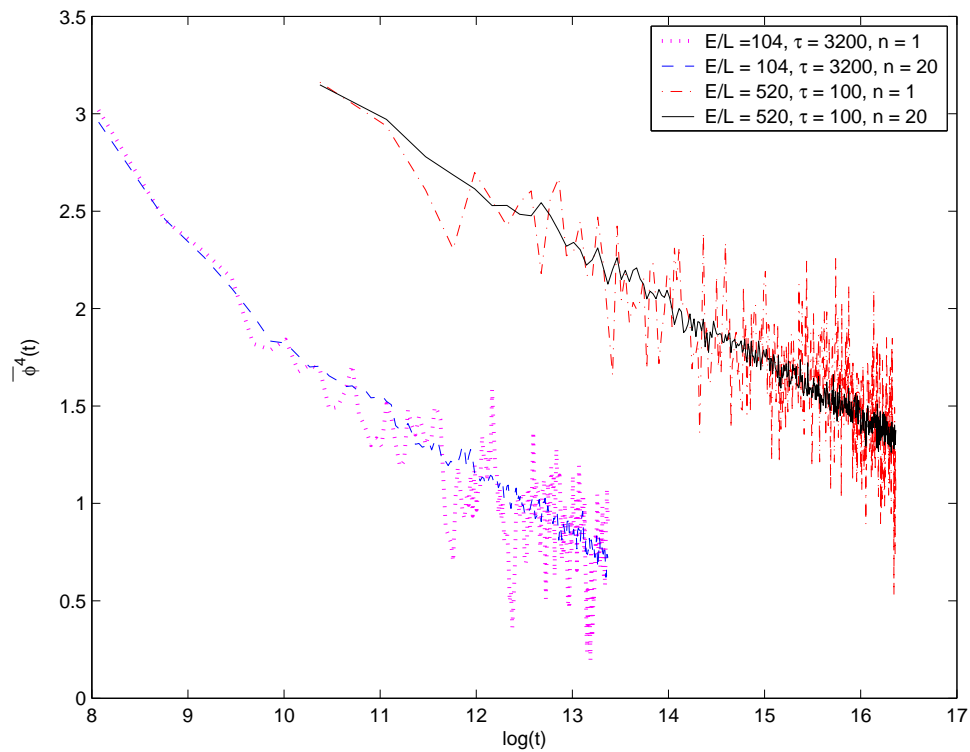}
\caption{\label{fluct} Comparison of different averagin methods.
$n$ is the
  number of initial conditions in each average.}
\end{figure}

In our numerical evolution the number of iterations of the
discrete system, eq.\eqref{evod}, is $t_{\rm max}/(2 \,a)$ when
evolving from $ t = 0 $ up to $ t = t_{\rm max} $. The computer
time increases for smaller spacing also because at fixed size $L$,
smaller $a$ implies larger $N=L/(2a)$. This implies a number of
operations of order $a^{-2}$. Moreover, if fast Fourier transforms
are performed every iteration, to average the power spectra over
the time intervals $[t,t+\tau]$ covering completely the time span
from $ t = 0 $ up to  $ t = t_{\rm max} $, then altogether the
computer time scales as $ a^{-2} \log\frac{1}{a} $ for small $a$.
Actually, when also averages over initial conditions are
performed, it turns out that $\tau$ need not be so large as to
provide a complete covering, since the average over initial
conditions is much more effective in reducing fluctuations than
averaging in time [see fig \ref{fluct}]. In other words, it is
sufficient to consider time intervals $[t,t+\tau]$ that cover only
a relatively small part of the full time span from $ t = 0 $ until
$ t = t_{\rm max}$. This implies that computer time really scales
just as $ a^{-2}$ for small $a$.

\medskip

As pointed out in sec. \ref{correti} we have two methods to
compute $|{\tilde\phi}_k(t)|^2$. Either, to extract the field
$\phi(x,t)$ from the lattice fields $F(n,s)$ and $G(n,s)$,
Fourier--transform it to ${\tilde\phi}_k(t)$ and then perform all
needed averages on $|{\tilde\phi}_k(t)|^2$. Or to directly compute
the averages of the correlations of $F(n,s)$ and $G(n,s)$ and
extract from them the correlations $\overline{\phi\phi}(x,t)$ and
$\overline{\pi\pi}(x,t)$. Thanks to the exceptional scaling
properties of the Fast Fourier Transform, the first method is
actually preferable to the second, if the space average in
eq.\eqref{phiphi} is really performed and very extensive time
averages are considered. However, as just explained, the time
averages can be greatly reduced when averaging on numerous enough
initial conditions.

\medskip

We have seen in the previous sections that the mean wavenumber
$\bk(t)$ plays a central role in the description of the UV
cascade. However, the numerical determination of the exponent
$\alpha$ and of the amplitude $h_1$ for the time behaviour of
$\bk(t)$ in the scaling window, eq.\eqref{bkh} or, even more
generally, in the determination of scaling function $h(s)$ in
eq.\eqref{bkh} is a nontrivial matter. This is due to the
characteristic fluctuations of the data that limit the control of
the sistematic effects caused by the initial transient, by the
scaling violations due to the UV cutoff and by the averaging over
time intervals when the time is not much larger than interval
length $\tau$. This type of fluctuations can be systematically
reduced by enlarging the set of initial conditions over which we
average, but are still present in a significant way for the number
(typically 20 to 30) of initial configurations we used.

\begin{figure}[ht]
\includegraphics[height=100mm]{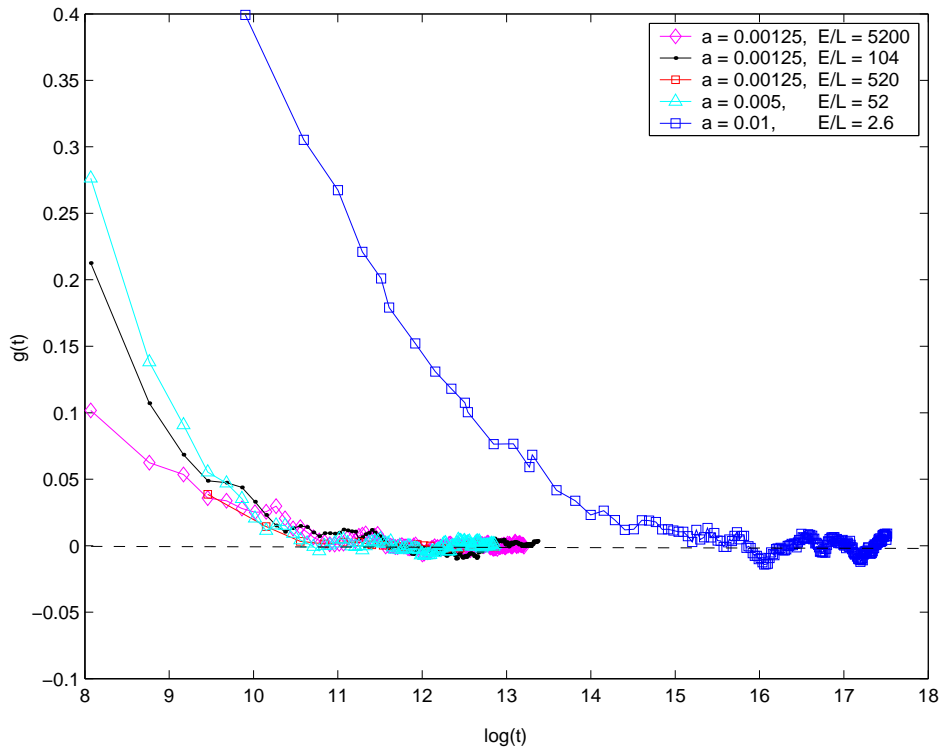}
\caption{\label{gfun} Numerical determinations of the function
$g(t)$ that describes the initial transient for several values of
$E/L$.}
\end{figure}

We identify the initial transient in $\bk(t)$ by a
self--consistent fitting procedure. Suppose a very good fit has
been obtained for the data points at times larger than a certain
time $\overline{t}$ (the actual fitting procedure is described
below). Then we can see if and how much the data at
$t<\overline{t}$ deviate from the fitting function. If the fit
starts  to worsen immediately before $\overline{t}$ and in a
steady manner towards $t=0$, then the transient cannot be fully
disentangled from the fitting errors and $\overline{t}$ needs to
be increased until we find that the fit remains quite good for a
while before $\overline{t}$ and then worsen in rather abrupt way.
The time at which this worsening occurs provides an alternative
estimate of the time scale $t_0$ introduced in sec. \ref{basico}
to signal the end of the transient and the start of the UV cascade
that will become universal for $t\gg t_0$. Using the
parametrization
\begin{equation*}
  \bk(t) = h_1\, t^\alpha \,[1 + g(t)]\;,
\end{equation*}
can then (approximatively) read $t_0$ out of the plot of $g(t)$
vs. $t$ [see fig \ref{gfun}]. We see that $\log t_0\lesssim 11$
(in agreement with what found in sec. \ref{basico} for other
observables) more or less independently on $E/L$ as long as
$E/L\ge 52$. When $E/L=2.6$, $t_0$ is much larger instead, $\log
t_0 \sim 15$, as found also for the other observables. Notice
that, when $E/L=2.6$, $\bk(t)$ is very well fitted for
$\overline{t} < t < t_{\rm max}$, with $\log\overline{t}\sim 14.5$
and by $\log t_{\rm max}\sim 17.5$ by the pure power law
\eqref{bkh} of the scaling window, but with a power $\alpha\sim
0.186$ definitely smaller than those at higher energy densities.
\medskip

Having disposed of the initial transient, we now turn to the
fitting procedure to determine $h_1$ and $\alpha$. As a matter of
fact, the simple, direct fitting of the numerical data of average
wavenumber $\bk(t)$ with eq.\eqref{bkh} is not quite precise, even
when the scaling window is identified. For instance, the almost
linear behaviour in the log--log scale for smaller values of the
spacing $a$ is evident from fig. \ref{bark}, but the actual values
of $\alpha$ and of $h_1$ are poorly fixed in this simple fitting
approach.

To give an idea of the situation, let us consider a specific
example. In a parallel run of 20 evolutions from $t=0$ to
$t=1.28\ldots \; 10^7$, with $a=0.0025$ and $E/L=520$, each one
starting from the same initial wavenumber $k_i$'s filled, but with
randomly chosen amplitude and phases [according to the discussion
in sec. \ref{incond}], a $20\times 400$ matrix $M$ of data for
$\bk(t)$ is obtained.  Each row of $M$ corresponds to a given
choice of initial conditions and the $n-$th column contains the
values of the mean wavenumber at $t=n*\Delta t$ for
$n=1,2,\ldots,400$ and $\Delta t=32000$. In this runs the time
average intervals are rather small ($\tau=100$) compared to the
data taking interval $\Delta t$ and so they cause a negligible
sistematic error. On the other hand the effects of the finite UV
cutoff are not so small [$\log \bk(t) $ bends downward as a
function of $\log t $] and this facts suggests to try a simple
least squares fit with
\begin{equation}\label{dev2}
  \bk(t) \simeq h_1  \; t^\alpha + h_2 \;  a \; t^{2\alpha} \; ,
\end{equation}
using $h_1$, $h_2$ and $\alpha$ as free fitting parameters. When
the fit is performed on the mean wavenumber extracted from the
power spectrum $\overline{|{\tilde\pi}|^2}(k,t)$ averaged over the
initial conditions, we obtain
\begin{equation}\label{fit1}
  h_1 = 4.3942 \;,\quad   h_2 = -7.4101 \;,\quad  \alpha=0.2401 \;.
\end{equation}
This differs very slightly from the result of the fit over the
column--averaged $M$, that is
\begin{equation}\label{fit2}
  h_1 = 4.4200 \;,\quad  h_2 = -7.4734 \;,\quad  \alpha=0.2397 \;.
\end{equation}
since $\bk(t)$ depends in a weak nonlinear way on
$\overline{|{\tilde\pi}|^2}(k,t)$. However, to give an estimate of
the statistical error on this figures, we should repeat the
procedure on each row of $M$ separately. This yields
\begin{equation}\label{fit3}
  h_1 = 4.7660 \pm 1.7855 \;,\quad h_2 = -1.9006 \pm 11.3948 \;,\quad
  \alpha = 0.2373\pm 0.0337 \;.
\end{equation}
Of course the mean values here do not coincide with the values in
either eq.\eqref{fit1} or eq.\eqref{fit2} since the fitting
parameters depend very nonlinearly on the data. The annoying point
however is the large statistical error on this figures, which
significantly exceeds the differences in the mean values. In
principle, relying on the standard error reductions in the mean of
independent random variables, we could assign error bars
eq.\eqref{fit1} or eq.\eqref{fit2} smaller by a factor
$1/\sqrt{20}$ with respect to those in eq.\eqref{fit3}. However,
at the same time the confidence level of each separate fit on the
rows of $M$ is much lower than in that yielding the estimates in
eq.\eqref{fit1} or eq.\eqref{fit2}, since the larger fluctuations
(more or less by the inverse factor $\sqrt{20}$) make much larger
the least squares of the fit. Alltogether this entail a relative
error ranging from $10\%$ for $h_1$ and $5\%$ for $\alpha$ (with
the most optimistic attitude) to $40\%$ for $h_1$ and $20\%$ for
$\alpha$ in a more conservative approach. The situations for $h_2$
is much worse, but it does not have the same relevance of $h_1$
and $\alpha$ in the continuum theory.

Interestingly enough, it turns out that the full application of
scaling hypothesis of eq.\eqref{bkh} is more efficient that direct
fitting in pinpointing the value of $h_i$ and especially $\alpha$.
That is, the numerical data of $a \; \bk(t)$, for different values
of $a$ at fixed $E/L$, collapse indeed very neatly on a single
curve when plotted versus $s=a \; t^\alpha$ (thus verifying the
scaling idea), if $\alpha$ has a specific value. Small changes of
$\alpha$, of relative order $10^{-2}$, spoil the collapse in a
evident and easily quantifiable fashion. In this way we determined
with few percent accuracy the values $\alpha=0.222$ for $E/L=104$,
$\alpha=0.231$ for $E/L=520$ and $\alpha=0.25$ for $E/L=5200$ [see
figs. \ref{hfun} and \ref{hfun5}]. Next, with $\alpha$ fixed, we
can repeat the fit with the expansion of $a \; \bk(t)$ in powers
of $a \; t^\alpha$ as in eqs.\eqref{dev2}, obtaining a much better
determination for $h_1$ and often also for $h_2$ [see table
\ref{htable}].  In particular, when the fit with fixed
$\alpha=0.231$ is considered in the example discussed above for
each row of the matrix $M$ separately, we obtain
\begin{equation*}
  h_1 = 4.98167 \pm 0.1389 \;,\quad h_2 = -8.3544 \pm 1.5271 \;,
\end{equation*}
to be compared with the fit in eq.\eqref{fit3} when $\alpha$ is a
fitting parameter.

We see that the parameters $h_i$ and $\alpha$ do depend on the
energy density $E/L=520$. $\alpha$ grows from $\alpha\sim 0.186$
when $E/L=2.6$ to $\alpha\sim 0.25$ when $E/L=5200$. We did not
attempt an analytic fit of $\alpha(E/L)$, but it does seem to
saturate fast at higher energy densities.

\bigskip
\vskip 2truecm
\begin{center}
\begin{table}
\setlength{\extrarowheight}{4pt}
\begin{tabular}{|c||c|c|}
\hline
 &\multicolumn{2}{c|}{$E/L = 104$, $\alpha=0.222$ }   \\\cline{2-3}
\raisebox{1.5ex}[0cm][0cm]{$a$}
 & $h_1$  & $h_2$  \\
\hline\hline
 0.01 & ~~3.167~~   & ~~$-3.393$~~  \\
\hline
 0.005 & 3.104  & $-3.305$ \\
\hline
 0.0025 & 2.921  & *   \\
\hline
 0.00125 & 2.945  & * \\
\hline
 0.000625 & 2.962  & *  \\
\hline
\end{tabular}
\begin{tabular}{|c|c|c|}
\hline
 \multicolumn{3}{|c|}{$E/L = 520$, $\alpha=0.231$}   \\\cline{1-3}
 $h_1$  & $h_2$ & $h_3$  \\
\hline\hline
 ~~5.084~~   & ~~$-8.116$~~  & ~~$-6.358$~~ \\
\hline
 5.120  & $-8.837$  & *  \\
\hline
 4.980  & $-8.331$  & *  \\
\hline
 4.641  & * & *  \\
\hline
 4.617  & * & *  \\
\hline
\end{tabular}
\begin{tabular}{|c|c|c|}
\hline
 \multicolumn{3}{|c|}{$E/L = 5200$, $\alpha=0.25$} \\\cline{1-3}
\hline
 $h_1$  & $h_2$ & $h_3$  \\
\hline\hline
 ~~9.094~~  & ~~$-35.27$~~  & ~~44.11~~  \\
\hline
 8.922  & $-27.37$  & *  \\
\hline
 8.726  & $-26.60$  & *  \\
\hline
 8.549  & $-29.65$ & * \\
\hline
 8.223  & * & *  \\
\hline
\end{tabular}
\caption{\label{htable} Numerical fits for the expansion
coefficients of
 the scaling function $h(s)$. Relative statistical errors on the data are
 of the size of few percent on $h_1$, but unavoidably larger for $h_2$ or
 $h_3$, since these correspond to corrections that are too small in
 most cases. When
 $h_2$ or $h_3$ are not included in the fit, an asterisk is shown.
 Another source of (sistematic) errors is the approximate validity of the
 scaling hypothesis eq.\eqref{bkh} for $a$ not small enough.}
\end{table}
\end{center}

{\bf Acknowledgements:} D. B. thanks the N.S.F. for partial
support through grants PHY-0242134, and the hospitality of LPTHE
where part of this work was carried out. C.  D. thanks the LPTHE
for the warm hospitality. H. J. d. V. thanks the Department of
Physics and Astronomy at the University of Pittsburgh for their
hospitality.

\end{document}